\documentclass[notitlepage,twocolumn]{report}
\usepackage{balance}

\usepackage{geometry}
\usepackage{rotating}
\usepackage{relsize}
\DeclareRobustCommand\V[1]{\ifmmode\mathbf{#1}\else$\mathbf{#1}$\fi}

\newcommand\hac[1]{\acsfont{#1}\xspace}
\newcommand*\acsfont[1]{\textsmaller{\scshape#1}}
\DeclareRobustCommand\SI{\hac{S1}}
\DeclareRobustCommand\SII{\hac{S2}}
\DeclareRobustCommand\SIII{\hac{S3}}
\DeclareRobustCommand\SIV{\hac{S4}}
\DeclareRobustCommand\SV{\hac{S5}}
\DeclareRobustCommand\SVI{\hac{S6}}
\DeclareRobustCommand\VS{\hac{VS}}

\usepackage{paper}

\renewcommand\secshow[3]{\acsfont{#3,~#2~\ref{#1}}}
\figbase{eps}
\seohbase{seoh-2003}

\let\Kvi=\Kv
\let\Cavi=\Cav
\let\Ki=\K
\let\Cai=\Ca
\let\Nai=\Na
\let\Hvi=\Hv
\renewcommand\Kv{\hac\Kvi}
\renewcommand\Cav{\hac\Cavi}

\renewcommand\Hv{\hac{\Hvi}}
\renewcommand\KvAP{\Kv{}\hac{AP}}
\newcommand\LU{\hac{LU}}
\renewcommand\K{\hac{\Ki}}
\renewcommand\Ca{\hac{\Cai}}
\renewcommand\Na{\hac{\Nai}}

\title{
  Theoretical Studies\\
  of Structure-Function Relationships\\
  in K$_{\text{V}}$ Channels:\\
  Electrostatics of the Voltage Sensor
}

\author{Alexander Peyser}

\usepackage{hyperref}
\begin{document}
\onecolumn
\maketitle

\begin{abstract}
  \parindent=0pt
  \parskip=\baselineskip
  \addtolength\parskip{0pt plus1pt minus1pt}

  \noindent Voltage-gated ion channels mediate electrical excitability
  of cellular membranes. Reduced models of the voltage sensor (\VS) of
  \Kv channels produce insight into the electrostatic physics
  underlying the response of the highly positively charged \SIV
  transmembrane domain to changes in membrane potential and other
  electrostatic parameters. By calculating the partition function
  computed from the electrostatic energy over translational and/or
  rotational degrees of freedom, I compute expectations of charge
  displacement, energetics, probability distributions of translation
  \& rotation and Maxwell stress for arrangements of \SIV positively
  charged residues and \SII \& \SIII negatively charged
  counter-charges; these computations can then be compared with
  experimental results to elucidate the role of various putative
  atomic level features of the \VS.

  A `paddle' model \citep{jiang:2003:xray} is rejected on
  electrostatic grounds, owing to unfavorable energetics, insufficient
  charge displacement and excessive Maxwell stress. On the
  other hand, a `sliding helix' model \citep{catterall:1986} with
  three local counter-charges, a protein dielectric coefficient of 4
  and a 2/3 interval of counter-charge positioning relative to the
  \SIV $\alpha$¬helix period of positive residues is electrostatically
  reasonable, comparing well with \emph{Shaker}
  \citep{seoh:1996}. Lack of counter-charges destabilizes the \SIV in
  the membrane; counter-charge interval helps determine the number and
  shape of energy barriers and troughs over the range of motion of the
  \SIV; and the local dielectric coefficient of the protein (\SII,
  \SIII \& \SIV) constrains the height of energy maxima relative to
  the energy troughs.

  These `sliding helix' models compare favorably with experimental
  results for single \& double mutant charge experiments on
  \emph{Shaker} by \citet{seoh:1996}. Single \SIV positive charge
  mutants are predicted quite well by this model; single \SII or \SIII
  negative counter-charge mutants are predicted less well; and double
  mutants for both an \SIV charge and an \SII or \SIII
  counter-charge are characterized least well by these electrostatic
  models (which do not include gating load, unlike their biological
  analogs). Further computational and experimental investigation of
  \SII \& \SIII counter-charge structure for voltage-gated ion
  channels is warranted.
\end{abstract}
\vspace{\stretch{1000}}

\tableofcontents
\listoffigures
\listofexternals

\columnsep=0.3in
\twocolumn
\chapter{Introduction}

\begin{figure*}
  \centering
  \includegraphics[width=\columnwidth]{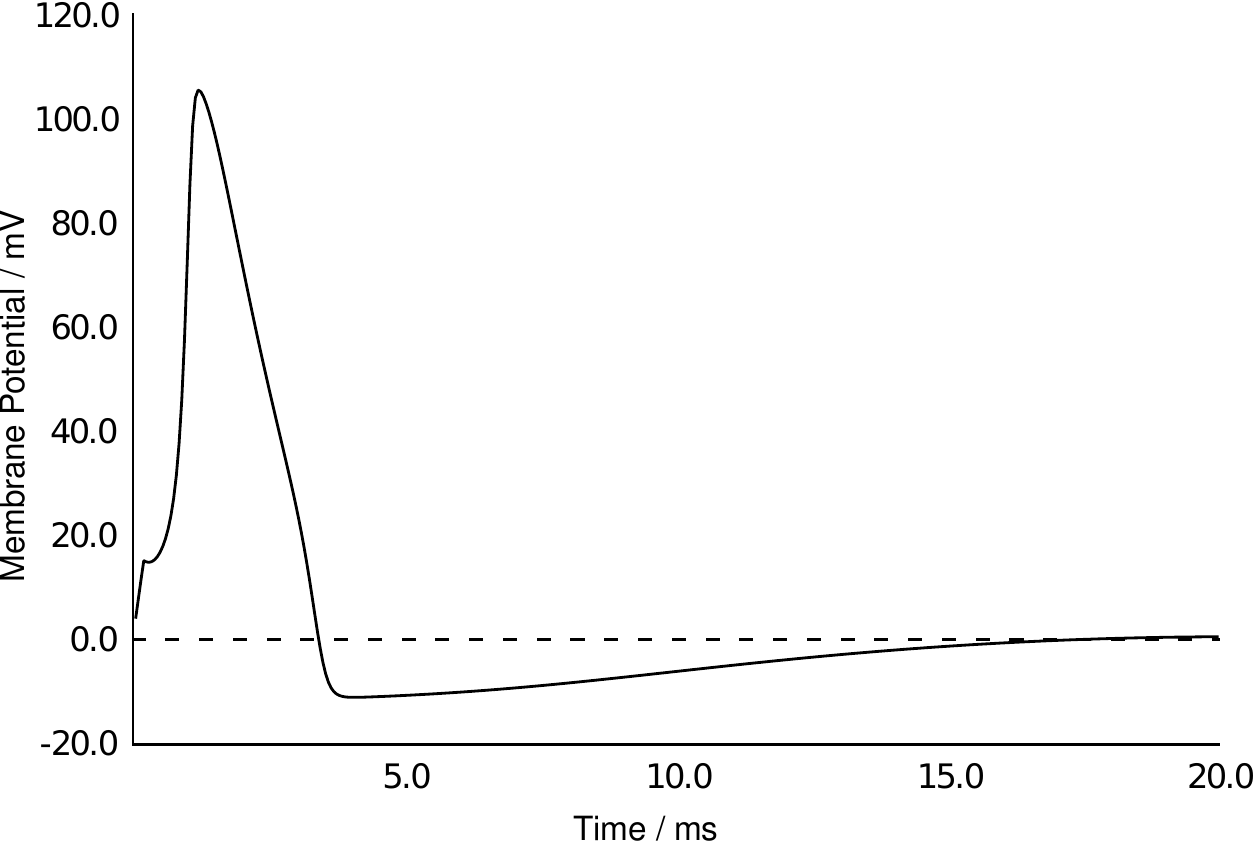}
  \caption[Computation of action potential]{\textbf{Computation of an
      action potential} according to the \citet{hodgkin:1952:quant}
    model (comparable to Fig.~13, upper curve, same paper). This curve
    is calculated with $I = C_{\textrm{\tiny M}} \dot V + \bar
    g_{\textrm{\tiny K}} n^4(V-V_{\textrm{\tiny K}}) + \bar
    g_{\textrm{\tiny Na}}m^3h(V-V_{\textrm{\tiny Na}})+\bar
    g_{\textrm{\tiny l}}(V-V_{\textrm{\tiny l}})$ [Eq.~26, same
    paper], using the same constants as the original paper for
    activation, inactivation, reversal potentials and
    conductances. The variables $n$, $m$ and $h$ represent the
    proportion of `activating or inactivating particles' for a channel
    in the activating position: $n$ for the activating \K channel
    particles; $m$ for the activating \Na channel particles; and $h$
    for the inactivating \Na channel particles.  Since the rate
    constants for the first derivative of these variables were voltage
    dependent, the physical interpretation was of a charge-carrying
    voltage sensor driving the opening and closing of ionic
    conductance.}
  \label{fig:HH}
\end{figure*}

\begin{figure*}
  \centering
  \includegraphics[width=\columnwidth]{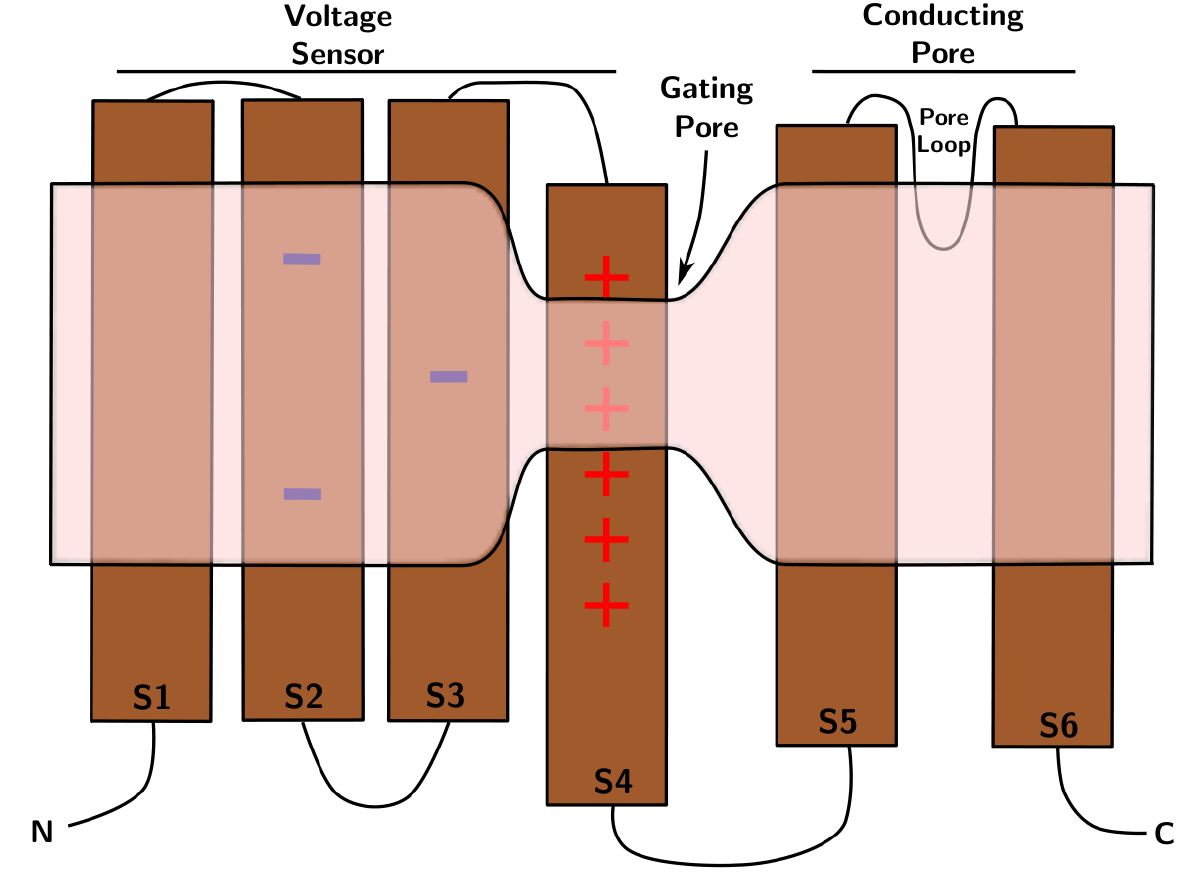}
  \caption[Topology]{\textbf{\Kv channel schematic topology:} a
    generic \Kv channel representation highlights the transmembrane
    domains of a single unit of the \Kv tetramer. On the left from the
    amino end of the protein are the voltage-sensor domains,
    \SI--\SIV. On the right are the conducting pore domains, \SV and
    \SVI. The lipid bilayer through which the protein is threaded is
    represented in pink. The red `$+$' symbols represent the excess
    positively charged amino acid residues on \SIV that are believed
    to cross the lipid bilayer at the invagination labeled `gating
    pore'. The blue `$-$' symbols represent excess negatively charged
    amino acid residues on the \SII and \SIII transmembrane domains.}
  \label{fig:topo}
\end{figure*}

\noindent Electrical excitability of cells is possible because the movement of a
few charges can control the flow of many charges. This principle --
amplification -- led \citet{hodgkin:1952:quant} to their theory of the
action potential in terms of electrically controlled membrane
conductances (for an example of a computation of their model, see
Fig.~\ref{fig:HH}). Such conductances have been localized to channel
proteins conducting \Na, \K, or \Ca ions. Besides a conductive port
(composed of transmembrane domains \SV and \SVI, see
Fig.~\ref{fig:topo}), the channels contain four transmembrane regions
(labeled \SI-\SIV starting from the amino end). In the \SIV region,
there are a total of three to seven positively charged amino acid
residues, each arrayed at every third amino acid position.  An
extensive electrophysiological data set exists on voltage-controlled
ionic conductance and the `gating current' attributable to charges
controlling the ionic port. A second extensive data set has emerged
from experiments measuring structure. Both sets provide essential
perspectives, but no direct means to assess physical interactions in
the structure and the significance of physical interactions for
function. In this thesis, I attempt to bridge these perspectives
computationally. I focus on electrostatics because the voltage-gated
ion channels are controlled by an applied electric field that acts on
intrinsic protein charges.

Voltage-gated \K channels (\Kv) are composed of two distinct
functional elements, a `voltage sensor' and a `pore domain'. As
described by \citet{lee:2005}, they appear to be ``membrane proteins
with separate, weakly attached membrane-spanning domains''. \Kv
channels may have evolved as the concatenation of two separate
proteins, one contributing a central tetrameric \K conducting pore and
the other contributing a weakly-attached peripheral voltage sensor
which transduces changes in transmembrane potential into a
flow-controlling action on the pore domain
\citep{kumanovics:2002}. The voltage sensor motif is homologous with
voltage-sensitive proton conducting channels (\Hv,
\citealp{ramsey:2006}). The physics underlying the well-known function
of the \K and related \Na and \Ca channels
\citep{hodgkin:1952:move,catterall:1988} in terms of the atomic
structures developed over the last 20 years
\citep{doyle:1998,jiang:2003:xray} has still not been fully
elucidated. Questions are still open regarding the precise
electrostatics, thermodynamics and distributions of relative
positions \& motions of charges at physiological temperatures.

Each voltage sensor comprises four largely helical secondary
structures. The \SIV \alpha¬helix bears a positively charged arginine
or lysine residue at every third position while the \SII and \SIII
helices bear a smaller number of negatively charged aspartate or
glutamate residues \citep{noda:1986}. The other residues of these
membrane-spanning segments are mostly hydrophobic. It has been
proposed that the transmembrane electric field moves the \SIV segment
through a combination of translation and rotation with respect to the
other helices and the lipid membrane (`sliding helix' hypothesis,
\citealt{catterall:1986}), or alternatively moves the \SIV segment in
association with part of the \SIII segment across the lipid much like
a large lipophilic ion (`paddle' hypothesis, \citealt{lee:2003}).

In both proposed mechanisms, multiple electric charges of the channel
protein move in the electric field of the membrane and therefore
produce a displacement current (`gating current',
\citealp{hodgkin:1952:quant,chandler:1965,armstrong:1973}), as well as
electric work which could drive gating of ionic conductance
\citep{hodgkin:1952:quant}. Translocation of these charges could be
facilitated by factors such as a local thinning of the lipid bilayer,
a reduction of the passage-way to a short `gating pore'
\citep{parsegian:1969,perozo:1993} or the provision of counter-charges
along the route of the moving \SIV charges by static residues of \SII
and \SIII. The existence of a short gating pore is indicated by the
accessibility of modified S4 positions to extra- or intracellular
cysteine reactants or protons \citep{yang:1996}. The relevance of
counter-charges has been indicated by neutralization mutants
\citep{seoh:1996} and coordination of S4 charges with residues of
opposite charge in recent crystallographic studies \citep{long:2007}.

\balance
An understanding of the natural design of the voltage sensor needs to
be based on a broad exploration of the components that have been
discovered experimentally. The importance of design elements that have
emerged from these experiments should be evaluated. How do such
elements determine sensor characteristics? How do their specific
evolutionary layouts bring about voltage sensor behavior as seen in
different channel types? And as an important control, can results of
experimental mutations be predicted?

\begin{figure*}
  \centering
  \includegraphics[width=\columnwidth]{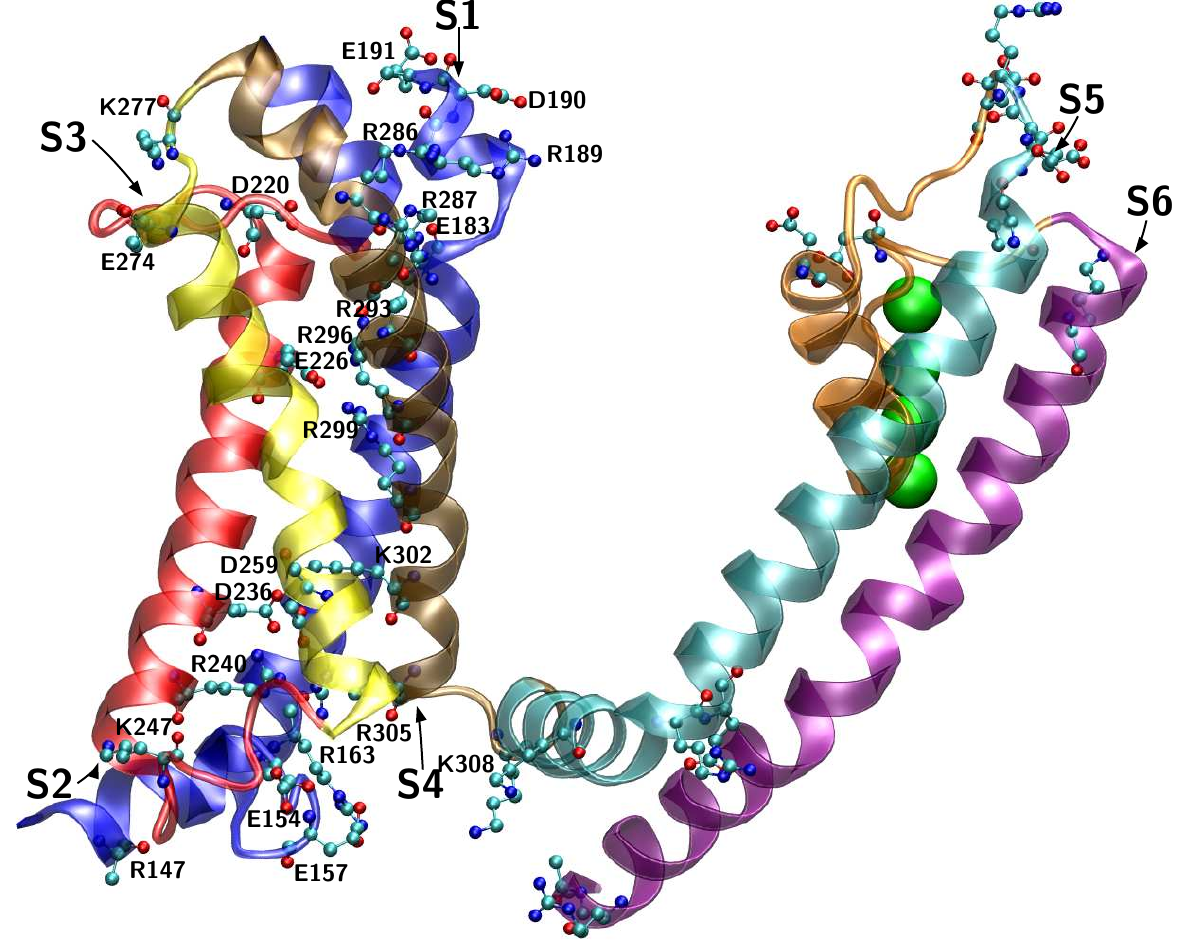}
  \caption[Crystal structure]{\textbf{Crystal structure of a chimeric}
    \Kv1.2 and \Kv2.1 \K channel \citep[PDB No.~2r94]{long:2007}. This
    single unit of the channel tetramer is composed of \Kv1.2 channel
    elements with \SIV and parts of \SIII replaced by homologous
    regions from \Kv2.1. The backbones of \SI--\SIV are represented by
    ribbons; charged residues (arginines, lysines, glutamates and
    aspartates) are represented in ball-and-stick format. The
    approximate \SI region ribbon is in blue; \SII is in red; \SIII is
    in yellow; \SIV is in brown; \SV is in cyan; \SVI is in purple;
    and the pore loop is in orange. \K ions near the pore loop are
    represented by green spheres. Charged residues which may be part
    of the voltage sensor (\SI--\SIV) are labeled by their residue
    number for this chimera. Graphic produced with VMD
    \citep{humphrey:1996}}
  \label{fig:crystal}
\end{figure*}

It is tempting to base exploration of a molecular device like the
voltage sensor solely on available information of atomic structures,
as represented in Fig.~\ref{fig:crystal}. The study of atomic
structures, however, is greatly limited in the case of the voltage
sensor. Crystallization destroys the natural (dielectric) environment
of the structure and in particular the electric field that the
structure is designed to detect. Atomic-level computations like those
based on Molecular Dynamics (\acsfont{MD}) are speculative to the
extent that they are based on questionable atomic
coordinates. Moreover, they are currently restricted to
explorations of sub-microsecond episodes of dynamics of a few chosen
initial configurations \citep{khalili-araghi:2010}.

In this thesis, I analyze reduced electrostatic models (see
\secref{concepts}) for the voltage sensor which can be thoroughly
specified: charged rigid bodies moving through piecewise homogeneous
dielectric domains. The sensor model is embedded in a simulation cell
mimicking a voltage-clamp setup including electrodes, allowing
macroscopically observed variables to be predicted from the
microscopic model. The electrostatics are solved self-consistently
with numerical methods that allow systematic
exploration. Specifically, my computations construct an
electrostatics-based partition function of charge configuration with
two degrees of freedom: rotation and translation of the S4
segment. Using this partition function, I compute the expectation of
random variables such as charge displacement in response to voltage
for comparison with experiment.  These measures have not been
accessible in other computational studies
\citep{bliznyuk:2004,chakrapani:2008,nishizawa:2008,khalili-araghi:2010}.

My results show that a voltage sensor (\VS) model involving a sliding
\SIV helix is realistic with respect to both gating charge and
electrostatic energetics. A crucial component of this model is
negative counter-charges arranged in close proximity to buried \SIV
charges. The electrostatic design of this \VS tolerates considerable
variation resulting in electrophysiologically interesting variations of
sensor characteristics. Simulations of \SIV, \SII and \SIII charge
mutants in this electrostatic model predict experimentally observed
patterns of charge displacement \citep{seoh:1996}.

\chapter{Methods}\nobalance
\seclabel{methods}{chapter}{Methods}

\noindent My approach to voltage sensor electrostatics has three major
elements: \acsfont{(1)} I use \emph{reduced} physical descriptions of
electrostatic components; \acsfont{(2)} I compute predictions that
correspond to \emph{experimental conditions and observables}; and
\acsfont{(3)} I develop a \emph{statistical-mechanical description} of
sensor behavior by exploring a wide range of sensor configurations.

\section{Concepts}
 \seclabel{concepts}{section}{Concepts}

\paragraph{Reduced descriptions.} By a `reduced model', I mean a model
where some details are reduced in resolution; for example, in
electrostatic models many charge distributions are reduced to
dielectric constants or tensors. The selection of which features are
to be reduced and which are to be represented at higher resolution is
an iterative problem involving the identification of the relevance of
those features to the measures of interest. All models that are not
calculated by \emph{ab initio} quantum mechanical calculations are
reduced in this sense. By explicitly structuring a model with multiple
tiers of resolution and identifying the relevance of those components
of the system to measures of interest, it is possible to distinguish
the dominant terms in the underlying physics.

Atomic matter is made of charged constituents, charged nuclei and
charged electrons \citep{feynman:1963:ii:1}.  The crucial charges for
this study are the uncompensated nuclear charges in the arginine or
lysine side chains of the \SIV region and the excess electronic
charges of the glutamate and aspartate side chain in the \SII and
\SIII regions \citep{creighton:1984:1,islas:1999}, out of the vast number
of nuclei and electrons in the system. Most of the `vast number' of
charges form neutral atoms or molecules, but at close range many
molecules reveal spatial asymmetries in their internal charge
distributions. Moreover, the electric field from other components can
distort the internal charge arrangements of molecules or groups that
are overall neutral into asymmetrical distributions of charge. Both
the rotational alignment of molecules with internal charge asymmetry
and the distortion of symmetrical charge distributions in molecules
are abstracted as `polarization.'  Polarization of matter in an
electric field `induces' charge that is hidden in the absence of the
field \citep[see Fig.~22]{griffiths:1999:4}. The most abundant
polarizable molecule in \VS system is the water of the solutions
bathing the membrane.

An \emph{ab initio} (quantum mechanical) description of the charged
nuclei and electrons in a channel protein, membrane, and embedding
ionic solutions is not possible --- approximations must be made.
Approximate physical descriptions can be made in multiple tiers of
resolution: either the results of lower tiers become parameters for
higher-tier theory, or independent experimental results produce
parameters for higher tiers. For example, an ion is an atom or
molecule in which the number of nuclear elementary charges is not
identical to the number of electron charges. A classical approximation
for an atomic ion is a point charge at the center of a sphere where
the effective diameter of the sphere that excludes other atoms can be
determined from crystallographic measurements. I use such a classical
description for formal charges explicitly included in models.

Polarization of neutral molecules and groups must also be described by
approximations. Voltage sensor charges buried in the membrane polarize
bath water. A reasonable starting point for a description of that
polarization is a continuum description of the water. A space element
containing polarizable molecules will exhibit polarization charges on
its surfaces when an electric field is present
\citep{boda:2004}. These surface charges represent the polarization of
the molecules in the space element: if these charges are included in a
computation of the electric field, the molecular polarization is
accounted for. The amount of polarization charge is proportional to
the strength of the applied field (over a range), and depends on the
atomic/molecular composition of the matter in the space element. In
the linear range, the polarization charge at the surfaces of the space
element is related to the field strength by a material coefficient (or
tensor for anisotropic polarization). Water polarization can be
described this way by one number, the dielectric
coefficient. Furthermore since polarization involves a rearrangement
of charges, polarization takes a finite amount of time to develop or
disappear. However, since voltage sensor relaxations are slow compared
to typical polarization relaxations, polarization can be approximated
as instantaneous. For instance, the rotational relaxation time of
water is on the order of \ten{-11}~s \citep{barthel:1995}, while
voltage sensor relaxation times are on the order of $>$~\ten{-5}~s
\citep{hille:2001:18,sigg:2003}.

The channel and membrane are bathed in electrolyte solutions on
either side. Electrostatic interactions involving bath ions are
reduced two orders of magnitude by the solvent (water) with respect to
the vacuum. This reduction is described by the dielectric coefficient
of water. Furthermore, ions screen one another to an ionic
concentration-dependent extent. Screening arrangements in solutions
form in nanoseconds \citep{barthel:1995}, which is much faster than
the time base for \VS motion --- like polarization, screening can be
approximately described as instantaneous.

\begin{figure*}
  \centering
  \linebox{
    \subfloat[\hspace{2em}\protect\shortstack{Simulation Environment\protect\\ for Sliding Helix}]{
      \label{fig:cell:helix}
        \includegraphics[width=\halfwidth]{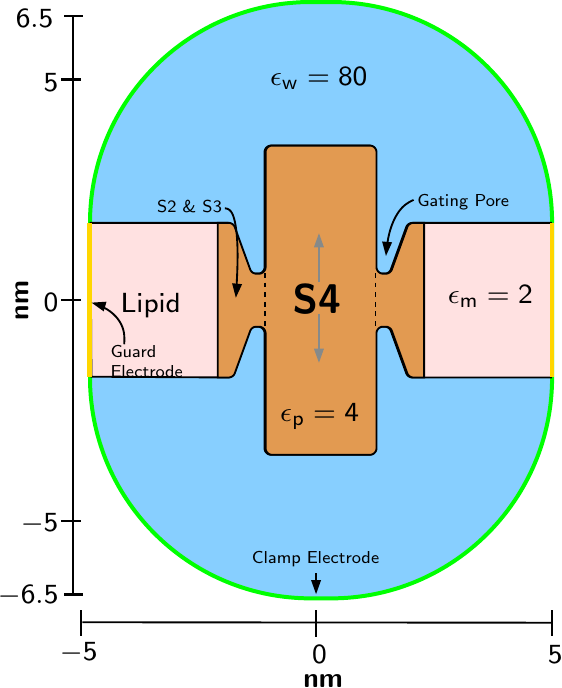}
    }
    \hspace{\stretch{1}}
    \subfloat[\hspace{2em}\protect\shortstack{Simulation Environment\protect\\ for Paddle}]{
      \label{fig:cell:paddle}
      \includegraphics[width=\halfwidth]{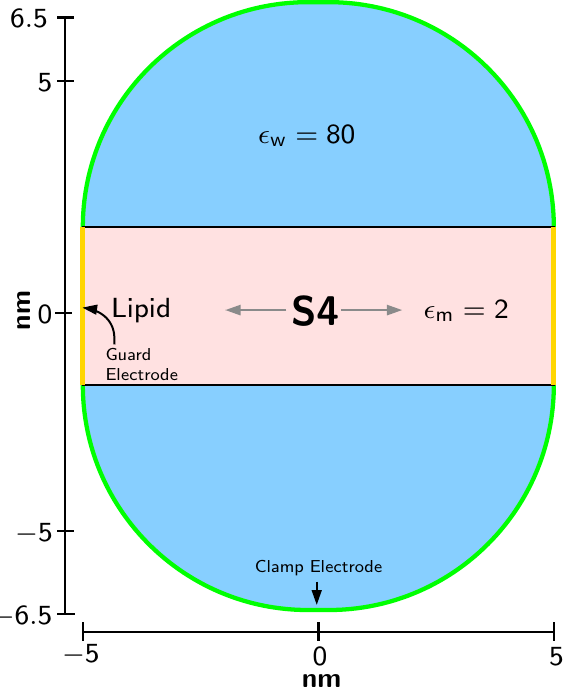}
    }
  }
  \caption[Simulation cells]{Cross-sections of the simulation cells
    for the sliding helix model~\subref{fig:cell:helix} and the paddle
    model~\subref{fig:cell:paddle}. The 3D setup is produced by
    rotating the cross-section about its vertical axis. The setup is
    bounded by two hemispherical bath electrodes (in green). Their
    potentials are maintained at prescribed values as the \SIV segment
    is moved (voltage clamp). The cylindrical electrode (in gold)
    completing the boundary is a guard. It is divided into rings
    maintained at potentials that are linearly graded between the
    potentials of the bath electrodes. The interior is
    divided into two aqueous baths (in blue) separated by a lipid
    membrane (in pink) and, in the sliding helix model, a \VS protein
    region (in brown) forming a gating pore around the sliding \SIV
    helix. In the paddle model, no distinction is made between
    membrane and protein --- the \SIV charges (not shown) are embedded
    in the membrane. The charge configuration of the sliding helix
    model is shown in Fig.~\ref{fig:tile:helix}.}
  \label{fig:cell}
\end{figure*}

Screening by bath ions is modeled in my simulation cells by placing
the electrodes closer to the membrane than would be done in an
experiment. A diffuse layer of screening counter-ions is
electrostatically equivalent to counter-charge smeared on a surface a
Debye length from the screened charge, which is the essential result
of the Debye-H\"uckel and Gouy-Chapman descriptions of screening
(\mbox{\citealp{debye:1923}}; \mbox{\citealp{gouy:1909}};
\mbox{\citealp{chapman:1913}}). By varying the distance between the membrane
and the screening electrode surface, variations of bath ionic strength
can be mimicked (for example, offsets of 3~nm and 0~nm bracket the
Debye lengths of dilute [3~mM] and infinitely concentrated
solutions\footnote{See Fig.~3 on pg.~3718 under \emph{Defining
    electrical coordinates and electrical travel} of
  \citet{nonner:2004:rs}}). In the simulations reported here I use the
electrode geometries of Fig.~\ref{fig:cell}, roughly equivalent to
bath solutions containing 3 mM salt. Experimental variation of bath
ionic strength has relatively small effects on experimental gating
currents \citep{islas:2001}. Indeed, control computations (not shown)
in which the water domain is removed from the simulation and the
electrodes are placed directly on the membrane and protein surfaces
(to mimic infinite ionic strength) yield results similar to those
obtained with the simulation cells of Fig.~\ref{fig:cell}.

\paragraph{Coupling microscopic \VS motion to macroscopic experiments.} 
Computer experiments provide insights into microscopic systems that
are difficult if not impossible to obtain by conventional
experiments. To estimate the functional competence of hypothetical
structures, experimental observables of function must be computed, and
conditions comparable to conventional experimental conditions must be
established. Gating current, the most direct observable reporting \VS
motion, is recorded experimentally while applying a prescribed voltage
across electrodes placed in the electrolytes bathing the membrane (a
voltage clamp). My computational setup is designed to establish a
voltage clamp and record the charge displaced by \VS motion in a
manner comparable to charge displacement recordings with macroscopic
electrodes.

\paragraph{From electrostatics to statistical mechanics.}
The primary results of my computations are the electrostatic potential
energy and gating charge corresponding to a particular location of the
formal charges of the voltage sensor model at a particular applied
voltage. (Another output is the Maxwell stress, see below in
\secref{maxwell}). The efficiency of my computational methods allows me
to compute these variables for a very large number of \VS
configurations and applied voltages, thus enabling me to elaborate a
statistical (thermodynamic) view of \VS configuration.

I explore a configuration space that includes the rotation of the \SIV
helix about its axis and the translation of the \SIV charges along
that axis at a fixed applied potential. A partition function in those
configurational degrees of freedom is constructed from the ensemble of
Boltzmann factors for each electrostatic potential energy. Using the
partition function, statistical expectations of equilibrium positions
and displaced charges are obtained. In this way, I determine both how
the model will configure at a particular membrane voltage, and along
which average configurations the model will re-configure as voltage is
varied in small increments -- these are equilibrium averages and not
trajectory calculations. The relation between gating charge and
voltage is predicted, allowing comparison with experimental
charge/voltage curves recorded from ensembles of voltage sensors.

\paragraph{The typical simulation cell.} A typical geometry for a
computational experiment is shown in
\ref{fig:cell}~\subref{fig:cell:helix}. The simulation environment is
represented by an axial cross-section of the radially symmetric
three-dimensional domain swept by rotating that cross-section about
its vertical axis. The hemispheric boundaries (in green) are electrode
surfaces kept at controlled electrical potentials. The blue zones
represent aqueous baths (with a dielectric coefficient
$\epsilon=$~80). The pink zone is a region of small dielectric
coefficient ($\epsilon=$~2) that represents the lipid membrane. The
brown zone represents both the non-\SIV components of the \VS protein
and the matrix of the \SIV helix (the central cylinder). The helix
crosses the membrane through a `gating pore' which extends the baths
into the region joining \SIV and the rest of the protein. The
dielectric coefficient of this protein region is varied in my
simulations to assess its importance for \VS motion. The dielectrics
in this reduced model are piecewise uniform and therefore have sharp
boundaries (solid black lines). Point charges representing \SIV and
other charges (not shown) are embedded in the region of protein
dielectric.

\VS motion is simulated in this geometry by \emph{moving only the \SIV
  charges} within the fixed cylinder bounding the \SIV
helix. Simulating \SIV motion this way (by moving charges and not
dielectric boundaries) greatly reduces the computational costs of
solving the electrostatics since solving for the effects of moving
dielectric boundaries requires recomputing the matrix inversion (see
\secref{matrix}). The dielectric geometry as defined is invariant in
terms of rotation. Only the end caps of the \SIV are not invariant in
terms of central-axis translation. If those ends of the biological
\SIV are translated, which is not well defined experimentally, then
the positionally fixed end-points of the \SIV in these simulations
would not capture the movement of surface charge on those \SIV ends.
However, the surface charge on those ends would move outside the
region of high electrical travel close to the gating pore
\citep[Fig.~1]{nonner:2004:rs}, therefore adding little to
either the charge displacement terms or the associated energy
terms. Charge/end-surface interaction terms could only become
significant at the extrema of \SIV charge motion along the central
axis.

\section{Computing the electrostatics}
\seclabel{computing}{section}{Computing the electrostatics}

My general approach for computing the electric field in this system is
to superpose the vectorial Coulombic fields of all charges. Likewise,
the electrical potential is computed by scalar superposition of the
Coulombic potentials of all charges \citep[section
3]{jackson:1999:0}. The contributions to the field made by the \SIV
and other formal charges at known positions are readily computed, but
the charges on the electrodes and the polarization charges at
dielectric interfaces are initially unknown. The determination of the
contribution of surface charges is a crucial (and computationally
expensive) element of solving the electrostatics. 

Since dielectrics are involved, a precise definition of `charge' is
needed for the following mathematical treatment. I follow the
nomenclature of \citet{jackson:1999}, who distinguishes between
`source' and `induced' charges. The formal charge on a side-chain of
the \SIV region (one proton charge) is a source charge, as are charges
placed by the external voltage clamp circuit on the
electrodes. Induced charges comprise those charges appearing on
dielectric boundaries in response to the fields of source charges and
other induced charges.  It is convenient to combine spatially
inseparable source and induced charges into `effective'
charges. Specifically, I assign to a point source charge~($q_s$)
embedded in a dielectric~($\epsilon$) an effective
charge~($q=q_s/\epsilon$), combining the source charge and the
polarization induced at the dielectric boundary around that source
charge~($-q_s(\epsilon-1)/\epsilon$). In other words, the effective
point charge is the sum of the source and induced charges, expressing
the fact that the field produced by a source charge in a dielectric is
$\epsilon$ times weaker than the field of the source charge in
vacuum. Electrode charges are also represented in computations by
effective charges, which eliminates the need to specify polarizable
matter outside the cell. Finally, induced charge on a dielectric
boundary separated from any source charge is formally treated as an
effective charge (lacking a source charge component). Thus, \emph{all}
charges are expressed in computations as effective charges. The field
computed from effective charges is identical to the superposition of
the fields of the source charges and their polarization charges. That
superposition of fields is the field that must be computed for the
computer experiments presented here.

\subsection{Electrical geometry made discrete}

\begin{figure*}[p]
  \centering
  \linebox{
    \subfloat[\hspace{1em}Side]{
      \label{fig:tile:helix:side}
      \includegraphics[width=\halfwidth]{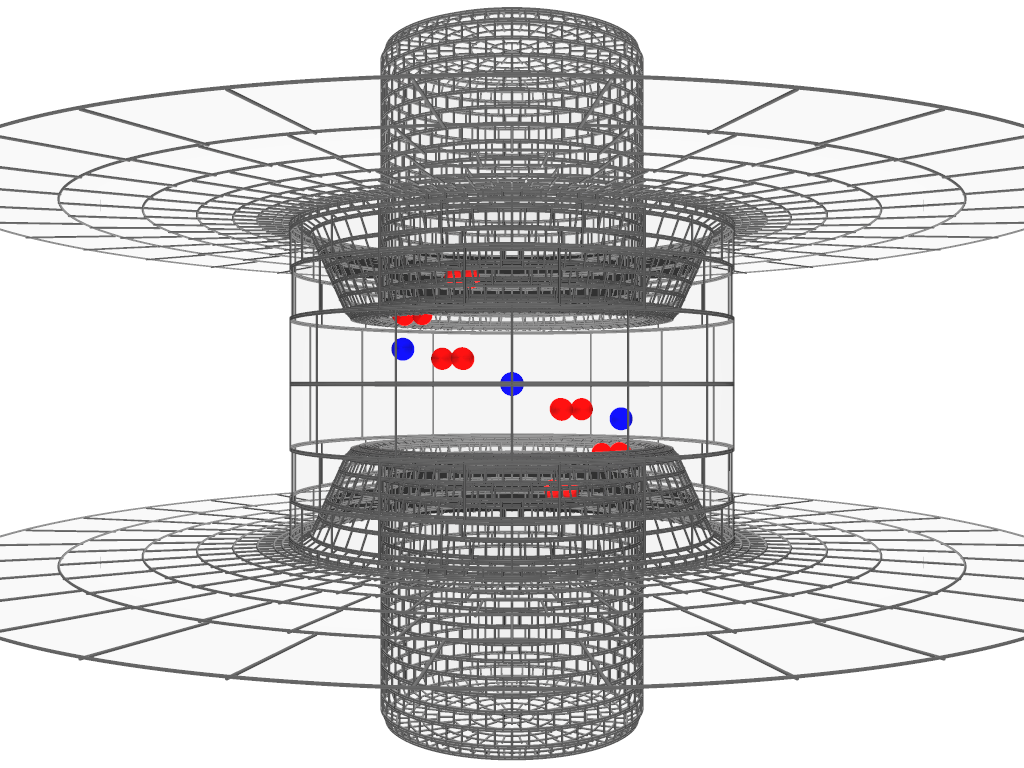}
    }
    \hspace{\stretch{1}}
    \subfloat[\hspace{1em}Bottom]{
      \label{fig:tile:helix:top}
      \includegraphics[width=\halfwidth]{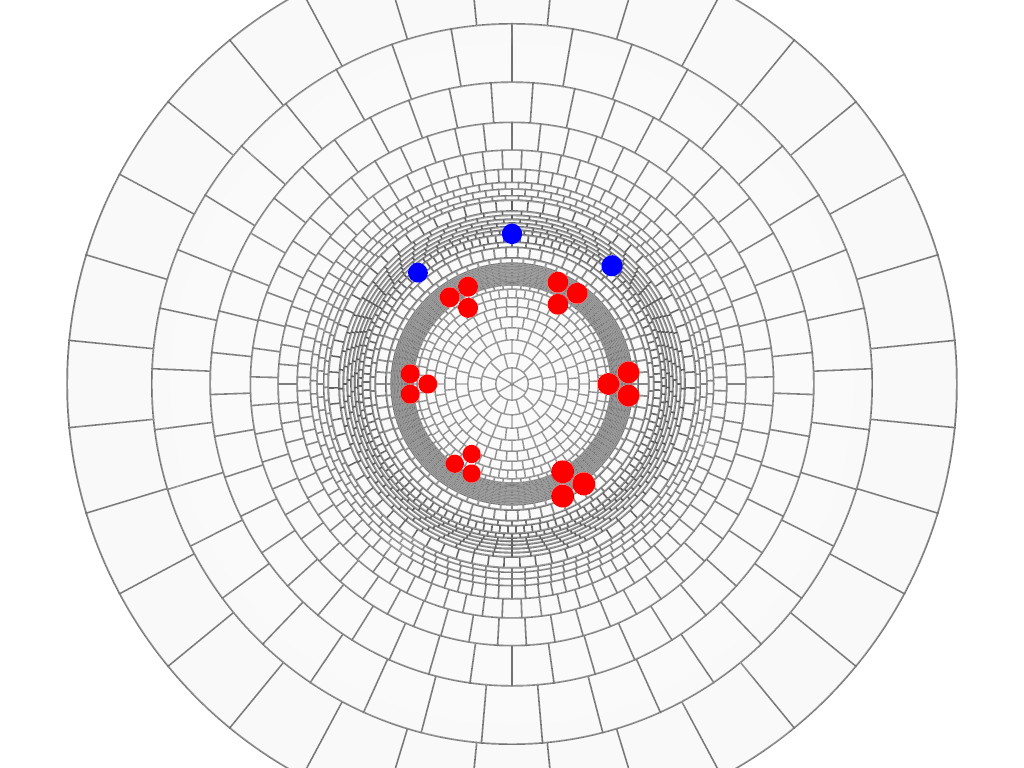}
    }
  }
  \caption[Source charges and dielectric boundaries: Sliding
  helix]{\textbf{Source charges and dielectric boundaries in a sliding
      helix model.} The sliding helix bears six triplets of 1/3~e$_0$
    point charges (shown as red balls for visibility) representing the
    guanidinium group of arginine residues. These charges are aligned
    on a superhelix that is oriented in the direction opposite to that
    of the \SIV helix. The \SIV charges are shown at the
    0~nm/0$^\circ$ position of their translational and rotational
    dimensions of motion. Counter-charges are -1~e$_0$ point charges
    (blue balls). They are aligned in a superhelix concentric with the
    superhelix of the positive charges (shown for the `2/3' spacing of
    counter-charges). The dielectric boundaries are shown as a 2D
    grid; the mesh is a dimensional representation of the surface
    tiling used in solving the electrostatics (tiling is fine near
    positions that are close to source charges). The change in average
    position of the \SIV charges under a gradual change of voltage
    from -100~mV to +100~mV can be seen in the supplementary movies:
    \moviefile{tile-helix-side.mp4} \& \moviefile{tile-helix-top.mp4}.}
  \label{fig:tile:helix}
\end{figure*}

\begin{figure*}[p]
  \centering
  \linebox{
    \subfloat[\hspace{1em}Side]{
      \label{fig:tile:paddle:side}
      \includegraphics[width=\halfwidth]{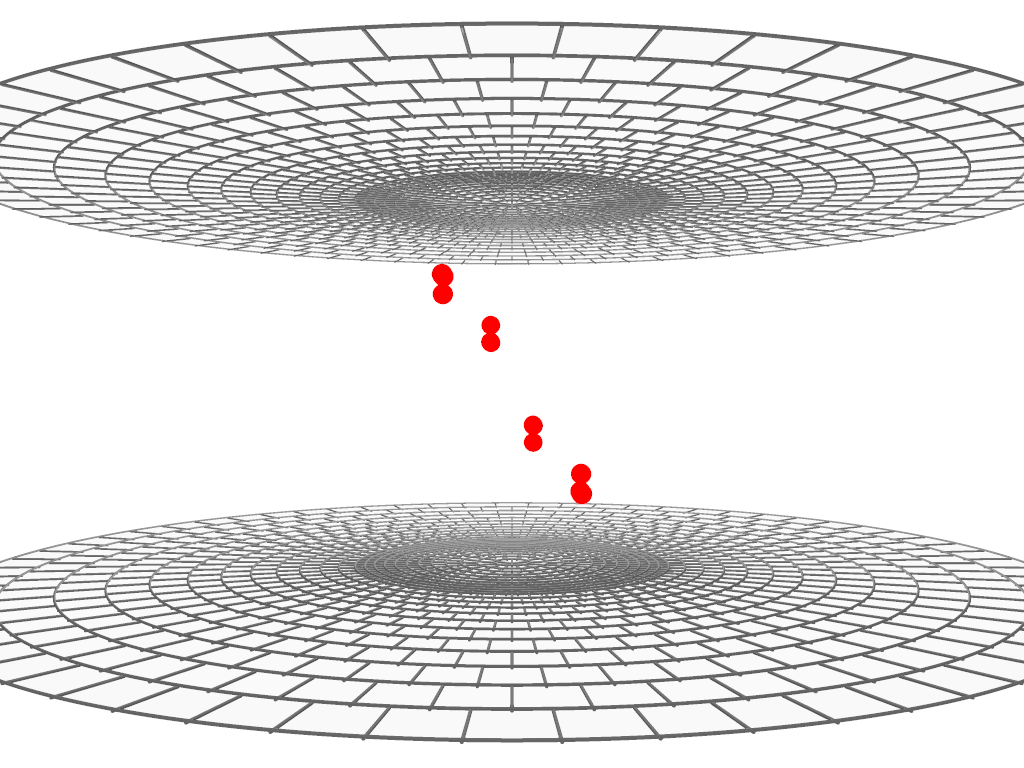}
    }
    \hspace{\stretch{1}}
    \subfloat[\hspace{1em}Bottom]{
      \label{fig:tile:paddle:top}
      \includegraphics[width=\halfwidth]{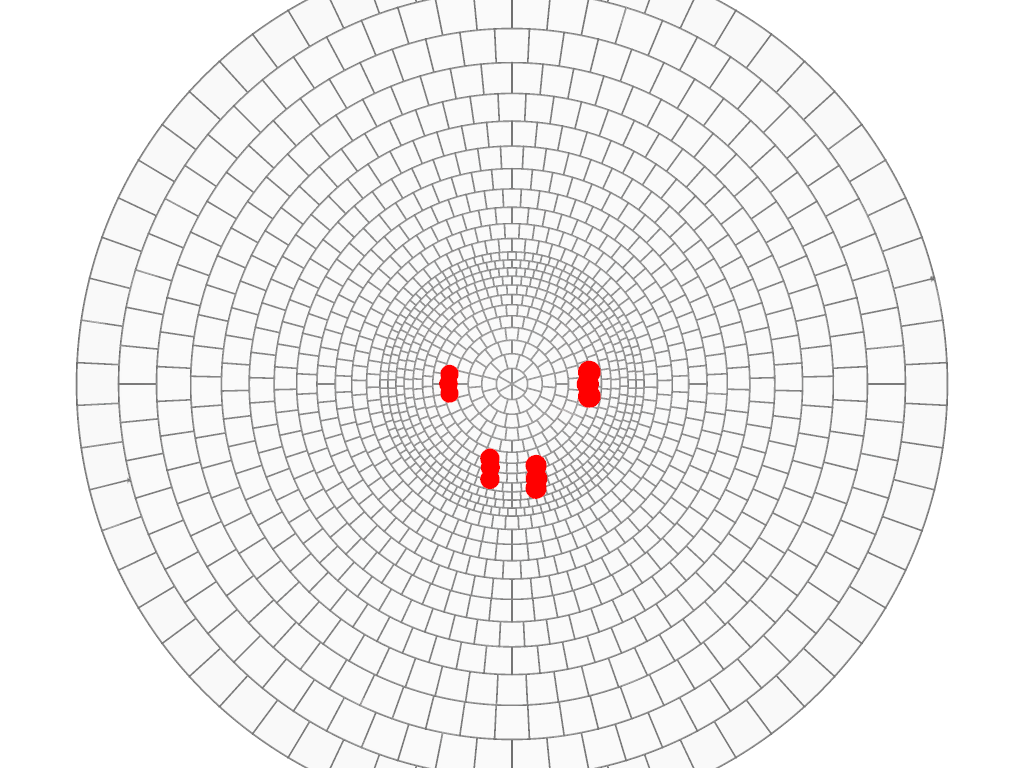}
    }
  }
  \caption[Source charges and dielectric boundaries:
  Paddle]{\textbf{Source charges and dielectric boundaries in a paddle
      model.} As in the sliding helix of Fig.~\ref{fig:tile:helix},
    the paddle bears four triplets of 1/3~e$_0$ point charges (shown
    as red balls for visibility) representing the guanidinium group of
    arginine residues. These charges are aligned on a superhelix that
    is oriented in the direction opposite to that of the \SIV
    helix. The \SIV charges are shown at the 0~nm/0$^\circ$ position
    of their translational and rotational dimensions of motion. The
    dielectric boundaries are shown as a 2D grid; the mesh is a
    dimensional representation of the surface tiling used in solving
    the electrostatics (tiling is fine near positions that are close
    to source charges). The change in average position of the \SIV
    charges under a gradual change of voltage from -100~mV to +100~mV
    can be seen in the supplementary movies:
    \moviefile{tile-paddle-side.mp4} \&
    \moviefile{tile-paddle-top.mp4}.}
  \label{fig:tile:paddle}
\end{figure*}

To compute the charges on the electrodes and dielectric boundaries,
the boundary surfaces are tiled into curved quadrangular surface
elements (Fig.~\ref{fig:tile:helix} for the sliding helix model \&
Fig.~\ref{fig:tile:paddle} for the paddle model). The size of
these surface elements is chosen such that the charge density present
on an element can be approximated as uniform on the element. Properly
choosing the tile size to conform to that approximation requires
numerical controls described later (\secref{gauss}). There is
one unknown to be determined for each surface element: its surface
charge. Below, I will show how one linear equation can be obtained per
surface element. Solving the system of these linear equations yields
all unknown charges, which typically involves 4000--10000 surface
elements of different sizes (and that number of unknowns) for
numerical accuracy (checked by Gauss' law to recover the total
integral number of charges within volumes surrounded by closed
surfaces, see \secref{gauss}).

The electrodes impose voltage clamp conditions. In the discrete
representation of the electrode surface, the potential at the center
of an electrode element~($i$) assumes a prescribed value~($V_i$). The
potential at the center of that element results from the superposition
of the potentials due to \emph{all} charges~($j$), including that of
element $i$ itself:
\begin{equation}
  \label{electrode}
  V_i = \frac{1}{4\pi\epsilon_o} \sum_j \frac{q_j}{|ßr_{ij}|}
\end{equation}
The definition of the distance $|ßr_{ij}|$ for a surface charge $j$
depends on the distance of that element from element $i$. The surface
charge of a distant element is combined into a single point charge
located at the charge center of that element, and $|ßr_{ij}|$ is
defined as the distance between those points. The surface charge of a
proximate element (in particular, element $j=i$ itself) is divided
into smaller charges obtained by sub-tiling element $j$ into a
number of subelements (typically 4~$\times$~4 for $j \neq i$, and
10~$\times$~10 for $j = i$).  Each subelement carries a fraction of
element $j$'s charge, and $1/|ßr_{ij}|$ is defined as equal to the
weighted average by subelement area of the inverse distances for all
subelements to the center of $i$. The subelements follow the
curvature of the surface element.

A correct representation of surface curvature and the curvature's
effects on charge distribution is crucial for numerical
accuracy. Inhomogeneities in charge distribution are particularly
problematic when source or other induced charges are close to the
surface element. If the element has curvature, the charge induced on
one part of the element will induce charge at close range on other
parts of the same element. Inaccuracies due to charge inhomogeneity
and surface curvature are limited by choosing an adequate initial
tiling and sub-tiling as needed.

The boundary condition describing the effect on the field due to
dielectric boundary elements can be expressed in two equivalent
ways. One way of describing the boundary condition is that the normal
components of the field strengths on each side of the boundary
($ßE^\bot_1$ and $ßE^\bot_2$) are inversely related to the dielectric
coefficients of each region \citep[40]{griffiths:1999:4}:
\begin{equation}
  \label{dielectric:epsilon}
  \epsilon_1ßE^\bot_1 = \epsilon_2ßE^\bot_2
\end{equation}
The other expression for relating normal field strengths says that the
field strengths differ by the field of the polarization charge induced
on the surface \citep[22]{jackson:1999:1}:
\begin{equation}
  \label{dielectric:sigma}
  ßE^\bot_1 + \frac\sigma{\epsilon_0}ßn = ßE^\bot_2
\end{equation}
where $ßn$ is the unit normal vector from the region of
$\epsilon=\epsilon_1$ to the region of
$\epsilon=\epsilon_2$. Furthermore the normal field strength
\emph{exactly} at the surface, without including the field due to
induced charge on that surface ($ßE^\bot$), is the average of the
normal field strength at an infinitesimal distance from the surface
including \emph{all} charges:
\begin{equation}
  \label{dielectric:diff}
  ßE^\bot = \frac{ßE^\bot_1 + ßE^\bot_2}{2}
\end{equation}
The difference between $ßE^\bot_1$ and $ßE^\bot_2$ arises by
superposition with the field of the polarization charge at the surface
which has a magnitude of $\sigma/2\epsilon_0$ in both normal
directions \citep[17]{griffiths:1999:2}.

Eqs.~\ref{dielectric:epsilon}--\ref{dielectric:diff} can be
combined into one expression relating the density of induced
polarization charge~($\sigma_i$) to the normal field strength on the
surface of dielectric boundary element $i$:
\begin{equation}
  \label{sigma}
  \sigma_i = \frac{2(\epsilon_1-\epsilon_2)}{\epsilon_1+\epsilon_2}
    \epsilon_oßE^\bot_i
\end{equation}
The component of the field strength normal to the tangent plane of
the boundary surface element is \citep[4]{jackson:1999:1}:
\begin{equation}
  \label{tangent}
  ßE^\bot_i = \sum_j \frac{q_j}{4\pi\epsilon_o} 
    \frac{(ßr_i - ßr_j)}{|ßr_i - ßr_j|^3}\cdot ßn_i
\end{equation}
If charge $j$ is the charge of a distant surface element, it is
treated as a point charge at $ßr_j$. Otherwise if element $j$ is a
sub-tiled surface element, the expression $(ßr_i-ßr_j)\cdot
ßn_i/|ßr_i-ßr_j|^3$ in \eqref{tangent} is replaced by its weighted
average (by subelement area) taken over all subelements of $j$. Note
that $q_j$ is the \emph{effective} charge of each entity.

Eqs.~\ref{electrode} and \ref{sigma} suffice to compute all initially
unknown surface (electrode or dielectric) charges. With these charges
known, \emph{all} charges in the system are known. The electric field
can then be computed for any location by superposition of the
Coulombic fields of individual charges (using appropriate
sub-tiling for nearby surface charges).

\subsection{Matrix inversion}
\seclabel{matrix}{subsection}{Matrix inversion}

My computer code calculates discretely tiled radially symmetric
surfaces or, in other words, boundary surfaces of shapes that can be
produced by turning a piece on a lathe, including hollow shapes
(cylindrically symmetrical). Techniques for tiling discretely
more general surfaces are known and could be used together with my
method for solving the electrostatics needed to explore more general
geometries than those in this study.

The electrostatic field is long-range --- the field strength at all
locations depends on all charges. Since the coefficient matrix for
those relations is dense, the computational method chosen for solving
the linear equations to determine surface charges is \LU decomposition
(\citealp{bowdler:1971}, implementation by
\citealp{whaley:1997}). Inverting an $N\times N$ matrix by \LU
decomposition requires order $N^3$ operations. Fortunately however,
the \LU decomposition needs to be executed only once for a given
surface geometry. If source charges are moved, the electrostatic
equations can be solved by repeated back-substitutions using the
inverse matrix, requiring order $N^2$ operations.

\subsection{Gauss box: A control for numerical accuracy}
\seclabel{gauss}{subsection}{Gauss box}

The error from approximating the surface charge distribution as
piecewise uniform can be assessed using Gauss' law. (The
approximation requires that the field strength at any point on a
surface element equals the field strength measured at the center of
charge of the element.) Gauss' law states that the integral of the
electrical flux normal to a closed surface over that closed surface is
proportional to the total source charge contained in the enclosed
volume \citep[39]{jackson:1999:4}:
\begin{equation}
  \label{gauss}
  \oint_{{\cal S}}\epsilon\epsilon_0ßE\cdot dßa
    = \sum_{i\in{\cal V}} q_{s,i}
\end{equation}
This conservation law holds for \emph{any} closed surface of
\emph{any} shape, and thus applies to any geometry of interest.

Since the density of polarization charge induced by a charge on a
dielectric boundary is particularly inhomogeneous when the charge is
close to the surface, the local surface region proximate to other
charges must be made into smaller discrete surface elements. On
the other hand, an excessive number of surface elements is
computationally expensive. Since \SIV charges move relative to
dielectric boundaries in my computer experiments, all \SIV positions
must be taken into account in designing surface tiling. The
adequacy of the surface tiling must be checked for all positions taken
by \SIV charges, which can be done by verifying Gauss' law for each
position of the \SIV.  For instance, the sum of all \SIV and
other source charges contained in the region of protein dielectric in
Fig.~\ref{fig:cell}~\subref{fig:cell:helix} must be accurately
recovered by summing the normal field strength multiplied by the
permittivity and the element area, over the surface bounding that
region ($\sum_{j\in{\cal S}} \epsilon_j\epsilon_0E^\bot_j\;a_j =
\sum_{i\in{\cal V}}q_{s,i}$, where $j$ varies over all elements of the
closed surface $\cal S$ and $i$ varies over all charges within the
volume $\cal V$ enclosed by $\cal S$).

My approach to solving the electrostatics differs from the more common
approach of solving Poisson's equation on a spatial grid. The approach
I use \citep{boda:2004} is based on relations describing boundary
conditions (making the method a `boundary element method',
\acsfont{BEM}). The resulting boundary integral equations are made
discrete on a surface grid. Owing to the relatively small
number of surface elements compared to the number of volume elements,
a full description of the charge distribution and therefore the
electric field can be stored in computer volatile memory. From this
information any desired electrostatic output variable can be
computed. Exhaustive \emph{a posteriori} tests of solution accuracy
are possible, such as verification of results by checking the
consistency of calculated surfaces charges against the integral number
of charges enclosed using Gauss' law.

\subsection{Computation of gating charge}
\seclabel{displacement}{subsection}{Computation of gating charge}

\SIV charge movements are restricted to a subrange of the distance
between the electrodes. The electrodes record a displacement current
due to the variations of the electrical field produced at the
electrodes by the \SIV charges. It is important to assess
this charge displacement because the displacement current (and
therefore its integral, the charge displacement) can be directly
observed in experiments: the displacement charge is the `gating
charge'.

The electrode charge displaced by the motion of \SIV charges is
assessed in independent ways to check numerical accuracy. One
method directly measures the integral of the displacement that reaches
the internal or external electrode. Convenient surfaces for
measuring this displacement are the internal and external dielectric
boundaries of the membrane and protein. Since the electric field
strength perpendicular to these surfaces has already been determined
in the computation of induced charge, the electrical fluxes through
these surfaces can be computed via integration of the known normal
field strength over the surface area.

A method to solve for charge displacement is provided by the
Ramo-Shockley theorem \citep{shockley:1938,ramo:1939}. This
method first solves the electric field determining the `electrical
distance' of a source charge, which is the field produced by the
electrodes in the absence of source charges inside the simulation
cell. The gating charge corresponding to an \SIV position in a
simulation including \emph{all} charges is given by scaling source
charges by their respective electrical travel:
\begin{equation}
  \label{ramo}
  Q = -\frac{1}{1\text{ volt}}\sum_jq_j[U(ßr'_j)-U(ßr''_j)]
\end{equation}
where $U$ is the potential with all source charges removed and the
external electrode fixed at $1$~volt; and $ßr_j'$ to $ßr_j''$ are the
endpoints of the trajectory of $q_j$.  This method has been described
for simulation cells like the ones used in this study
\citep{nonner:2004:rs}. An important implication of this theorem is
that the gating charge contributed by an \SIV charge is exclusively
determined by the position of the \SIV charge in the simulated
system. It is entirely independent of other source charges, fixed or
moving, that exist in the system.

Another implication of the Ramo-Shockley theorem is used in the
computation of potential energy (described below in
\secref{potential}). The Ramo-Shockley theorem implies that applying a
voltage to the electrodes modifies the electrical potential at the
location of a charge by the fraction of applied voltage corresponding
to the electrical distance of that charge ($U(ßr_j)/\text{1~V}$). This
is true regardless of the presence of other charges and of dielectrics
in the system.

\subsection{Electrostatic potential energy}
\seclabel{potential}{subsection}{Electrostatic potential energy}

The charges of the \VS move in an electric field that originates from
other intrinsic charges, charges induced on dielectric boundaries, and
charges delivered from an external battery to the electrodes to
establish a voltage clamp (or under natural conditions from the
charges that produce the action potential). There are therefore
both internal and external sources of electrical work. The biological
purpose of the voltage sensor is to transduce electrical work from
this environment into mechanical or other work that is applied to
other parts of the channel, in particular toward operating the gate of
the conducting pore.

The \VS can be viewed as traveling in a force field that is probed
by fixing the positions of the \VS while measuring the force on the
\VS, both translational and rotational. Integrating the force along
the path of movement gives the required work for moving the \VS
between the endpoints of the probed path:
\begin{equation}
  \label{deltawork}
  \Delta W = \sum_s q_s \int_{ßr_{s,\mathrm{old}}}^{ßr_{s,\mathrm{new}}}ßE(ßr_s)\cdot dßr_s
\end{equation}
where the summation is over the source charges ($q_s$) moving with the
\VS, and $ßE$ is the field produced by all charges of the system,
excluding $q_s$ itself. Note that the electrode charge is continually
updated by the voltage clamp as the \VS moves along the path.

This method of computing work requires that an entire path be scanned
at small intervals for a path integral. It is computationally
expensive and generates cumulative numerical errors. A more efficient
method based on a direct calculation of work is desirable, one that
only needs to probe the endpoints of the path. However, I have used
the path integration of force as a useful control for other methods
that compute work. 

\def\Vz{^{V_{\mathrm{m}}\negthinspace=0}} %
A highly efficient method to compute work can be constructed using the
Ramo-Shockley theorem. This method requires the computation of the
configurational energy when the electrodes are grounded (hence
$V_m=0$) and the charge displacement for a given configuration of
source charges. Computation of the charge displacement is described in
\secref{displacement}, and the method for computing configurational
energy when $V_m=0$ follows here.

The configurational energy considered in my simulations is the
electrostatic energy (including all implicit polarization stress).
The configurational energy at a given potential is the interaction
energy of every charge in the system with: the source charges; induced
dielectric polarization charges (including implicit microscopic stress
from polarization); and the charges on the electrodes:
\begin{equation}
  \label{eq:wtot}
  W_{\textrm{config}} = W_s + W_{diel} + W_{el}
\end{equation}
where the interaction energy (with all charges) of: source charges is
$W_s$; induced charges for dielectrics is $W_{diel}$; and electrode
charges is $W_{el}$.

The configurational energy can also be decomposed into the total
electrostatic interaction of all explicit charges in the system plus
the implicit mechanical energy of twisting and stretching the
polarized molecules:
\begin{equation}
  \label{eq:wtot:2}
  \begin{split}
  W_{\textrm{config}} & = W_{electrostatic} + W_{diel,stress} \\
  & = \frac{1}{2}\sum_jq_jV(ßr_j) + W_{diel,stress}    
  \end{split}
\end{equation}
The total electrostatic energy of a discrete charge distribution is
$\frac{1}{2}\sum_j q_jV(ßr_j)$, where $q_j$ are all the source and
induced charges \emph{in vacuo} including dielectric and electrode
charges, and $V(ßr_j)$ is the potential due to all charges not located
at $ßr_j$ \citep{griffiths:1999:2}. For dielectric polarization
charges, the additional mechanical term $W_{diel,stress}$ corresponds
to the mechanical work of twisting and stretching polarized molecules,
including any implicit electric fields involved in these deformations.

The configurational energy when the electrodes are grounded
$\left(W_{\textrm{config}}\Vz\right)$ is then:
\begin{equation}
  \label{eq:eq:wv0}
  \begin{split}    
  W_{\textrm{config}}\Vz =
  \underbrace{\frac{1}{2}\sum_sq_sV(ßr_s)}_{W_s}
  & + \underbrace{
    \overbrace{\frac{1}{2}\sum_dq_dV(ßr_d)}^{W_{diel,electrical}}
    + W_{diel,mech}}_{W_{diel}\equiv 0} \\
  & + \underbrace{\frac{1}{2}\sum_{el}q_{el}V(ßr_{el})}_{W_{el}=0}
  \end{split}
\end{equation}
where $V(ßr)$ is the potential due to all charges not located at $ßr$,
$q_s$ are the source (not effective) point charges, $q_d$ are all
induced polarization charges on dielectric boundaries and surrounding
source charges, and $q_{el}$ are all charges on electrodes. The work to
polarize dielectric boundaries is $W_{diel}\equiv 0$ since there must
be no net electrical and mechanical work on induced charges at
equilibrium \citep[pg.~193]{griffiths:1999:4}. Additionally, no work
is required to place charges at the electrodes ($W_{el}=0$) when the
electrodes are grounded: $V(ßr_{el}) = 0$. 

\paragraph{\normalfont\itshape{Therefore,}} the total configurational
energy when the electrodes are grounded is:
\begin{equation}
  \label{work}
  W_{\textrm{config}}\Vz = \frac{1}{2}\sum_sq_sV(ßr_i)
\end{equation}
In other words, the total configurational energy is equal to the work
to place the source point charges in the potential field produced by
all charges, including those produced by dielectric polarization and
charges on the electrodes.

\paragraph{The Ramo-Shockley based method} used for the efficient
computation of electrostatic potential energy is based on the fact
that the Ramo-Shockley theorem allows the dissection of the total
electrostatic work (including configurational and displacement
energies) into two components $\Delta W_1$ and $\Delta W_2$, computed
separately as follows \citep{he:2001}:
\begin{enumerate*}
\item With zero voltage applied to the electrodes, compute the
  configurational energies for the given old and new positions of the
  \VS charges. If the position of VS charges change in terms of
  electrical distance between these two geometrical positions,
  displacement charge flows from one electrode to the other.

  Since both electrodes are at the same potential, no work is involved
  in that charge displacement. Hence,
\begin{equation}
    \Delta W_1 = 
    W_{\mathrm{config, new}}\Vz - W_{\mathrm{config, old}}\Vz
\end{equation}

\item Apply the desired voltage $V_{\mathrm{m}}$ between the
  electrodes. This will modify the potential energy of the VS
  charges $q_i$ by
  \begin{equation}
    \Delta W_2 = - QV_{\mathrm{m}}
  \end{equation}
  where $Q$ is the charge displaced at the electrodes when moving from
  the old to the new position (see Eq.~\ref{ramo}).
\end{enumerate*}

\noindent Therefore, the total potential energy
change sensed by the VS is:
\begin{equation}
\label{potential}
\Delta W = W_{\mathrm{config, new}}\Vz
           - W_{\mathrm{config, old}}\Vz
           - QV_{\mathrm{m}}
\end{equation}
This work defines the `electrostatic potential energy landscape' in
which the \VS operates under an applied voltage. Note that the
configurational energy only needs to be determined for one applied
voltage, 0~mV. Landscapes for any applied voltage can then be computed
once the displacement charge for the \SIV charge position at that
applied voltage is known. The displacement charge itself is
efficiently computed using the Ramo-Shockley theorem as described
previously (Eq.~\ref{ramo}). Note that for all graphs, electrostatic
potential energy of the \VS will be reported relative to the 0~nm
translation position (and 0$^\circ$ rotation for potential energy
landscapes).

\subsection{Maxwell stress}
\seclabel{maxwell}{subsection}{Maxwell stress} 

Charges buried in the protein polarize water in the baths and attract
induced polarization charges. The polarization charges are induced on
the water surface, and their attraction produces a pressure (normal
component of the Maxwell stress) that tends to bring water toward the
charges buried in the protein. If a lipid bilayer is uniformly charged
to 400~mV, the electrostrictive pressure across the bilayer is about
0.3~MPa ($3$~atm). This condition typically breaks the bilayer by
hypothetically stabilizing the formation and expansion of
transmembrane pores \citep{melikov:2001,troiano:1998}.

The normal component of the Maxwell stress (pressure, ${\cal P}$)
acting at a dielectric surface is the product of the induced surface
charge density ($\sigma$) and the normal component of the electric
field at that surface ($ßE^\bot$):
\begin{equation}
  \label{pressure}
  {\cal P}=\sigmaßE^\bot
\end{equation}
Both quantities on the right hand side are computed in my
electrostatic analysis of the system (Eq.~\ref{sigma}). Maxwell stress
may stabilize or destabilize structures composed of charges embedded
in a weak dielectric. Therefore, I assess the strength and distribution
of the normal component of Maxwell stress for \VS models.

\subsection{Statistical mechanics}
\seclabel{mechanics}{subsection}{Statistical mechanics}

A partition function is computed from the electrostatic potential
energy. The configuration space has two degrees of freedom in \VS
motion: rotation and translation. The partition function (``the key
principle of statistical mechanics,'' \citealt{feynman:1988:1}) in
discrete configuration space is:
\begin{equation}
\mathcal{Q} = \sum_{i,j} e^{-E_{ij}/k_BT}
\label{eq:partitionfct}
\end{equation}
where $i$ and $j$ are the indices of the rotational and translational
discrete positions; $E_{ij}$ is the electrostatic potential energy of
the \VS in configuration $(i,j)$; $k_B$ is the Boltzmann constant; and
$T$ is absolute temperature.

Movement of \SIV charges is restricted in my studies to the rotational
range of $-180^\circ$ to $+180^\circ$ and a typical translation range
of $-1.925$~nm to $+1.925$~nm relative to the central position. Each
degree of freedom is made discrete in 50 increments resulting in a
total of 2500 energy computations for the discrete partition function.

With this partition function known, the probability of a configuration
is:
\begin{equation}
  \label{eq:dist}
  P_{ij} = \frac{1}{\mathcal{Q}}e^{-E_{ij}/k_BT}
\end{equation}
and the expectation value of a random variable $X$ is:
\begin{equation}
\langle X\rangle = \sum_{i,j} X_{ij}P_{ij} 
= \frac{1}{\mathcal{Q}}\sum_{i,j} X_{ij} e^{-E_{ij}/k_BT}
\label{eq:randvar}
\end{equation}
The random variables of interest are the rotational and translational
positions, the associated gating charge and the Maxwell stress. These
are computed for 1~mV steps in the range from -100~mV to +100~mV of
membrane voltage.

I also compute the expectation of electrostatic potential energy for
the \VS at the particular translational positions with rotational
equilibrium established at each translational position. The
statistical weights are then given by the rotational partition
function for that translational position.

\section{Details of the simulation environment}

The model comprises a region of inhomogeneous dielectrics surrounded
by an eggshell-shaped system of electrodes
(Fig.~\ref{fig:cell}). The electrode eggshell is composed of
three regions: two half-spheres (of radius $r=$ 5.0~nm) interconnected
by an open cylinder (of radius $r=$~5.0~nm and height
$d=$~3.0~nm). The intracellular hemisphere is at a fixed
potential $\psi/2$, where $\psi$ is the applied transmembrane
voltage; the extracellular hemisphere electrode is
anti-symmetrically fixed at $-\psi/2$. The cylindrical electrode joining
the hemispherical electrodes is subdivided into bands, each of which
is held at the potential $-\psi * z/d$, where $z$ is the height of the
center of the band relative to the midpoint of the system, and $d$ is
the total length of the cylinder. Therefore the potential on the
cylinder varies linearly from $-\psi/2$ at the junction with the
extracellular electrode, to 0 at the midpoint, to $\psi/2$ at the
junction with the intracellular electrode.

The space enclosed by the electrodes is divided into four volumes: the
two aqueous baths, the membrane and the protein. The membrane is a
disk of the same height as the cylinder electrode (3.0~nm) and has a
radius of 5.0~nm. It has a fixed dielectric coefficient
$\epsilon_m=$~2, representing its lipid composition. The protein
region is located in the center of the simulation environment and
spans the membrane (its dielectric is described by a varied dielectric
coefficient, $\epsilon_p$); for the paddle model,
$\epsilon_{\textrm{p}} = \epsilon_{\textrm{m}}$. The space
between the upper surface of the membrane/protein and the
extracellular electrode and the space between the lower surface and
the intracellular electrode is an aqueous region with a fixed
dielectric coefficient of $\epsilon_w=$~80.

For all statistical mechanics calculations, simulation temperature is
fixed at 30$^\circ$C (see
eqs.~\ref{eq:partitionfct}¬\ref{eq:randvar}).

\subsection{Sliding helix model}\balance
\seclabel{details:helix}{subsection}{Sliding helix model}

Embedded at the center of the membrane is the protein region (of
$\epsilon_p$) representing the \SIV and the surrounding transmembrane
domains; $\epsilon_p$ is varied by experiment, but for most cases is
set at 4. The protein region is radially symmetric, of radius 2.15~nm.

Moving radially inward from the membrane juncture, there is the
`gating pore' from 1.966~nm to 1.266~nm. The surface smoothly dips
from a depth of $\pm$~1.5~nm at the juncture with the membrane to the
total height of counter charges plus 0.3~nm on either side (between
$\pm$~0.6~nm to $\pm$~0.975~nm); the \SII and \SIII counter-charges
are placed equally close to the protein/water interface for all
variations. The pore is smoothed by rounded corners of radius 0.15~nm.

At the center is the surface of the \SIV proper, of radius 1.266~nm
and length 6.5~nm. The \alpha¬helix lies within this envelope. The
\SIV charges are at a radius of 1.0~nm, each split into 3 charges of
1/3~$e_o$ on a circle of radius 0.122~nm centered around the charge
position (Fig.~\ref{fig:tile:helix}). This reflects the structure of
arginine residues' guanido group charge distribution. The \SIV
charges are separated in the $z$-direction by 0.45~nm and in the
$xy$-plane by 60$^\circ$ (each sixth residue completes a full turn
counter to the direction of the \alpha-helix). There are 6 \SIV
charges, some of which are eliminated or replaced with dipoles in
specific computations to simulate charge-neutralization mutants.

Counter-charges are located in the protein dielectric on a curve of
1.4~nm radius, concentric to the curve of the \SIV
charges. Counter-charges are generally spaced at angular intervals
different from those of the \SIV charges; the interval ratios of
counter- and \SIV charges are referred to by fractions between 1/2 and
3/2. There are three counter-charges, centered at the midpoint of the
membrane. To simulate experiments with charge neutralization mutants,
some of these counter-charges are either eliminated for specific
computations or replaced with dipoles.

\begin{figure*}
  \centering
  \includegraphics{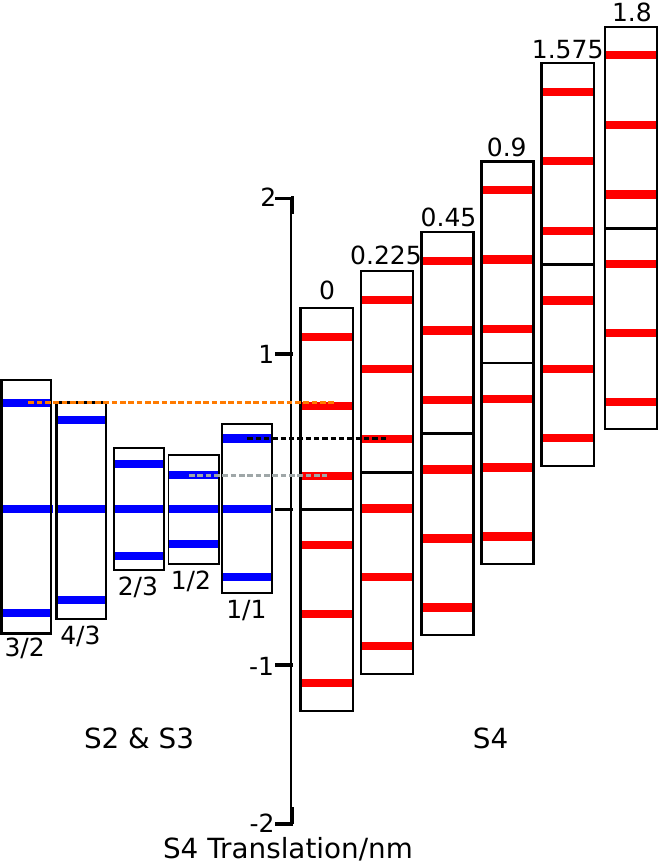}
  \caption[Configurations of charges in sliding helix
  models]{\textbf{Configurations of \SIV charges and counter-charges
      in sliding helix simulations.} Translational positions of \SIV
    charges are marked in \emph{red} and of counter-charges in
    \emph{blue}. \SIV charges are spaced at a uniform and invariant
    interval. Counter-charge interval is varied between simulations
    and is specified by its ratio with the \SIV charge interval
    (labels below the columns). The \SIV helix undergoes translation
    such that its charges line up with counter-charges to varying
    extents and at varying periods. This is illustrated on the right
    for a number of \SIV positions (translation indicated at the top
    of columns). The same map applies to the rotational dimension
    since the rotational intervals between charges are kept in a fixed
    proportion to the translational intervals (60$^\circ$ and 0.45~nm
    respectively for 1/1).}
  \label{fig:offset}
\end{figure*}

The relationship of \SIV charges and counter-charges for a given
translation is given in Fig.~\ref{fig:offset}. In figures mapping
translation to applied voltage (such as
Fig.~\ref{fig:helix:stat}~\subref*{fig:helix:stat:z}), the
translational axis is labeled in terms of that offset --- the distance
between the \emph{center} of the \SIV charges and the \emph{center} of
the model membrane and counter-charges along the translational axis of
motion, which is the vertical coordinate in Fig.~\ref{fig:cell}.

\paragraph{Dipole mutants.} To simulate charge neutralization mutants
from charged residues to glutamine or asparagine residues, two methods
are used: simple elimination of a net charge as described previously
or replacement of the charge by a dipole. The dipole is centered at the
same position as the center of the original \SIV charge or \SII \&
\SIII counter-charge. The orientation of the dipole is radial,
with the negatively charged end pointing towards the central ($z$)
axis of the simulation setup and the positive end pointing away from
the central axis. The charges of the dipole are separated by 0.27~nm
and the magnitude of each charge is 1/2 e$_0$
\citep{pauling:1960:8,lozano-casal:2008}. This dipole representation
of the mutant produces favorable interactions not present in the
representation by simple charge deletion.

\subsection{Paddle model}

In simulations of the `paddle' model, the \SIV region is entirely
buried in the membrane region with the axis of the helix parallel to
the membrane plane. The number of \SIV charges is four (as in
\KvAP). Since the radius of the \alpha¬helix is 1.122~nm, the membrane
thickness is extended by 0.5~nm to 3.5~nm, allowing translational
motion as well as rotational motion of the \SIV helix within the
membrane. I explore the range of motion possible without the emergence
of the \SIV region into the baths.

\chapter{Results and Discussion}\nobalance

\noindent Can an electrostatically viable voltage sensor model be
constructed on the basis of proposed models? If such a model is at
hand, its sensitivity to model parameter variation can be explored in
order to understand why the model works. Verifiable predictions can be
made about the electrostatic consequences of charge mutants. I will
consider three criteria in evaluating \VS models:
\begin{enumerate}
\item The \VS model produces adequate gating charge. For \Shaker \K
  channels \citep{jan:1977,hoshi:1990}, a total gating charge movement
  of $>$~3~$e_0$ per channel subunit is expected
  \citep{schoppa:1992,sigg:1994,aggarwal:1996}. When the
  membrane potential changes from a large negative to a large positive
  value, this charge determines how large the change in electric
  potential energy is for an individual \VS.

\item The model \VS moves at the time scale of gating. Since such
  motion occurs in condensed matter, its rate is limited by friction
  and possibly energy barriers. Consider the passage of a single \K
  ion across the pore of a \K channel: in a large-conductance
  \Ca-activated \K channel, a typical passage time is on the order of
  30~ns (corresponding to $\approx$~20~pA conducted by a queue of 3
  \K). If the \VS experiences at least as much friction as the
  friction in \K motion through a channel pore (even though the \VS is
  considerably larger than \K) and the \VS completes its motion within
  3~ms \citep{islas:1999}, then it can not encounter electrostatic
  energy barriers greater than 11.5~kT ($\approx$~0.3~eV).

\item The model \VS provides adequate force for operating parts of the
  channel molecule constituting the gate of the pore. This force is
  related to the number of \VS charges that are in the membrane
  electric field at a given moment. That number of charges is also a
  determinant of the slope of the gating-charge/voltage (Q/V)
  relationship. The \VS model therefore must predict the slope
  of the experimental Q/V curve.
\end{enumerate}
My computations simulate an individual, isolated \VS. The \VS is
therefore simulated under `zero-load' conditions: it does not drive
the gating machinery that a real \VS would drive when integrated in
the channel. Since experimental gating currents have been recorded
only from whole channels, the \VS performance parameters derived from
these experiments likely need to be exceeded by a viable \VS model
studied under zero-load conditions. Specifically, the total charge
movement, slope of the Q/V curve and rate of motion are expected to be
reduced in a \VS operating under its natural load.

In the following simulations, the \SIV helix is modeled as a solid
body with embedded positive charges which move with two degrees of
freedom, translation and rotation about the helix axis.
Counter-charges are kept in fixed positions. These constraints are a
first step toward understanding the electrostatics of the \VS. If the
\VS is deformed in addition to the translation and rotation of \SIV,
then total gating charge, internal friction, energy barriers and force
developed will be affected.

\section{A `paddle' model}

In the original version of the paddle model \citep{lee:2003}, the \SIV
helix was embedded in the membrane lipid (like a paddle in water) and
the proposed motion was like the transfer of a large lipophilic ion
between the two lipid/water interfaces. This model has been found
electrostatically implausible in a previous computational study
because the electrostatic work required to translocate the
multiply-charged \SIV helix across the weak dielectric of the lipid is
very large \citep{grabe:2004}. The original paddle model has been
modified since its inception; a recently proposed version
\citep{tao:2010} has gained features of the sliding helix models that
I have analyzed and which are described below (\secref{helix}), while
losing paddle-like features. This section presents computations of the
original paddle model to assess how unfavorable electrostatic features
produce unfavorable consequences for the stability and function of a
\VS design.

\begin{figure*}
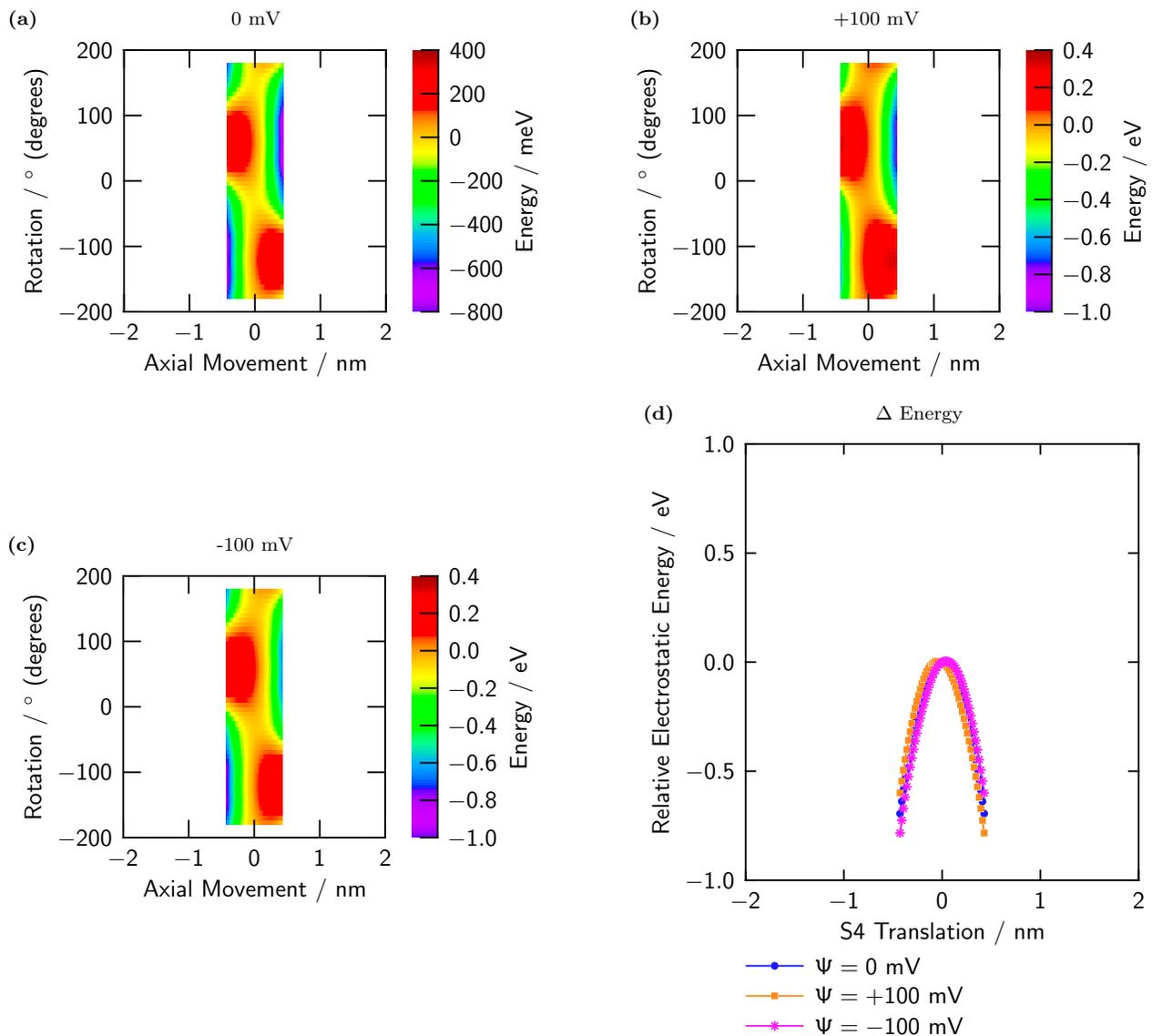

  \centering
  \figtype{paddlemp_QR}
  \setfiggroup{2d}{potential}
  \linebox{
    \subfloat[0~mV]{
      \label{fig:paddle:0}
      \label{fig:paddle:first}
      \getfiggraph{3_1_0_0_0}{0/e}
    }
    \hspace{\stretch{1}}
    \subfloat[+100~mV]{
      \label{fig:paddle:+100}
      \getfiggraph{3_1_0_0_0}{100/e}
    }
  }
  \linebox{
    \hskip-0.5em
    \valignbox{
      \subfloat[-100~mV]{
        \label{fig:paddle:-100}
        \label{fig:paddle:last}
        \getfiggraph{3_1_0_0_0}{-100/e}
      }
    }
    \hspace{\stretch{1}}
    \valignbox{
      \subfloat[\hspace{3em}$\Delta$ Energy]{
        \label{fig:paddle:dE}
        \getfiggraph{3_1_0_0_0}{denergy}
      }
    }
  } 

  \caption[Paddle: Stability in membrane]{\textbf{A paddle
      configuration is electrostatically bistable.}
    Panels~(\subref*{fig:paddle:first}-\subref*{fig:paddle:last}):
    false-color maps of electrostatic potential energy of the \VS
    scanned over two degrees of freedom for three applied
    voltages. The \SIV helix axis is parallel to the membrane plane
    and translated in the direction normal to the membrane plane. The
    \SIV segment is buried in the lipid in all scanned translational
    positions. Rotation is about the \SIV axis. Electrostatic
    potential energy strongly favors positions near the bath
    interfaces at all applied voltages.  Panel~\subref{fig:paddle:dE}:
    Rotation-averaged electrostatic potential energy versus
    translation for three applied voltages (averaging is based on the
    rotational partition function). Since electrostatic potential
    energy depends on rotational position
    (panel~\subref*{fig:paddle:0}), allowing the \SIV helix to rotate
    minimizes energy. Nevertheless, the profile of averaged energy is
    parabolic, and applied membrane voltage does not remove the large
    barrier to \SIV translation. Note that
    panels~(\subref*{fig:paddle:first}-\subref*{fig:paddle:last}) show
    the electrostatic potential energy of the \VS relative to that at
    translation 0~nm and rotation 0$^\circ$ of that graph
    (Eq. \ref{potential}), while panel~\subref{fig:paddle:dE} shows
    electrostatic potential energy of the \VS relative to translation
    0~nm (expectation over the rotational degree of freedom).}
  \label{fig:paddle}
\end{figure*}

I have simulated an \SIV helix whose movements are computationally
restricted to not extend the helix beyond the boundaries of the lipid
membrane. Electrostatic potential energy of the \VS (with respect to
the central position) is computed while varying the position in two
degrees of freedom. The \SIV helix is translated between the two
membrane boundaries with its axis kept parallel to the boundaries and
allowed to rotate fully about its axis. Electrostatic potential energy
maps for three different applied membrane voltages are shown in
Fig.~\ref{fig:paddle}~(\subref*{fig:paddle:0}-\subref*{fig:paddle:-100}),
with energy represented in false color. Note that these energy maps
are similar for all applied voltages, as if applied voltage has a
relatively small effect relative to the contributions of other
simulation parameters.

Panel~\subref{fig:paddle:dE} of Fig.~\ref{fig:paddle} displays the
mean electrostatic potential energy of the \VS as a function of
translation. This energy is computed by averaging over all
rotational angles of the paddle using the statistical weights of the
rotational partition function (see \secref{mechanics}). In other
words, the energy for each angle is used in a Boltzmann factor to
statistically weigh that energy to derive an overall expectation
energy for that translational position; graphically, one point for a
curve at specific potential in panel~\subref{fig:paddle:dE} is the
expectation value calculated from the matching column (by translation)
of the respective potential energy graph of
panels~(\subref*{fig:paddle:first}-\subref*{fig:paddle:last}).

\balance
The energy in \subref{fig:paddle:dE} has a large maximum when the \SIV
axis is in the center of the membrane, more than $0.5$~eV above the
energy at the extreme points of the scanned range of translation
(\citealt{grabe:2004} scanned a wider range of translation and
obtained an even larger variation of energy). The variation of
electrostatic potential energy due to the varied applied membrane
voltage are indeed small compared to the large electrostatic
barrier. The energy profile makes this \VS in essence a bistable
structure. When in one extreme position, the \SIV is very unlikely to
ever flip to the other position.

\figtype{paddlemp_QR}

My energy computations with this paddle model also reveal that
translating the helix produces strong rotational forces. At the end
points of the $\pm$~0.5~nm translations from the center, electrostatic
potential energy of the \VS is minimal in different rotational
positions. This applies for all tested membrane voltages. The
differences between favorable and unfavorable rotational positions
approach $1$~eV, making it highly unlikely that the \SIV segment could
undergo translation without rolling by about half a turn. A proposed
model in which the \SIV helix moves by translation alone would produce
energetically unstable configurations. If the helix is allowed to
follow electrostatic force in the rotational degree of freedom, the
electrostatics of this paddle model antagonize \VS function
(Fig.~\ref{fig:paddle}~\subref*{fig:paddle:dE}).

\begin{figure*}
  \vskip-3\baselineskip
  \centering
  \linebox{
    \subfloat[Internal at 0~mV]{
      \label{fig:paddle:press:0:int}
      \includegraphics[width=\halfwidth]{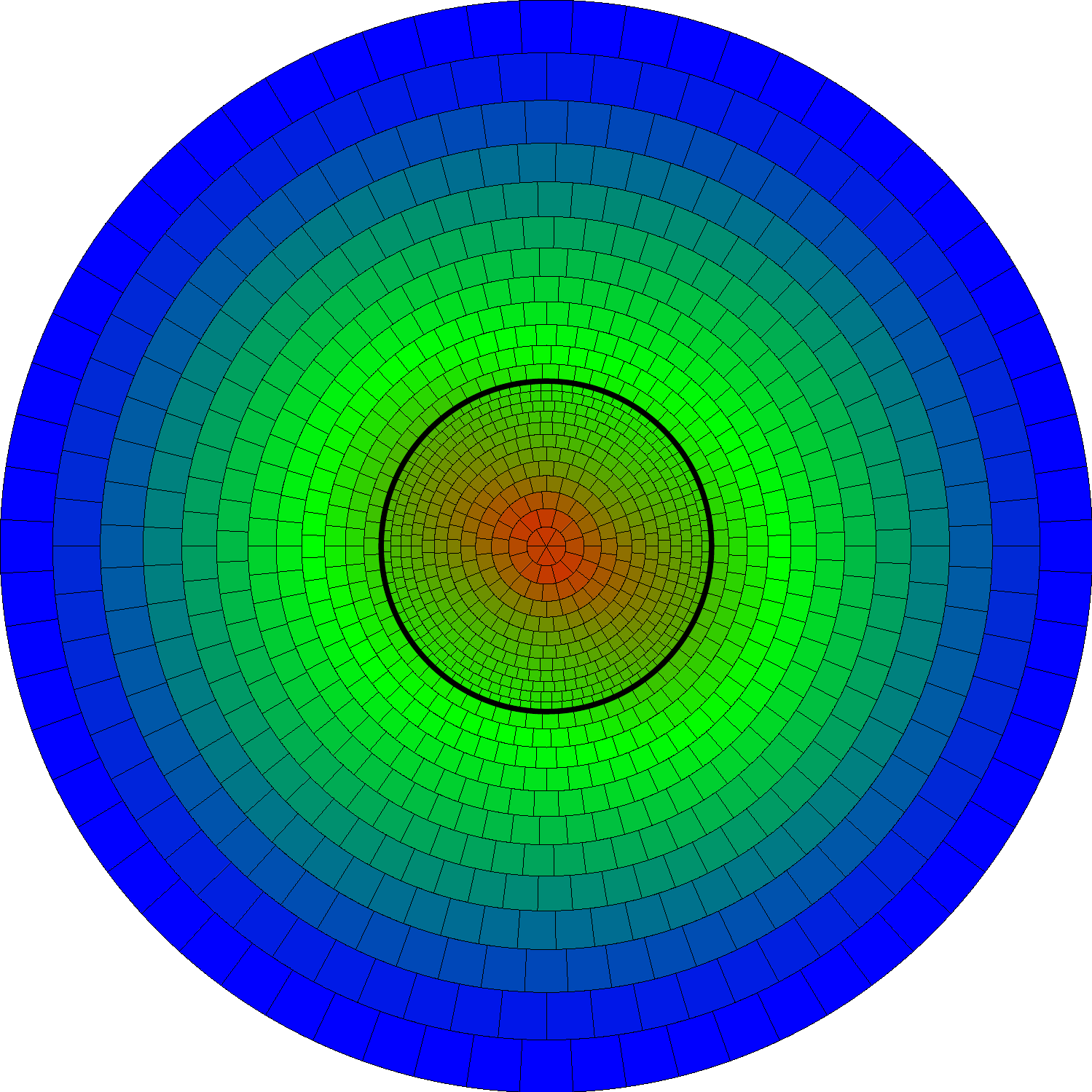}
    }
    \hspace{\stretch{1}}
    \subfloat[External at 0~mV]{
      \label{fig:paddle:press:0:ext}
      \includegraphics[width=\halfwidth]{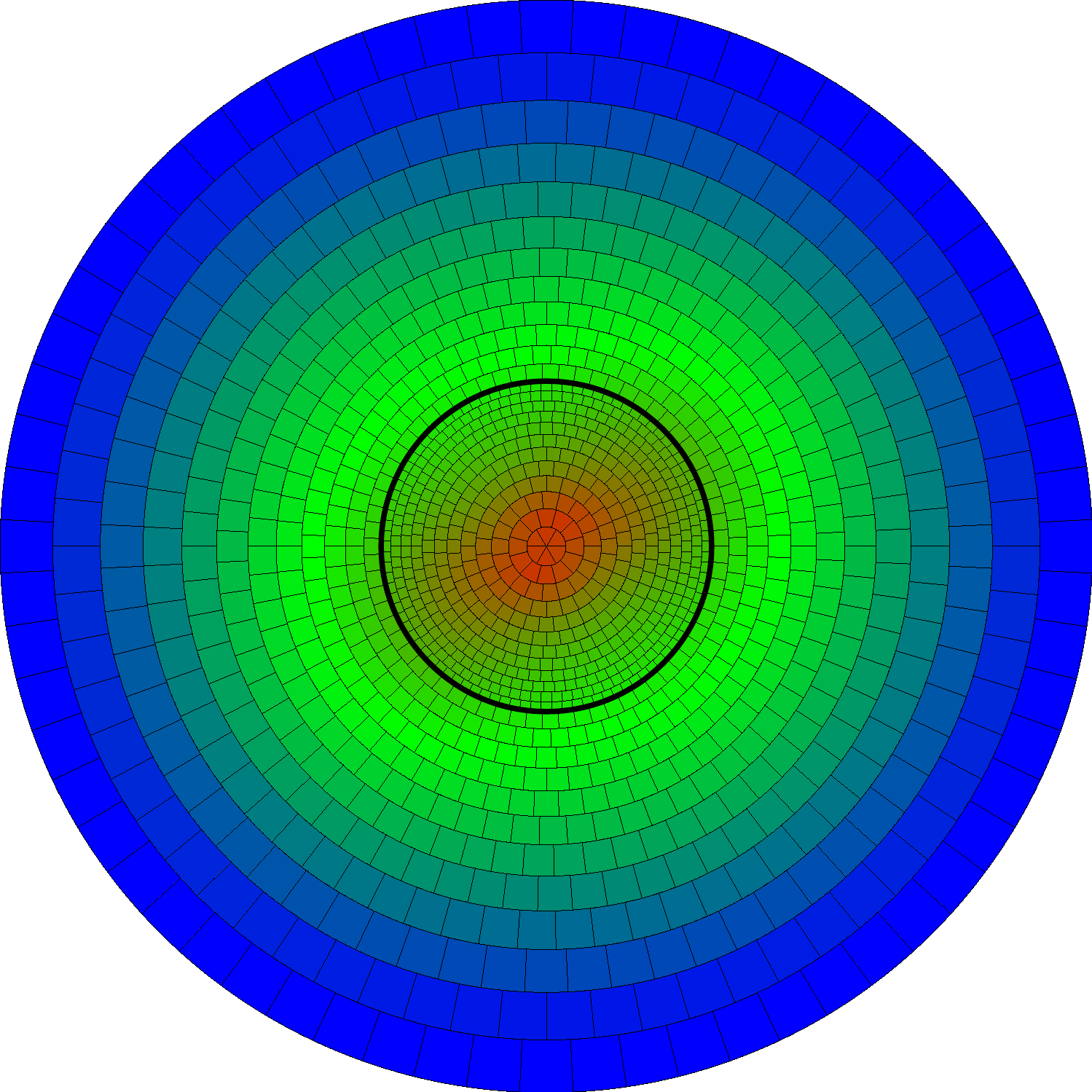}
    }
  }
  \linebox{
    \subfloat[Internal at +100~mV]{
      \label{fig:paddle:press:100:int}
      \includegraphics[width=\halfwidth]{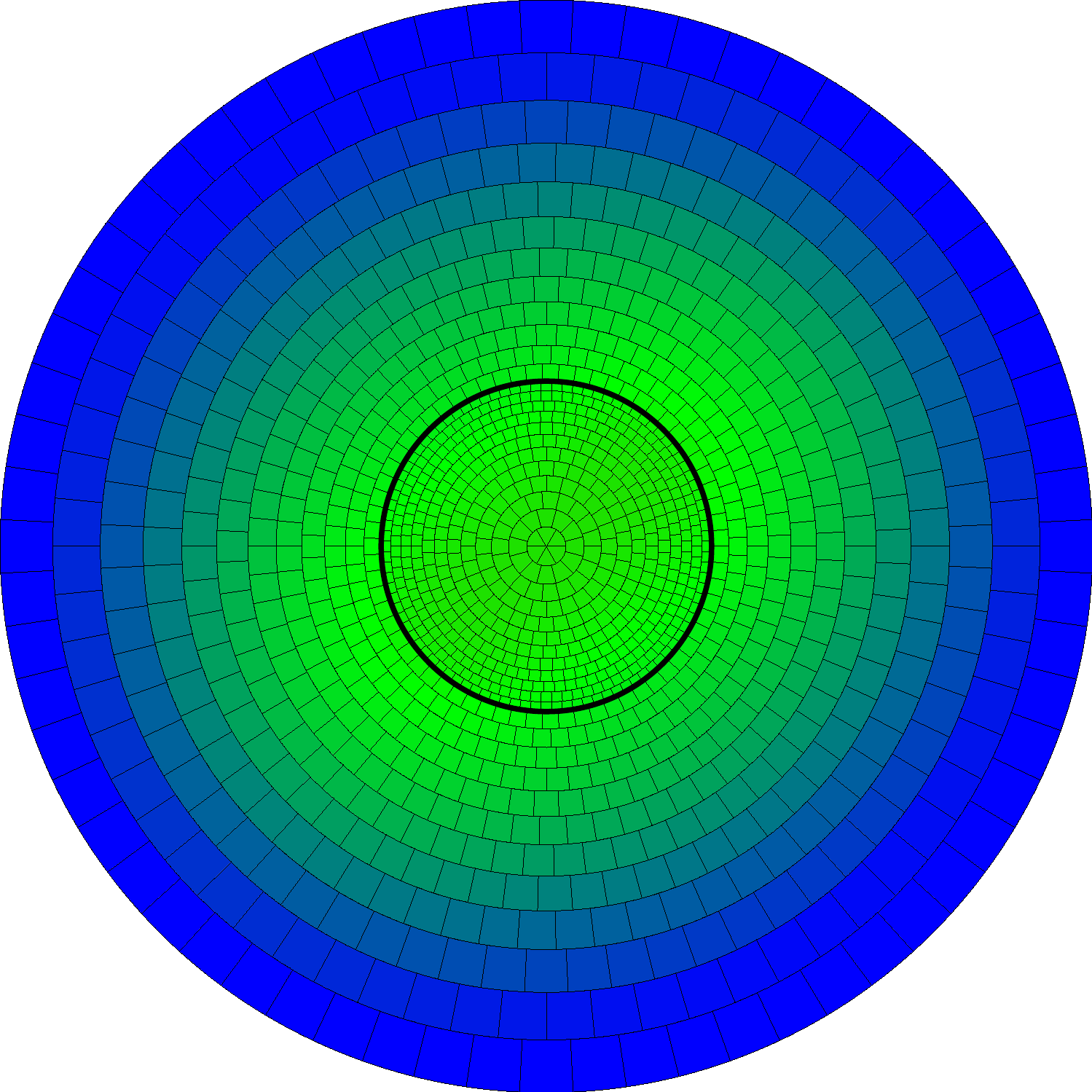}
    }
    \hspace{\stretch{1}}
    \subfloat[External at +100~mV]{
      \label{fig:paddle:press:100:ext}
      \includegraphics[width=\halfwidth]{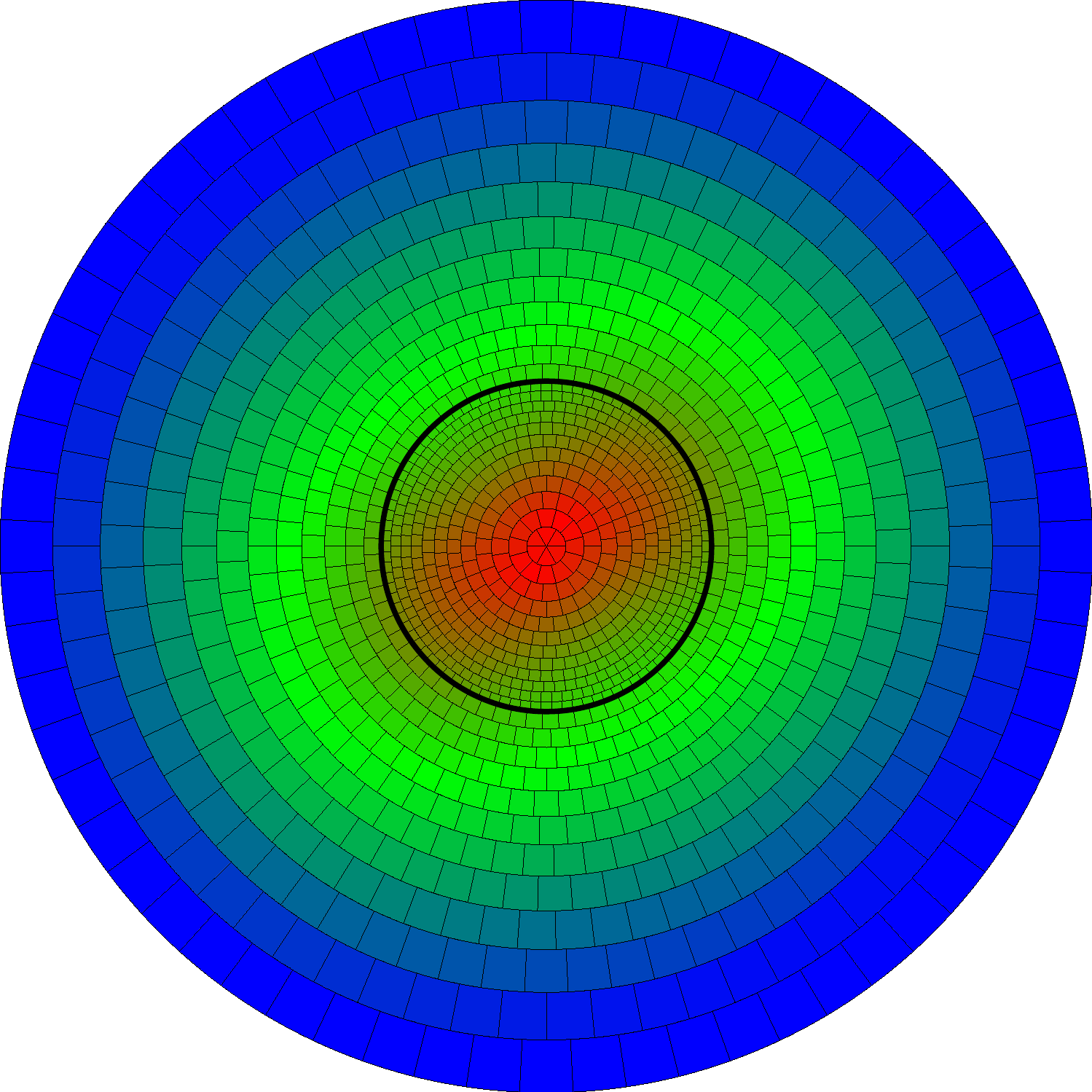}
    }
  }
  \caption[Paddle: Maxwell stress]{\textbf{A paddle configuration
      is mechanically unstable.}  False-color maps of the Maxwell
    stress acting on the water/lipid interfaces. This stress tends to
    pull the water boundary toward the \SIV charges. Logarithmic color
    scale goes from \ten{2}~Pa (\emph{blue}) through \ten{5}~Pa
    (\emph{green}) to 2.5~$\times$~\ten{8}~Pa (\emph{red}). The
    surface area shown is 10~nm in diameter. (An electrostrictive
    pressure of $\approx$~3~$\times$~\ten{5}~Pa is known to break a lipid
    bilayer.)  The position in the membrane of the \SIV segment is
    controlled by the applied membrane voltage (0 or +100~mV), and the
    Maxwell stress shown is the expectation of the Maxwell stress
    (based on the translational/rotational partition function). A
    supplementary movie (\moviefile{flat-paddle.mp4}) shows how Maxwell
    stress varies as applied voltage is varied between -100 and
    +100~mV.}
  \label{fig:paddle:press}
\end{figure*}

Lipid bilayers tend to break down when voltages larger than 400 mV are
applied \citep{melikov:2001,troiano:1998}. The electrostrictive
pressure across the bilayer under those conditions is
$\approx$~0.3~MPa (3~atm). In the paddle model, electrical charges are
located inside the lipid membrane and create strong electric
fields. These fields polarize the water adjacent to the membrane,
inducing charge at the lipid/water boundaries. \SIV and induced
charges attract one another; therefore, the lipid/water boundaries are
attracted toward the \SIV charges. This is the Maxwell stress. I have
computed Maxwell stress on the lipid/water interface (see
\secref{maxwell}) to see how this stress relates to the
electrostrictive pressure known to break lipid bilayers. For this
paddle model, the Maxwell stress on membrane boundary regions near the
\SIV charges is very large. Fig.~\ref{fig:paddle:press} shows the
pressure distributions on the internal and external bath boundaries
with a logarithmic false-color scale (pressures range from \ten{2}~Pa
in blue to 2.5~$\times$~\ten{8}~Pa in red). The peak pressure on the
bath boundaries is much larger than a safe electrostrictive stress,
even if the \SIV axis is centered in the membrane (the minimal Maxwell
stress configuration). The likely consequence of the large Maxwell
stresses of the paddle model is that the lipid retreats and thus
exposes the charged surface of the helix to both baths. Such a
configuration would not function as voltage sensor.

\newpage\nobalance\enlargethispage{\baselineskip}
\section{A `sliding helix' model}
\seclabel{helix}{section}{A `sliding helix' model}

The sliding helix models investigated here are \VS models in which the
axis of the \SIV helix is oriented perpendicularly to the plane of the
membrane. Two independent kinds of motion are allowed: translation
along the axis of the \SIV helix and rotation about the axis. No
particular trajectory in these degrees of freedom is prescribed. Thus
both motions envisaged for the `helical screw' hypothesis of \SIV
motion are possible but are not \emph{a priori} coupled to one another
as the term `screw' would imply. A concentric invagination of the
protein dielectric around the \SIV helix forms a `gating pore'. In
addition to the \SIV positive charges, three negative point charges
are present in the protein domain. For the specific version of the
sliding helix that I consider in this section, these are aligned in a
spiral pattern concentric to that of the \SIV charges, but the angular
and translational intervals between the counter-charges are chosen to
be two-thirds (2/3) that of the \SIV charge interval (i.e., 40$^\circ$
and 0.3~nm). The counter-charge positions are fixed. The
dielectric of the protein is represented by a dielectric coefficient
of 4. These parameters have been chosen via an iterative process to
identify the envelope of parameters that are physically reasonable and
consonant with known biology.

\begin{figure*}
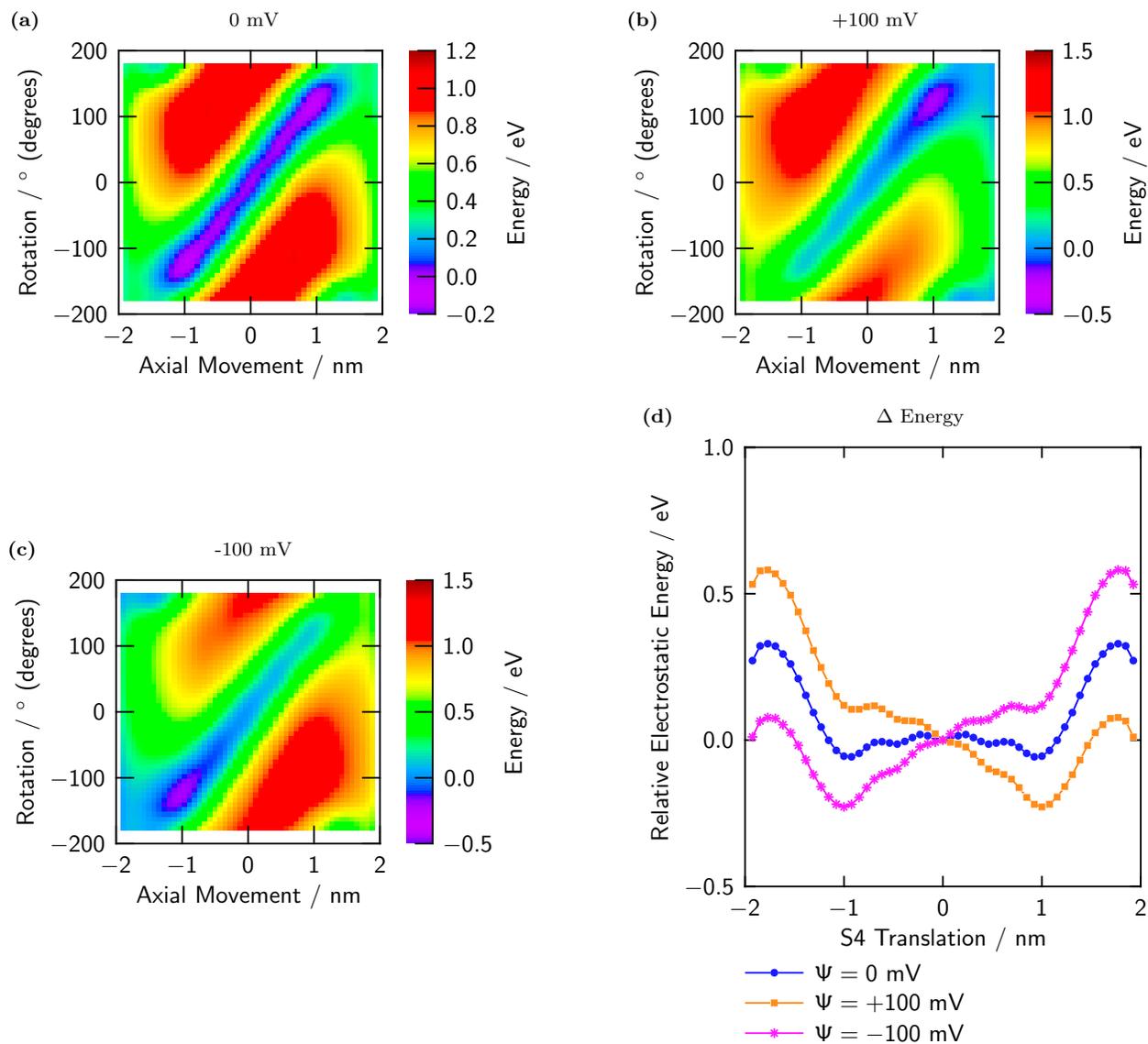

  \centering
  \figtype{s4mp_QR}
  \setfiggroup{2d}{potential}
  \linebox{
    \subfloat[0~mV]{
      \label{fig:helix:0}
      \getfiggraph{1_2_3_8_0_0}{0/e}
    }
    \hspace{\stretch{1}}
    \subfloat[+100~mV]{
      \label{fig:helix:+100}
      \getfiggraph{1_2_3_8_0_0}{100/e}
    }
  }
  \linebox{
    \hskip-0.5em
    \valignbox{
      \subfloat[-100~mV]{
        \label{fig:helix:-100}
        \getfiggraph{1_2_3_8_0_0}{-100/e}
      }
    }
    \hspace{\stretch{1}}
    \valignbox{
      \subfloat[\hspace{3em}$\Delta$ Energy]{
        \label{fig:helix:dE}
        \getfiggraph{1_2_3_8_0_0}{denergy}
      }
    }
  }

  \caption[Sliding helix: Potential energy landscapes]{\textbf{A
      sliding helix configuration has electrostatics suited for a
      \VS.}  Panels~(\subref*{fig:helix:0}-\subref*{fig:helix:-100}):
    false-color maps of electrostatic potential energy of the \VS
    scanned over two degrees of freedom, for three applied
    voltages. The \SIV helix has 6 positive charges, the three
    counter-charges are spaced at the 2/3 interval, and the protein
    dielectric coefficient is 4. The electrostatic potential energy
    map for 0~mV applied voltage forms a trough favorable to combined
    translational/rotational (`screw') motion of the \SIV
    helix. Applied voltages of -100 or +100~mV convert the energy
    trough into a pit at one end of the trough seen with
    0~mV. Panel~\subref{fig:helix:dE}: Rotation-averaged electrostatic
    potential energy versus translation for three applied voltages
    (averaging is based on the rotational partition function). Since
    electrostatic potential energy depends on rotational position
    (panel~\subref*{fig:helix:0}), allowing the \SIV helix to rotate
    minimizes energy. The averaged energy forms a trough that tends to
    restrict translation at both ends but is almost flat over
    intermediate translations (thus allowing diffusive motion of the
    \SIV helix). Applied voltage tilts this profile (promoting
    drift/diffusion of the \SIV helix). Note that
    panels~(\subref*{fig:helix:0}-\subref*{fig:helix:-100}) show the
    potential energy of the \VS relative to translation 0~nm and
    rotation 0$^\circ$, while panel~\subref{fig:helix:dE} shows the
    potential energy of the \VS relative to translation 0~nm
    (expectation over the rotational degree of freedom).}
  \label{fig:helix:land}
\end{figure*}

The landscape of electrostatic potential energy
(Fig.~\ref{fig:helix:land}) is very different from that computed for
the paddle model. When a membrane voltage of 0~mV is applied, a trough
of electrostatic potential energy tends to confine the \SIV charges to
a range of positions which can be reached by moving the \SIV helix
like a screw (Fig.~\ref{fig:helix:land}~\subref*{fig:helix:0}). Thus
the \SIV segment in this model tends to be electrostatically stable in
its environment (rather than being strongly driven towards the baths,
like the paddle considered before). The energy trough is quite shallow
however, so additional stabilizing features are required to ensure
long-term stability.

The bottom of the energy trough is nearly flat, allowing the helical
screw to visit a wide range of positions with nearly uniform
probability. When a strong positive or negative voltage is applied to
the membrane, the energy trough is shortened to a deep pit at either
end of the \SIV range of screw motion
(Figs.~\ref{fig:helix:land}~\subref*{fig:helix:+100} \&
\subref*{fig:helix:-100}). Panel~\subref{fig:helix:dE} shows
the expected electrostatic potential energy for each translational
position (a statistical average over the rotational degree of freedom
based on the rotational partition function, see
\secref{mechanics}). There are no significant energy barriers to
translation. The applied membrane voltage simply tilts the flat bottom
of the energy trough. Altogether, the electrostatic energetics of this
model are consistent with a screw motion --- the rotation is a
physical consequence of the electrostatics. This voltage sensor
strongly resists either exclusively translational or exclusively
rotational motion.

\begin{figure*}
  \centering
  \vskip-5\baselineskip
  \linebox{
    \subfloat[Internal at 0~mV]{
      \label{fig:helix:press:0:int}
      \includegraphics[width=\halfwidth]{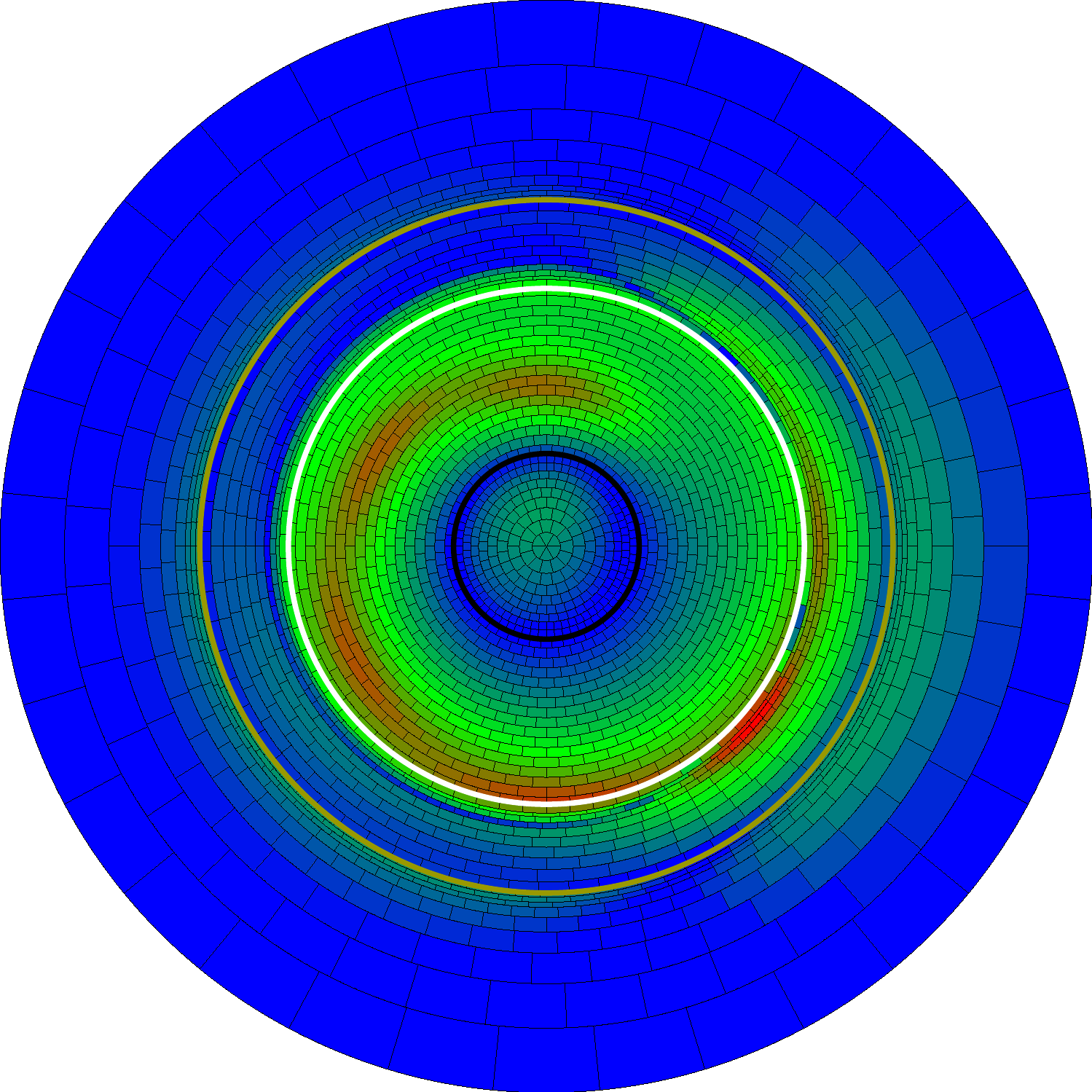}
    }
    \hspace{\stretch{1}}
    \subfloat[External at 0~mV]{
      \label{fig:helix:press:0:ext}
      \includegraphics[width=\halfwidth]{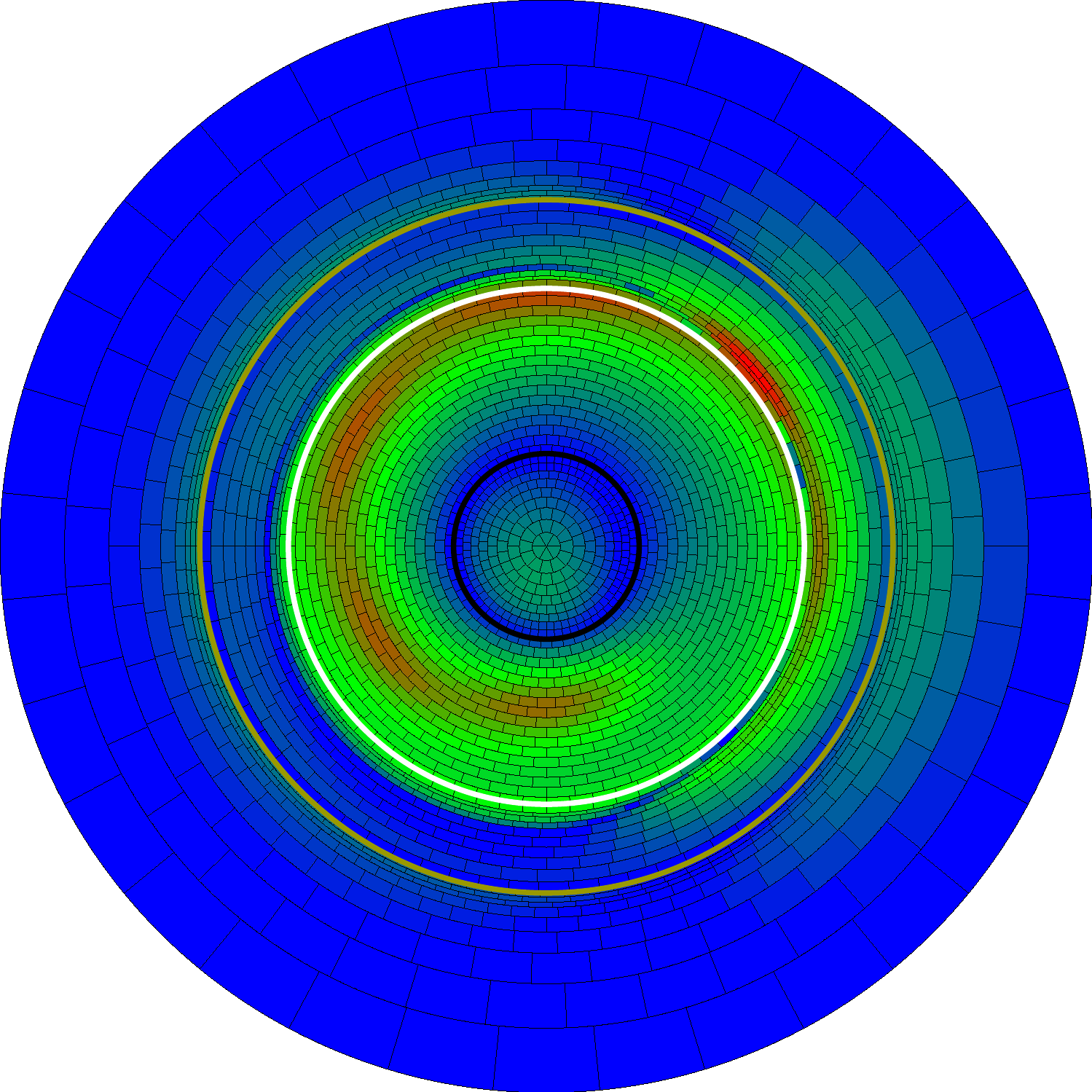}
    }
  }
  \linebox{
    \subfloat[Internal at +100~mV]{
      \label{fig:helix:press:100:int}
      \includegraphics[width=\halfwidth]{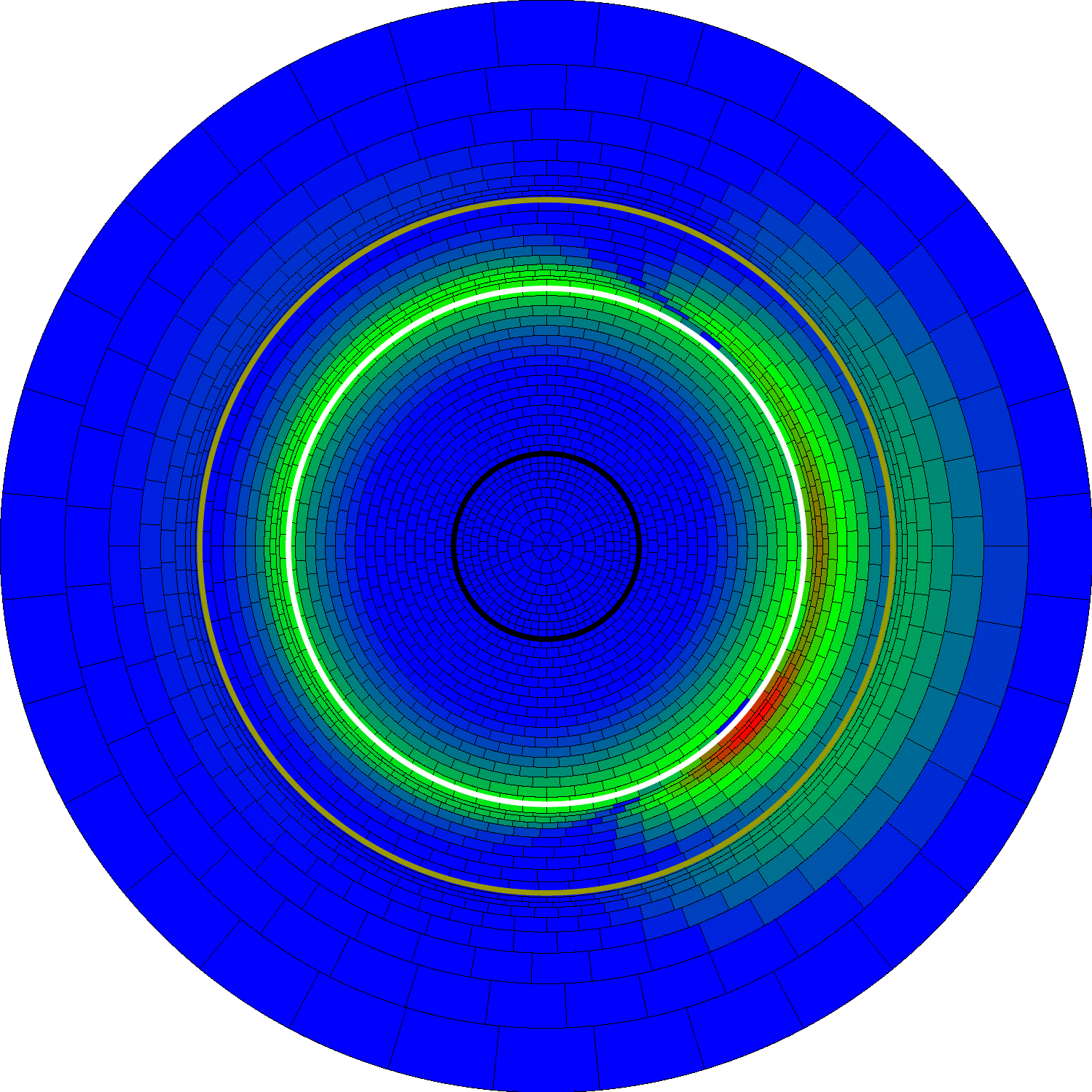}
    }
    \hspace{\stretch{1}}
    \subfloat[External at +100~mV]{
      \label{fig:helix:press:100:ext}
      \includegraphics[width=\halfwidth]{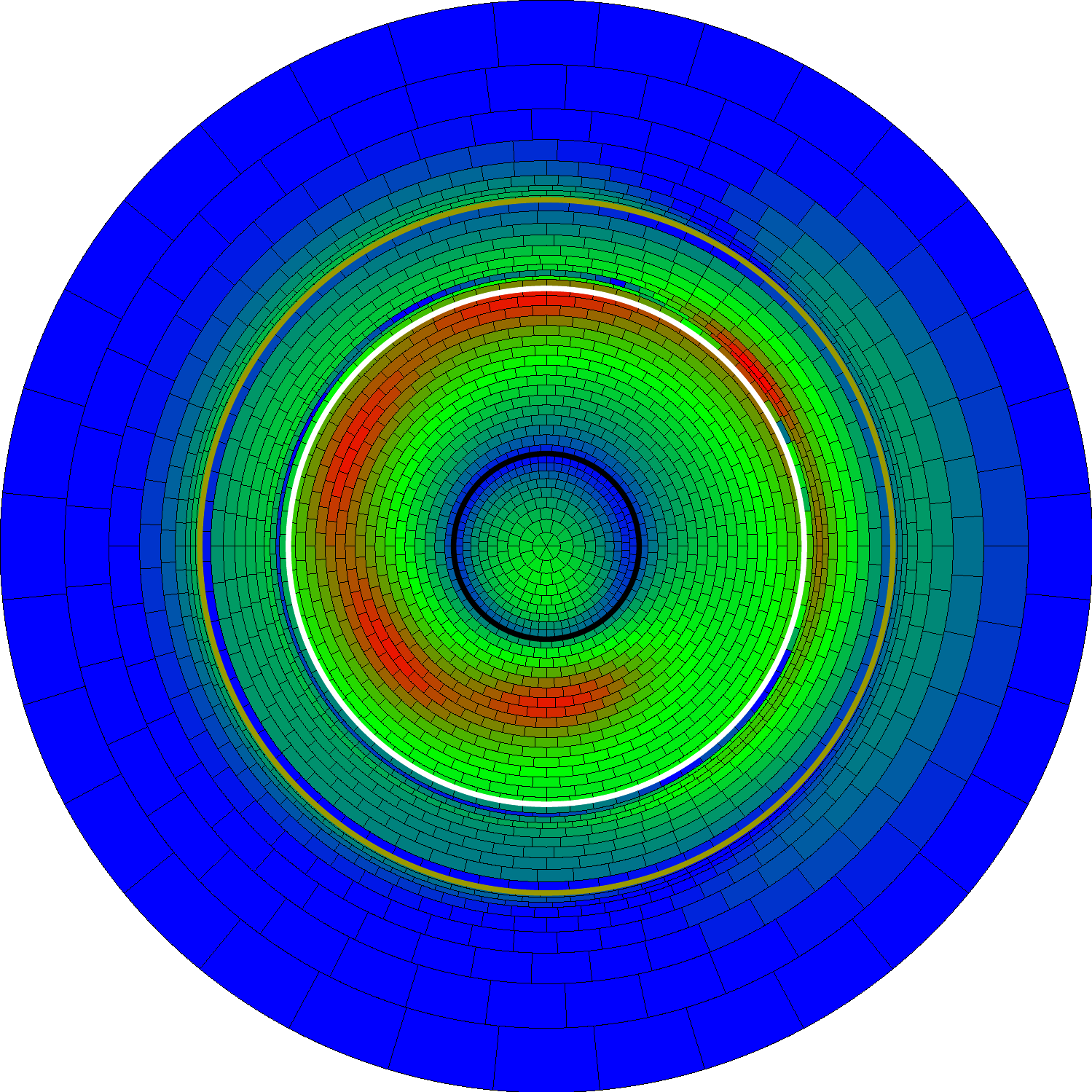}
    }
  }
  \caption[Sliding helix: Maxwell stress]{\textbf{A sliding helix
      configuration is mechanically stable.}  False-color maps of the
    Maxwell stress acting on the bath interfaces. The gating pore lies
    between the \emph{yellow} and \emph{white} rings; the top of the
    \SIV segment is delimited by the \emph{black} ring. The Maxwell
    stress tends to pull the water boundary toward the \SIV
    charges. Logarithmic color scale from \ten{2}~Pa (\emph{blue})
    through \ten{5}~Pa (\emph{green}) to 2.5~$\times$~\ten{8}~Pa
    (\emph{red}). The curved surface of the membrane/protein bath
    interface (10~nm in diameter) is projected into a plane
    (preserving path length in walking from the center to the
    periphery). High pressures occur where charges face water and in
    locations at the bottom of the gating pore (stabilizing the gating
    pore near buried counter-charges). The same model parameters are used
    as in Fig.~\ref{fig:helix:land}. The position in the membrane of
    the \SIV segment is controlled by the applied membrane voltage (0
    or +100~mV), and the Maxwell stress shown is the expectation of
    the Maxwell stress (based on the translational/rotational
    partition function). A supplementary movie
    (\moviefile{flat-s4.mp4}) shows how Maxwell stress varies as
    applied voltage is varied between -100 and +100~mV.}
 \label{fig:helix:press}
\end{figure*}

\enlargethispage{\baselineskip}
As I do for the paddle model, I assess the mechanical stability of the
dielectric geometry of the sliding helix model using the computed
Maxwell stress. Fig.~\ref{fig:helix:press} shows the pressure
distribution on the protein/water interface (note that the invaginated
interface has been mapped onto a plane, see figure legend). Both the
intra- and extracellular pressure distributions are shown for a
simulation with applied voltages of 0~mV or +100~mV.

The Maxwell stress for the sliding helix model is largest where
charges are close to the bath interface. Large stresses appear at the
water interfaces of the \SIV helix where charges face a bath. These
stresses do not compromise mechanical stability since the charged
groups are in direct contact with water. The surface region lining the
gating pore receives a moderate Maxwell stress, except for one angular
region located at the bottom of the gating pore close to the innermost
and outermost counter-charges (the gating pore is located between the
white and yellow rings in Fig.~\ref{fig:helix:press}).  There, local
Maxwell stress is on the order of \ten{8}~Pa. This magnitude of stress
provides a physical cause for the invagination of the water boundary
into a nano-scale gating pore, which requires work against the surface
tension of the water/protein interface. The existence of a gating pore
is thus made plausible by the electrostatics, although the gating pore
postulated in the model is not a computed consequence of the physics
at this level of modeling. The Maxwell stress due to the near-surface
counter-charge is narrowly localized so that it, by itself, would not
produce a full-circular gating pore like that assumed in the
model. On the other hand, \SIV gating charges could help
stabilize a larger gating pore on the side(s) where they are exposed.

Since this particular sliding helix model has desirable properties, I
compute the partition function over the rotational and translational
degrees of freedom for stepwise varied applied voltages. The
expectation values of rotational and translational positions and of
the predicted gating charge can thereby be determined as functions
of voltage. The relationship between displaced gating charge and
voltage is a prediction of the gating charge per \VS displaced when a
voltage is applied to an ensemble of channels in an experiment.

\begin{figure*}
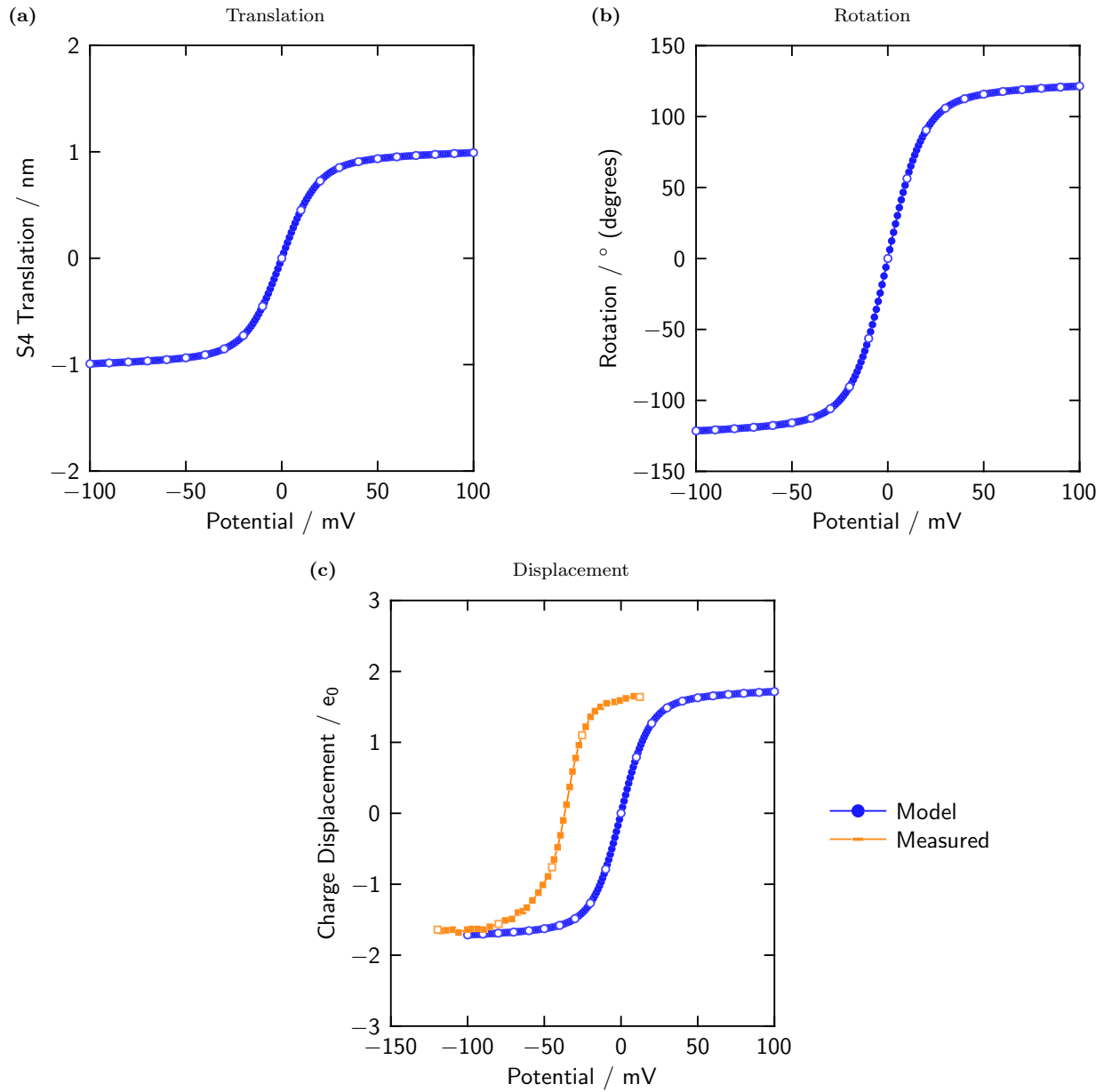

  \centering
  \figtype{s4mp_QR}
  \setfiggroup{3d}{potential}
  \linebox{
    \subfloat[\hspace{3em}Translation]{
      \label{fig:helix:stat:z}
      \getfiggraph{1_2_3_8_0_0}{ZV}
    }
    \hspace{\stretch{1}}
    \subfloat[\hspace{3.5em}Rotation]{
      \label{fig:helix:stat:phi}
      \getfiggraph{1_2_3_8_0_0}{phiV}
    }
  }
  \setfiggroup{3d}{over-leg}
  \hspace*{\stretch{1}}
  \valignbox{
    \subfloat[\hspace{2.5em}Displacement]{
      \label{fig:helix:stat:disp}
      \getfiggraph{1_2_3_8_0_0.seoh}{displacementV}
    }
  }
  \valignbox{
    \subfloat{
      \getfiggraph{1_2_3_8_0_0.seoh}{displacementV-leg}
    }
  }
  \hspace*{\stretch{0.5}}
  \caption[Sliding helix: Action of voltage]{\textbf{The action of
      voltage on a sliding helix \VS:} Expectations of
    translation~\subref{fig:helix:stat:z},
    rotation~\subref{fig:helix:stat:phi} and displaced gating
    charge~\subref{fig:helix:stat:disp} in response to varied applied
    voltage. The expectations of these random variables are computed
    using the electrostatic partition function for the two degrees of
    freedom. The same parameters for the sliding helix model are used
    as in Fig.~\ref{fig:helix:land}. Voltage moves the mean position
    of the \VS in a screw-like trajectory and displaces gating charge
    of the proper magnitude with the proper slope (an experimental
    charge/voltage curve \citep{seoh:1996} for \Shaker \Kv channels is
    shown as the \emph{orange} line in
    panel~\subref*{fig:helix:stat:disp}).}
  \label{fig:helix:stat}
\end{figure*}

Figs.~\ref{fig:helix:stat}~\subref{fig:helix:stat:z} \&
\subref{fig:helix:stat:phi} show the expectation of position for the
model \VS at given voltages. The positions follow the trajectory of a
screw motion. Varying the voltage gradually from -100~mV to +100~mV
drives the expected position of the \VS over the range suggested by
the potential energy landscapes in Fig.~\ref{fig:helix:land}. The
amount of gating charge displaced over that range of motion
(Fig.~\ref{fig:helix:stat}~\subref*{fig:helix:stat:disp}, blue line)
exceeds 3~$e_0$ per \VS, which is close to the total gating charge
measured per \VS in \Shaker \K channels (orange line). Since a \VS
driving coupled gating machinery might have a smaller range of motion,
it is reasonable to expect a model of an uncoupled \VS to produce at
least as much charge per \VS as that observed in channels.

Charge is displaced in the model over a voltage range symmetrical with
respect to 0~mV of applied voltage, whereas the experimental charge
displacement is centered about a negative voltage. The slopes
of the two charge displacement curves are quite similar; the chief
difference between the model \VS and the real \VS is an offset between
the charge/voltage curves. The real \VS is integrated into a channel
and drives the activation gating of the channel. My model voltage
sensor is isolated and drives no load. If the voltage offset between
the two charge/voltage relations is due to the gating work that the
real voltage sensor does on the rest of the channel, then:
\acsfont{(1)} the gating work is applied in closing the channel, and
\acsfont{(2)} the counter-force exerted by the gate onto the \VS is
approximately constant over the range of \VS travel (in contrast to an
elastic counter-force).

The sliding helix model presented fulfills the criteria for a viable
\VS model as listed above. Therefore in this study I will use those
model parameters as a basis of further exploration of \VS electrostatics.

  \section{Which model features are important for 
    voltage-sensing by a sliding helix?}

\noindent
Sliding helix models have many features that can be
parametrized and studied: counter-charge position \& number, protein
dielectric, gating pore size \& shape and membrane thickness, for
example. In this section I present three features of interest that
show sensitivity and a significant effect on \VS function. In
particular, the existence of counter-charges, their spacing and the
local dielectric through which they interact with \SIV charges are
strong determinants of the viability of a \VS model.

\subsection{Counter-charges eliminate the induced-charge barrier}
\seclabel{barrier}{subsection}{Counter-charges eliminate the
  induced-charge barrier}

\begin{figure*}
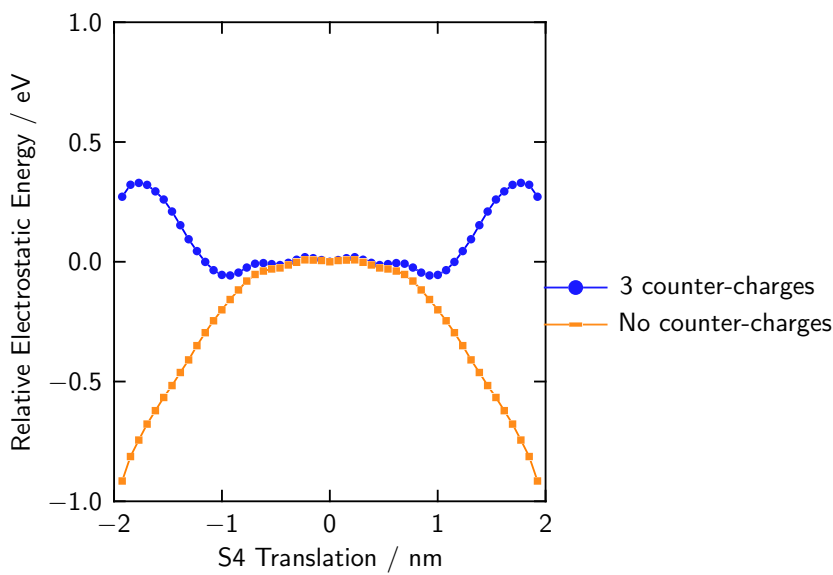

  \centering
  \figtype{s4mp_QR}
  \setfiggroup{2d}{over}
  \valignbox{
    \subfloat{
      \getfiggraph{1.2.3.8.0.0_1.2.0.8.0.0}{denergy}
    }
  }
  \makebox[2em]{
    \hspace{6em}
    \valignbox{
      \subfloat{
        \getfiggraph{1.2.3.8.0.0_1.2.0.8.0.0}{denergy-leg}
      }
    }
  }
  \caption[Counter-charges: Energetically
  required]{\textbf{Counter-charges are required for a sliding helix
      to function as \VS.}  Electrostatic potential energy of the \VS
    averaged over rotation and relative to that at translation zero
    versus translation. The \emph{blue} line describes the sliding
    helix model of Fig.~\ref{fig:helix:land} which includes three
    counter-charges. The \emph{orange} line describes the same model
    but with all counter-charges deleted. The energy profile then
    becomes parabolic like that of a paddle model (compare Fig.~\ref
    {fig:paddle}~\subref*{fig:paddle:dE}).}
  \label{fig:counter:dE}
\end{figure*}

The gating pore reduces the length of the \SIV segment exposed to the
weak dielectric separating the two baths. The sliding helix model
studied in the previous section also includes three negative
counter-charges in the region of weak dielectric. Either feature is
expected to reduce the energetic cost of moving the \SIV charges from
the baths into the region of weak dielectric. The role of the
counter-charges can be assessed by deleting them from the
model. Fig.~\ref{fig:counter:dE} shows electrostatic potential energy
of the \VS versus translation for the models with and without three
counter-charges. Removing the counter-charges produces a large
electrostatic barrier, much like that computed for the paddle model
described above (Fig.~\ref{fig:paddle}~\subref*{fig:paddle:dE}, note
the distinct shape in the mid-range, however).

The \SIV charges of the sliding helix model induce a substantial
charge on the bath interfaces. The induced charges are negative,
attracting the \SIV charges toward the baths and thereby destabilizing
the buried \SIV charges. With the buried positive \SIV charges
neutralized by negative counter-charges, the charges induced on the
bath interfaces are greatly reduced, creating a trough of
electrostatic potential energy. These computations indicate that an
appropriate number of counter-charges are required if the sliding
helix model is to function as a \VS. The gating pore alone does not
lower the induced-charge barrier to the extent needed for the sliding
helix models to function as a \VS.

Another consequence of the deletion of all counter-charges from the
model is that all rotational positions now have equal electrostatic
potential energy. Thus the \SIV segment in the model no longer
operates like a screw. If the delivery of torque is important for
operating the gate of the real channel, this would add another
consequence to neutralizing mutations of \VS counter-charges.

\subsection{Counter-charge spacing matters}

The counter-charges of the described sliding helix models are arranged
following the spiral curve on which the \SIV charges are
positioned. The intervals of the \SIV and counter-charges, however,
differed for the previously described model (\secref{barrier}): there,
the counter-charges were spaced at 2/3 the interval between \SIV
charges. This precludes simultaneous alignments of more than one \SIV
charge with a counter-charge (see Fig.~\ref{fig:offset}). In order to
see how much the close-range electrostatic interactions of charge and
counter-charge affect electrostatic potential energy of the \VS, I
have computed the consequences of counter-charge intervals of 1/2,
1/1, 4/3, and 3/2 times the interval of \SIV charges. A later section
(\secref{mutants}) will present computations of less regular charge
spacings obtained by deleting charges at certain positions of a
periodic pattern.

\begin{figure*}
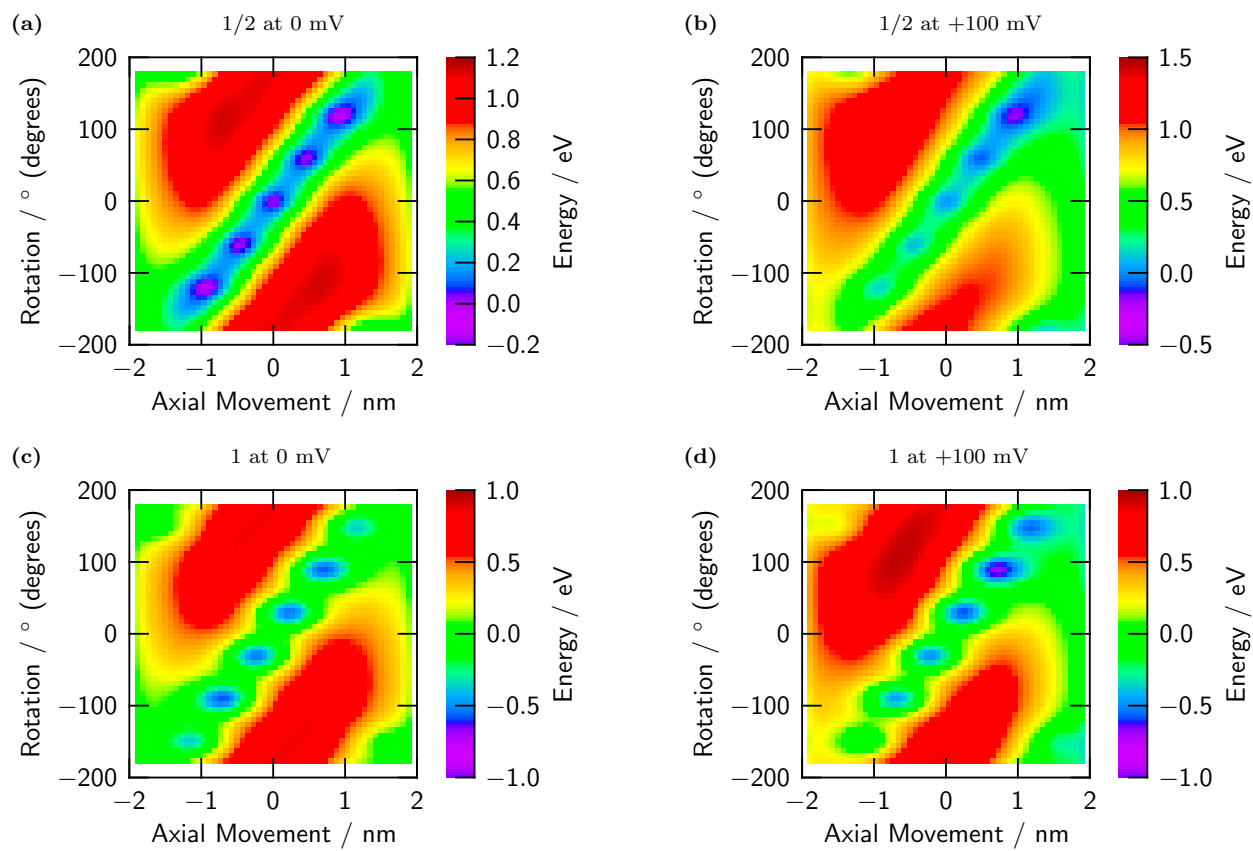

  \centering
  \figtype{s4mp_QR}
  \setfiggroup{2d}{potential}
  \linebox{
    \subfloat[1/2 at 0~mV]{
      \label{fig:pitch:1/2:0mV}
      \getfiggraph{0_2_3_8_0_0}{0/e}
    }
    \hspace{\stretch{1}}
    \subfloat[1/2 at +100~mV]{
      \label{fig:pitch:1/2:+100mV}
      \getfiggraph{0_2_3_8_0_0}{100/e}
    }
  }
  \linebox{
    \subfloat[1 at 0~mV]{
      \label{fig:pitch:1:0mV}
      \getfiggraph{2_2_3_8_0_0}{0/e}
    }
    \hspace{\stretch{1}}
    \subfloat[1 at +100~mV]{
      \label{fig:pitch:1:+100mV}
      \getfiggraph{2_2_3_8_0_0}{100/e}
    } }
  \caption[Counter-charge spacing: Potential energy landscapes
  (Pt.~1)]{\textbf{Counter-charge spacing controls electrostatic
      potential energy landscape (Pt.~1).} The spacing is specified as
    the ratio of counter-charge spacing to \SIV charge spacing (the
    ratio applies to both the rotational and translational
    spacing). The protein dielectric coefficient is 4. Energy of each
    configuration is represented relative to translation 0~nm,
    rotation 0$^\circ$. Note differences in scale.
    
    \textbf{Figure Continued in Pt.~2.}}
  \label{fig:pitch:land}
\end{figure*}
\begin{figure*}
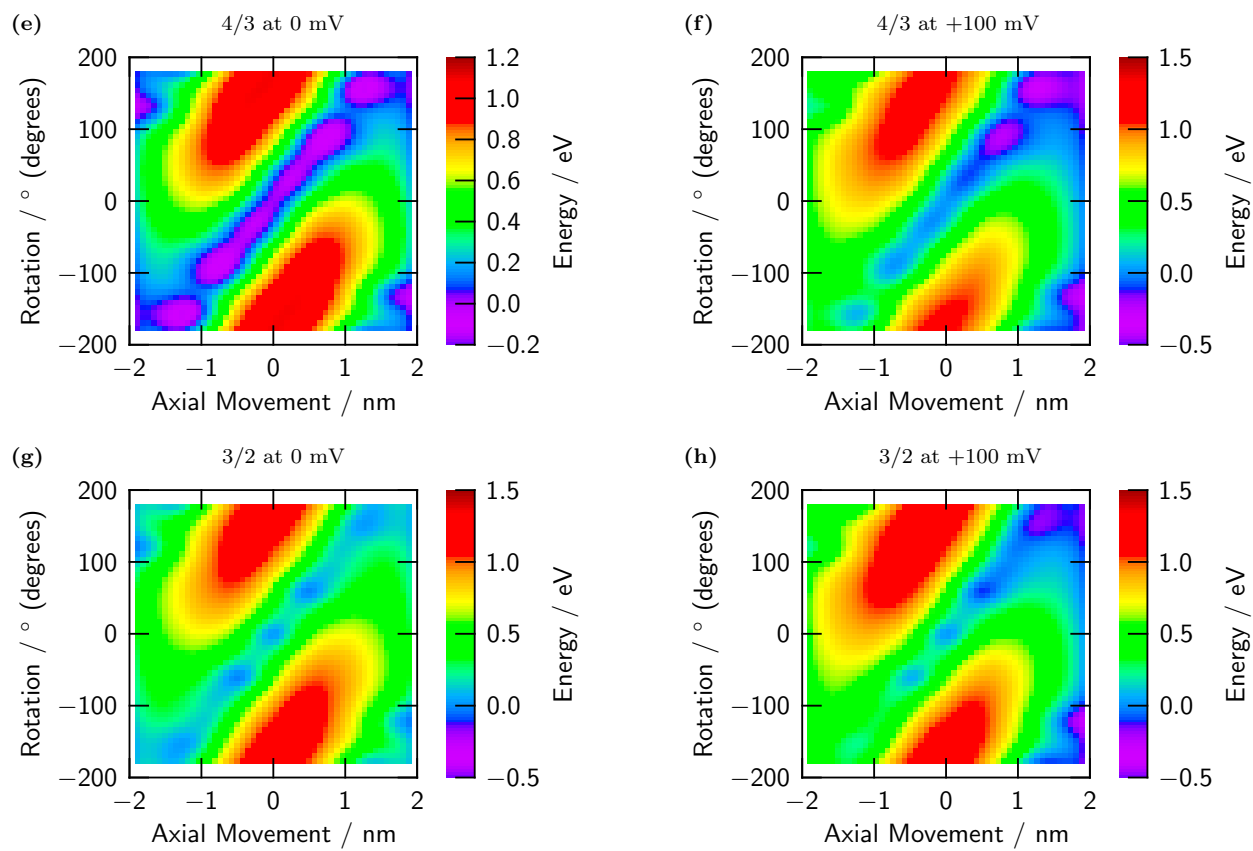

  \centering
  \figtype{s4mp_QR}
  \setfiggroup{2d}{potential}
  \ContinuedFloat
  \linebox{
    \subfloat[4/3 at 0~mV]{
      \label{fig:pitch:4/3:0mV}
      \getfiggraph{4_2_3_8_0_0}{0/e}
    }
    \hspace{\stretch{1}}
    \subfloat[4/3 at +100~mV]{
      \label{fig:pitch:4/3:+100mV}
      \getfiggraph{4_2_3_8_0_0}{100/e}
    }
  }
  \linebox{
    \subfloat[3/2 at 0~mV]{
      \label{fig:pitch:3/2:0mV}
      \getfiggraph{3_2_3_8_0_0}{0/e}
    }
    \hspace{\stretch{1}}
    \subfloat[3/2 at +100~mV]{
      \label{fig:pitch:3/2:+100mV}
      \getfiggraph{3_2_3_8_0_0}{100/e}
    }
  }
  \caption[Counter-charge spacing: Potential energy landscapes
  (Pt.~2)]{\textbf{Counter-charge spacing controls electrostatic
      potential energy landscape (Pt.~2).}

    \textbf{Figure continued from Pt.~1.}
  }
\end{figure*}

\begin{figure*}
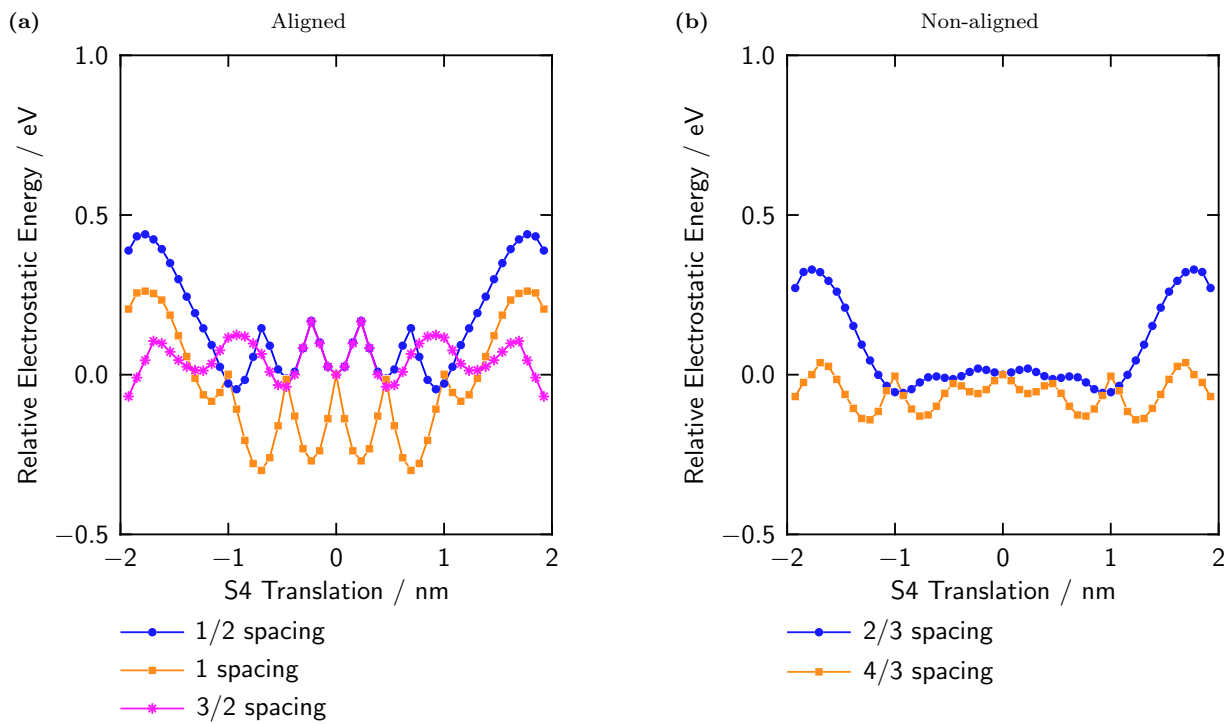

  \centering
  \figtype{s4mp_QR}
  \setfiggroup{2d}{over}
  \linebox{
    \subfloat[\hspace{3em}Aligned]{
      \label{fig:pitch:dE:reson}
      \getfiggraph{0.2.3.8.0.0_2.2.3.8.0.0_3.2.3.8.0.0}{denergy}
    }
    \hspace{\stretch{1}}
    \subfloat[\hspace{3em}Non-aligned]{
      \label{fig:pitch:dE:resoff}
      \getfiggraph{1.2.3.8.0.0_4.2.3.8.0.0}{denergy}
    }
  }
  \caption[Counter-charge spacing: Electrostatic barriers \& wells]{
    \textbf{Counter-charge alignment with \SIV charges creates
      electrostatic barriers and wells.} Electrostatic potential
    energy of the \VS versus translation, averaged over rotation and
    relative to translation 0~nm. The spacing of counter-charges is
    varied. The protein dielectric coefficient is 4. The spacings of
    1/2, 1/1, and 3/2 allow two counter-charges to align with
    corresponding \SIV charges in certain positions, thereby
    generating ripples of electrostatic potential energy
    (panel~\subref*{fig:pitch:dE:reson}). The spacings that allow only
    one counter-charge to align at a time with an \SIV charge produce
    smoother profiles of energy
    (panel~\subref*{fig:pitch:dE:resoff}).}
  \label{fig:pitch:dE}
\end{figure*}

Fig.~\ref{fig:pitch:land} shows maps of electrostatic potential energy
in the translational and rotational degrees of freedom, for both 0~mV
and +100~mV applied voltage. The potential energy of the \VS expected
when rotation is free is plotted versus translation in
Fig.~\ref{fig:pitch:dE} with 0~mV curves for different spacings
superimposed. The 1/2, 1/1, and 3/2 intervals yield energy landscapes
more hilly than those of the 2/3 or 4/3 intervals; the \SIV segment of
these models with counter-charges spaced at 1/2, 1/1 or 3/2 intervals
tend to dwell in more discrete positions in energy valleys. When a
strong voltage of +100~mV is applied in these models, some of the
energy valleys persist as discrete features.

\begin{figure*}
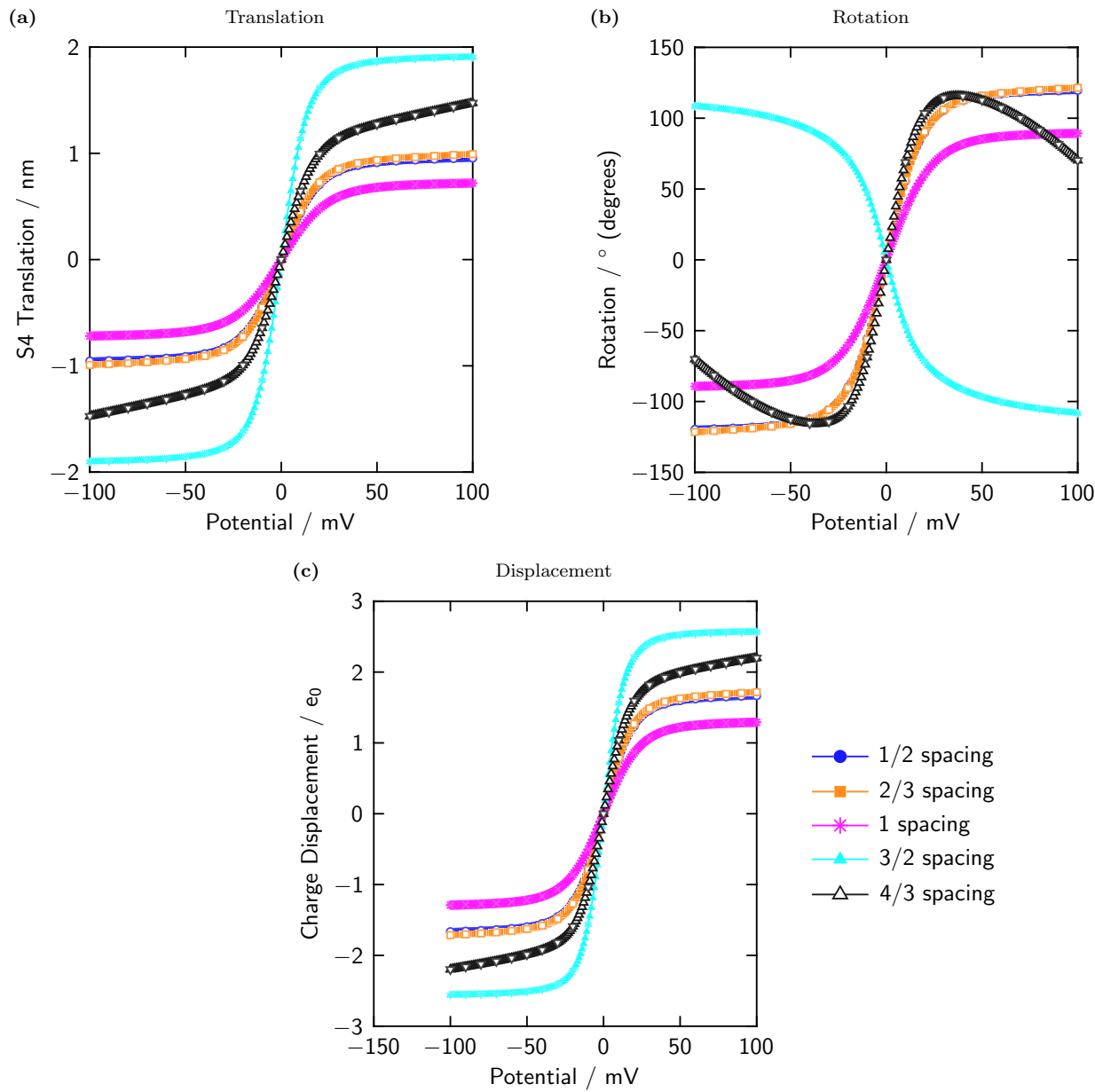

  \centering
  \figtype{s4mp_QR}
  \setfiggroup{3d}{over-leg}
  \linebox{
    \subfloat[\hspace{3em}Translation]{
      \label{fig:pitch:stat:z}
      \getfiggraph{0.2.3.8.0.0_1.2.3.8.0.0_2.2.3.8.0.0_3.2.3.8.0.0_4.2.3.8.0.0}{ZV}
    }
    \hspace{\stretch{1}}
    \subfloat[\hspace{3.5em}Rotation]{
      \label{fig:pitch:stat:phi}
      \getfiggraph{0.2.3.8.0.0_1.2.3.8.0.0_2.2.3.8.0.0_3.2.3.8.0.0_4.2.3.8.0.0}{phiV}
    }
  }
  \setfiggroup{3d}{over-leg}
  \hspace*{\stretch{0.5}}
  \valignbox{
    \subfloat[\hspace{2.5em}Displacement]{
      \label{fig:pitch:stat:disp}
      \getfiggraph{0.2.3.8.0.0_1.2.3.8.0.0_2.2.3.8.0.0_3.2.3.8.0.0_4.2.3.8.0.0}{displacementV}
    }
  }
  \valignbox{
    \subfloat{
      \getfiggraph{0.2.3.8.0.0_1.2.3.8.0.0_2.2.3.8.0.0_3.2.3.8.0.0_4.2.3.8.0.0}{displacementV-leg}
    }
  }
  \hspace*{\stretch{0.25}}
  \caption[Counter-charge spacing: Statistical
  mechanics]{\textbf{Counter-charge spacing is important for \VS
      response to voltage:} Expectation values of
    translation~\subref{fig:pitch:stat:z},
    rotation~\subref{fig:pitch:stat:phi} and displaced gating
    charge~\subref{fig:pitch:stat:disp} in response to varied applied
    voltage. The expectations for these random variables are computed
    using the electrostatic partition functions for the two degrees of
    freedom. The same sliding helix models are used as in
    Figs.~\ref{fig:pitch:land} \& \ref{fig:pitch:dE}. Counter-charge
    spacing controls the extent of \SIV charge motion, direction of
    rotation, as well as magnitude of gating charge and shape of the
    charge/voltage relation. (Note that 1/2 interval in \emph{blue}
    falls closely on top of 2/3 interval in \emph{orange}).}
  \label{fig:pitch:stat}
\end{figure*}

What are the consequences of these different potential energy
landscapes for the gating charge displaced in response to applied
voltage? How does the expected position of the \SIV region respond to
voltage?  Fig.~\ref{fig:pitch:stat} shows the expectation
translation~\subref{fig:pitch:stat:z},
rotation~\subref{fig:pitch:stat:phi} and charge/voltage
relation~\subref{fig:pitch:stat:disp} based on the partition function
over rotational and translational degrees of freedom. The charge
voltage relations vary in steepness and total displaced gating charge
even though the numbers of \SIV charges and counter-charges are fixed,
as is the dielectric environment. These variations of the
charge/voltage curve must originate from the range of travel produced
by the applied voltage as well as the distribution of \SIV positions
among more or less discrete locations. Predicting these variations
requires numerical analysis of the electrostatics.

The expectation values for rotation and translation at different
voltages (Figs.~\ref{fig:pitch:stat}~\subref*{fig:pitch:stat:z} \&
\subref*{fig:pitch:stat:phi}) reveal a monotonic increase in
translation as voltage is increased; however, non-monotonic variations
of rotation occur in some cases. Thus, for the 3/2 counter-charge
interval, the rotation is in the opposite direction to that seen for
other intervals. Translational motion (and ability to produce
translational force) is robust, while rotational motion (and ability
to produce torque) is sensitive to counter-charge
alignment. Nevertheless, the \SIV segment is expected to rotate for
all tested counter-charge intervals.

Varying a single parameter of counter-charge configuration has strong
effects on function in these sliding helix \VS models. Counter-charges
and their arrangement are crucial for building a working \VS.

  \subsection{Electrical polarizability of the protein controls effective
    gating charge}

\begin{sidewaysfigure*}
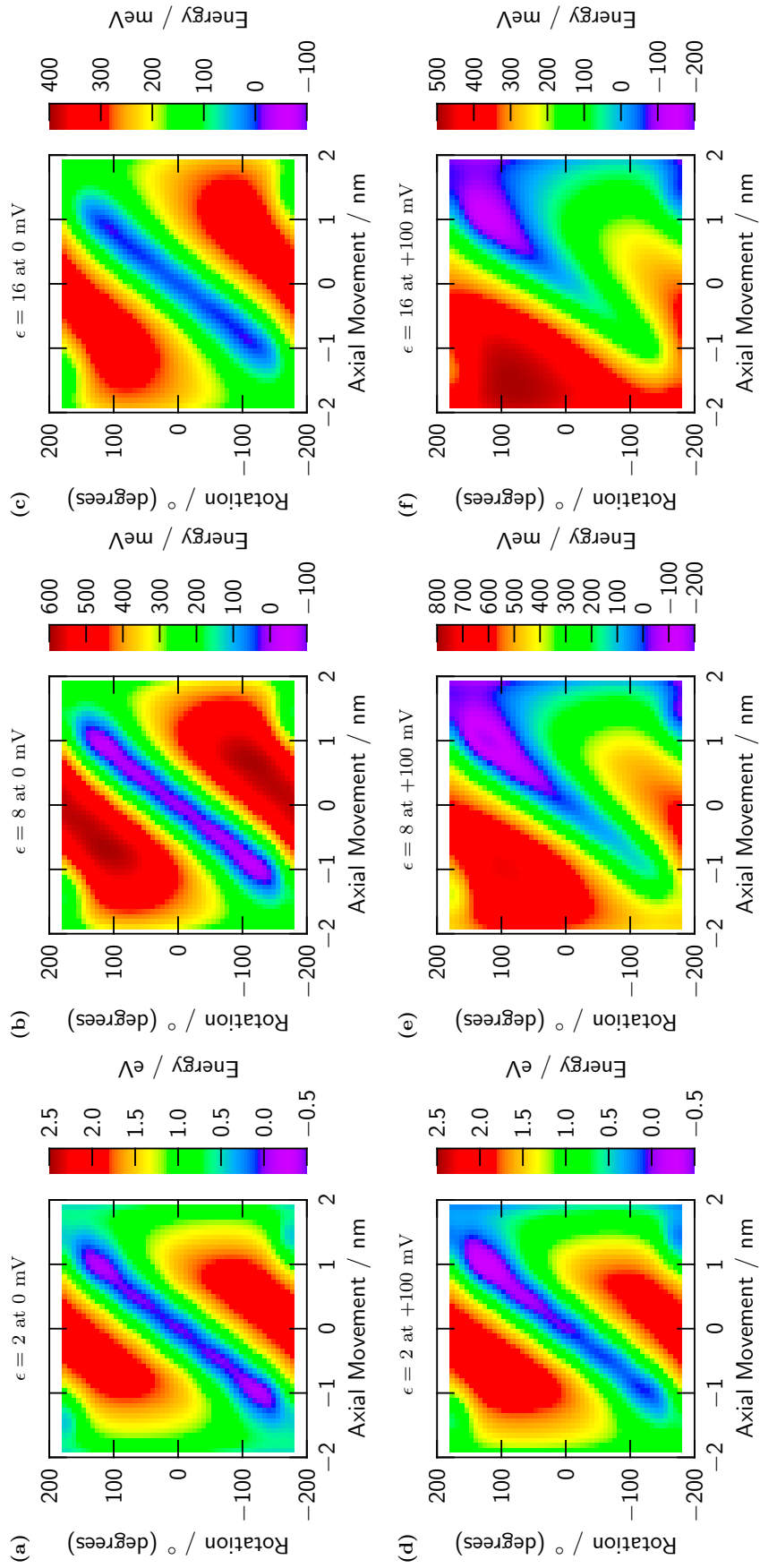

  \figtype{s4mp_QR}
  \setfiggroup{2d}{potential}
  \linebox{
    \subfloat[$\epsilon=$~2 at 0~mV]{
      \label{fig:eps:2:0}
      \getfiggraph{1_1_3_8_0_0}{0/e}
    }
    \hspace{\stretch{1}}
    \subfloat[$\epsilon=$~8 at 0~mV]{
      \label{fig:eps:8:0}
      \getfiggraph{1_3_3_8_0_0}{0/e}
    }
    \subfloat[$\epsilon=$~16 at 0~mV]{
      \label{fig:eps:16:0}
      \getfiggraph{1_7_3_8_0_0}{0/e}
    }
  }
  \linebox{
    \subfloat[$\epsilon=$~2 at +100~mV]{
      \label{fig:eps:2:100}
      \getfiggraph{1_1_3_8_0_0}{100/e}
    }
    \hspace{\stretch{1}}
    \subfloat[$\epsilon=$~8 at +100~mV]{
      \label{fig:eps:8:100}
      \getfiggraph{1_3_3_8_0_0}{100/e}
    }
    \subfloat[$\epsilon=$~16 at +100~mV]{
      \label{fig:eps:16:100}
      \getfiggraph{1_7_3_8_0_0}{100/e}
    }
  }

  \begin{center}\begin{minipage}{0.7\linewidth}
      \caption[Dielectric coefficient: Potential energy
      landscapes]{\textbf{Protein dielectric coefficient constrains
          the accessible range of motion.}  From left to right
        (panels~\subref*{fig:eps:2:0}-\subref*{fig:eps:16:0}) the
        dielectric coefficient $\epsilon_p$ of \SIV and the
        surrounding protein matrix varies over 2, 8 \& 16 for models
        voltage-clamped at 0~mV with a 2/3 interval (4 is presented in
        Fig.~\ref{fig:helix:land}). Likewise for +100~mV, $\epsilon_p$
        is varied over 2, 8 \& 16 in
        panels~(\subref*{fig:eps:2:100}-\subref*{fig:eps:16:100}). The
        potential energy of each \VS configuration relative to
        translation 0~nm and rotation 0$^\circ$ is represented in
        false color; note the difference in scales. At $\epsilon_p=$~2
        the range between maxima and minima are $\approx$~3~eV, while
        at $\epsilon_p=$~16 this difference is reduced to
        $\approx$~0.5~eV \& 0.7~eV.}
      \label{fig:eps:land}
  \end{minipage}\end{center}
\end{sidewaysfigure*}

\noindent
Simulations were conducted in which the dielectric coefficient of the
protein region (including the \SIV segment) was varied over the values 2,
4, 8 and 16. Landscapes of electrostatic potential energy for
dielectric coefficients 2, 8 and 16 are shown in
Fig.~\ref{fig:eps:land}; the potential energy landscape for a
dielectric coefficient of 4 was presented in
Fig.~\ref{fig:helix:land}. The general effect of increasing the
dielectric coefficient is to moderate energy variations. Broader
ranges of rotation and translation become accessible (note the varying
energy scales between these graphs).

\begin{figure*}
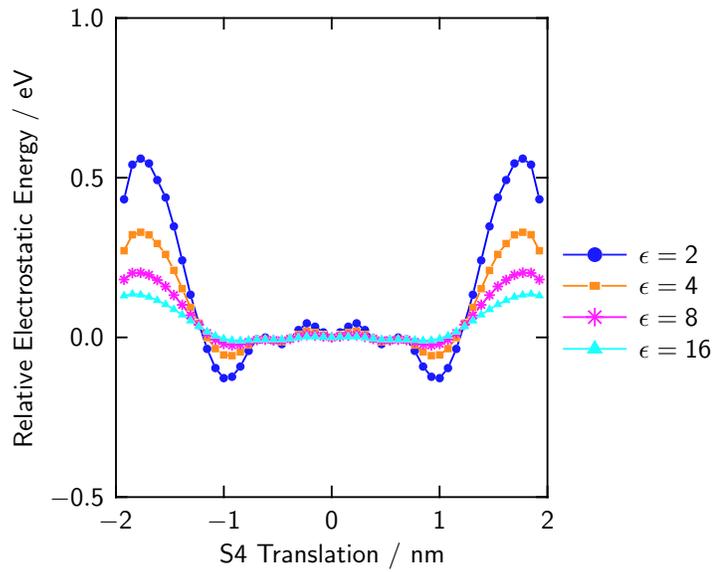

  \centering
  \figtype{s4mp_QR}
  \setfiggroup{2d}{over}
  \valignbox{
    \subfloat{
      \getfiggraph{1.1.3.8.0.0_1.2.3.8.0.0_1.3.3.8.0.0_1.7.3.8.0.0}{denergy}
    }
  }
  \makebox[2em]{
    \hspace{2em}
    \valignbox{
      \subfloat{
        \getfiggraph{1.1.3.8.0.0_1.2.3.8.0.0_1.3.3.8.0.0_1.7.3.8.0.0}{denergy-leg}
      }
    }
  }
  \caption[Dielectric coefficient: Size of potential energy
  barriers]{\textbf{Protein dielectric coefficient constrains the size
      of energy barriers distinguishing stable configurations.} The
    electrostatic potential energy at 0~mV and 2/3 interval relative
    to the 0 translational position is depicted for each translational
    position from the rotational partition function, as $\epsilon_p$
    is varied over 2 (\emph{blue}), 4 (\emph{orange}), 8
    (\emph{magenta}) \& 16 (\emph{cyan}).  The data for
    $\epsilon_p=$~4 is also presented as the blue curve in
    Fig.~\ref{fig:helix:land}~\subref{fig:helix:dE}. As $\epsilon_p$
    is increased, the energetic barrier distinguishing $\pm$~1~nm from
    each other falls, as well as the barriers constraining \SIV to the
    membrane.}
  \label{fig:eps:dE}
\end{figure*}

\noindent The expectation values of energy (based on the rotational partition
function) are shown versus translation in Fig.~\ref{fig:eps:dE}. As the
dielectric coefficient is decreased, the shallow trough of energy
becomes deeper and a pattern of wells and barriers emerges. In
particular, significant wells develop around $\pm$~1~nm of
translation.

\begin{figure*}
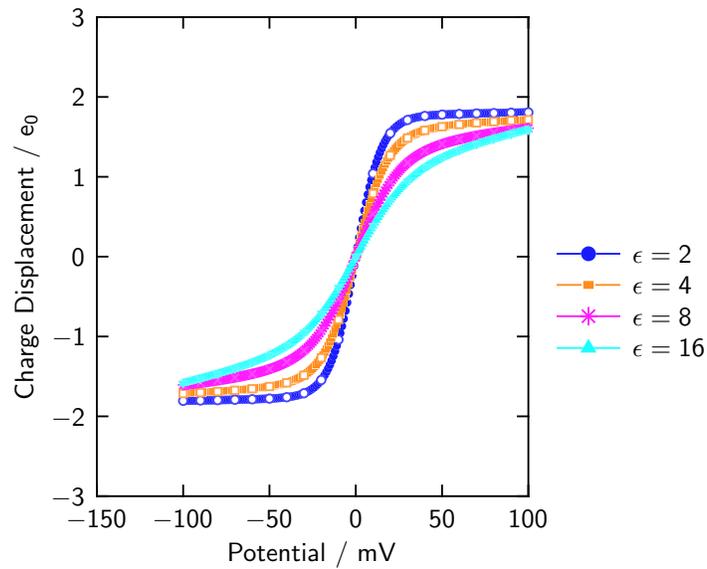

  \centering
  \figtype{s4mp_QR}
  \setfiggroup{3d}{over}
  \valignbox{
    \subfloat{
      \getfiggraph{1.1.3.8.0.0_1.2.3.8.0.0_1.3.3.8.0.0_1.7.3.8.0.0}{displacementV}
    }
  }
  \makebox[0pt]{
    \hspace{4em}
    \valignbox{
      \subfloat{
        \getfiggraph{1.1.3.8.0.0_1.2.3.8.0.0_1.3.3.8.0.0_1.7.3.8.0.0}{displacementV-leg}
      }
    }
  }
  \caption[Dielectric coefficient: Displacement]{\textbf{Protein
      dielectric coefficient constrains the distribution of charge
      displacement.}  As $\epsilon_p$ is varied over 2 (\emph{blue}),
    4 (\emph{orange}), 8 (\emph{magenta}) \& 16 (\emph{cyan}), the
    maximum slope of the charge displacement curve is reduced, while
    total charge displacement is not. The charge displacement for
    $\epsilon_p$ has been previously presented versus experimental
    results in Fig.~\ref{fig:helix:stat}~\subref{fig:helix:stat:disp},
    \emph{blue} curve. All curves were calculated with a 2/3 interval
    at 0~mV.}
  \label{fig:eps:disp}
\end{figure*}

The changes in potential energy landscape have interesting
consequences for the relation between charge displacement and voltage
(Fig.~\ref{fig:eps:disp}). Reduction of the protein dielectric
coefficient increases the slope of the charge/voltage relation. These
simulations were all calculated with the same dielectric geometry,
protein charges and counter-charges in the model. Nevertheless, the
`effective gating charge' of the \VS as assessed by the steepness at
the midpoint of the charge/voltage curve increases as the protein
dielectric coefficient is reduced. The reason for this effect of the
dielectric coefficient is evident in Fig.~\ref{fig:eps:dE}. With a
dielectric coefficient of 2, there are two crisp energy minima at
translations $\pm$~1~nm, leading the \SIV segment to dwell
preferentially near these two positions. With a dielectric coefficient
of 16, however, there is no significant energy variation (at 0~mV) at
any position within the energy trough, so that no \SIV position is
preferred. Hence, the distribution of the \SIV segment in the
translational degree of freedom varies from a virtual `two-state
Boltzmann distribution' to a distribution within a space of uniform
potential energy. The uniform-energy distribution of charge approaches
hyperbolic rather than exponential asymptotic behavior at extreme
voltages. The midpoint slopes of the charge/voltage curves can be
analytically determined; the midpoint slope of the charge/voltage
curve is three times greater for the two-state case than for the
uniform-energy case \citep{neumcke:1978:gating}. In this manner, the
same structural charges produce up to a three-fold varying effective
gating charge as the protein dielectric coefficient (and therefore the
potential energy landscape) is varied. The force that can be delivered
by \SIV charge movement is therefore constrained by the local
dielectric coefficient.

\section{Charge mutations act via charge/counter-charge
    interactions}
\seclabel{mutants}{section}{Charge mutations}

\noindent
\Citet{seoh:1996} reported the results of 9 neutralization mutants
(over 8 residues) of \Shaker \K channels plus the wildtype in terms of
open probability and charge displacement per channel. In single
mutants, one of the four outer \SIV positive residues or one of three
negative charges on the \SII and \SIII transmembrane segments were
neutralized. In addition, two double mutants were investigated. These
mutations produce a complex pattern of change in the charge/voltage
curves, including reductions of total gating charge, shift, and
alteration of slope and shape. Predicting such patterns is a challenge
for a physical model.

For these comparisons, I use 3 counter-charges set at a 2/3 interval
to test whether the apparent \VS-like behavior of that model as
investigated above responds like a biological \VS to physiological
extremes. The positions of these counter-charges in the biological \VS
is ambiguous. Unlike \SIV charges which are regularly arrayed on a
single transmembrane domain that is $\alpha$¬helical in nature,
counter-charges are irregularly placed on multiple helices connected
by amorphous linking regions and arranged at various orientations and
relative positions to each other
\citep{jiang:2003:xray,long:2005:structure,long:2007,tao:2010},
recalling that each of these examples is crystallographic data from a
small subset of the full ensemble of conformations of the \VS. Motions
of the \SI--\SIII regions of the \VS are experimentally
undetermined.

The charged residues of the \VS have many points of rotation and
extend relatively far from the $\alpha$¬helical backbones to which
they are attached (particularly arginine and glutamate residues,
\citealt{creighton:1984:1}); for example, in \citet[Fig.~4]{long:2007},
four charge/counter-charge pairs are shown in direct contact despite
the \SIV appearing to form an angle (and therefore creating a gap)
with the \SI, \SII and \SIII domains. Additionally, there exists
ambiguity regarding which charges are permanently situated in the
intra- and extra-cellular solutions and which ones are within the
gating pore regions where the ratio of electrical travel to
geometrical travel is larger \citep[Fig.~2]{nonner:2004:rs}.\enlargethispage{\baselineskip}

However, it is known that three or four counter-charges on the \SII
and \SIII transmembrane domains are highly conserved depending on
which \Kv channels are included
\citep{islas:1999,jiang:2003:xray,tao:2010}, and that neutralization
of three counter-charges have profound effects on \VS behavior
\citep{planells-cases:1995,papazian:1995,seoh:1996}. From the results
shown in \secref{barrier}, three counter-charges are sufficient to
produce \VS-like behavior, and fewer charges fail to stabilize the
sliding helix in the membrane (data not shown for single and double
counter-charge simulations).

The biologically possible configuration space is of high
dimensionality, so as a test of the variety of conditions under which
the previously elucidated model can function I begin with a reduced
model of counter-charge positions and mutations. As described in
\secref{details:helix}, charge and counter-charge mutants are modeled
both as a simple elimination of the point-charges associated with the
residues and by the point-charges' replacement by a dipole to
represent the polarizability of glutamine and asparagine residues. The
envelope contained by the curves representing these two extreme models
of each mutation can be used to investigate the robustness of
components of this model (charge elimination is a
maximally energetically unfavorable model, whereas the chosen dipole
representation is minimally energetically unfavorable).

I have calculated the behavior of models with corresponding charge
deletions, testing to what extent the experimental charge/voltage
curves are predicted. This requires assigning the \SII and \SIII
negative charges neutralized by \citet{seoh:1996} to counter-charge
positions in the sliding helix models. The residues $E283$, $D316$,
and $E293$ are assigned to the outermost, central, and innermost
counter-charge respectively. The counter-charges are spaced at the 2/3
interval, and the protein dielectric coefficient is set to 4. In these
models when one of the counter-charges is removed, the trough of
electrostatic potential energy that confines \SIV charge motion (see
Fig.~\ref{fig:counter:dE}) becomes inverted; therefore the \SIV
segment is no longer stable in the weak dielectric. On the other hand,
mutants with one \SII or \SIII negative residue neutralized are
functional, indicating that their \SIV segments are not
dislocated. Therefore, I confine the \SIV segment of the models
presented in this section within the translational range of
$\pm$~1.5~nm. Charge/voltage curves are computed using the partition
function resulting from the electrostatic potential energy of the \VS
sampled over the rotational and translational degrees of freedom.

\enlargethispage{\baselineskip}
Figs.~\ref{fig:m-first} through \ref{fig:m-last} summarize the
results. Panel~\subref{fig:m:exp} of each figure shows the
experimental charge/voltage curves for the wildtype channel
(\emph{orange squares}) and the mutant (\emph{blue circles}), as
reported in Fig.~2~(A-J) of
\citet{seoh:1996}. Panel~\subref{fig:m:the} of each figure shows the
corresponding curves computed for the model. The experimental
charge/voltage curves are vertically aligned so that their midpoints
correspond to zero displaced charge. There are no horizontal
alignments or normalizations except in Fig.~\ref{fig:m0}
(\citealp{seoh:1996} report in their Fig.~2~E a normalized
charge/voltage curve for mutant $R362Q$). All other curves represent
charge per \VS.

In comparing the computed and experimental results, it is useful to
consider differences in the size of wildtype and mutant charges,
shifts between wildtype and mutant voltage dependencies, and slope and
shape changes between wildtype and mutant. My computations apply to an
isolated \VS, while the experiments were done on channels; therefore,
the `idle' \VS is being compared to the naturally gate-coupled \VS. To
the extent that the coupling to the gate restricts \SIV charge motion
and requires work to be done by the \VS, it is expected that the
gating charge of the model \VS is greater than that of the channel
\VS, and that the charge/voltage curves are shifted with respect to
one another. Because \VS model and channel charge/voltage curves are
expected to be shifted with respect to one another, it is useful to
focus on the differences between mutant and wildtype curves. To what
extent does the model account for the mutation-versus-wildtype
changes?

\newpage
\subsection{Mutation of a positive charge}\enlargethispage{\baselineskip}

Model and experimental results of successively mutating one \SIV
charge are shown in Figs.~\ref{fig:m:s4-first} through
\ref{fig:m:s4-last}, starting with the outermost arginine residue. In
real \Shaker channels, these mutations produce substantial changes to
the charge/voltage relation including reduction of total charge,
changes of slope, deformations of the charge/voltage relation and
shifts along the voltage axis. The experimental pattern of change
varies from mutation to mutation. The predictions from the model
reflect the varying experimental patterns very well. Only one
qualitative difference is seen: the model does not predict the shift
toward negative voltages seen in the $R368N$ mutant
(Fig.~\ref{fig:m2}).

\newtoks\figures
\addtotoks\figures{
\begin{figure*}
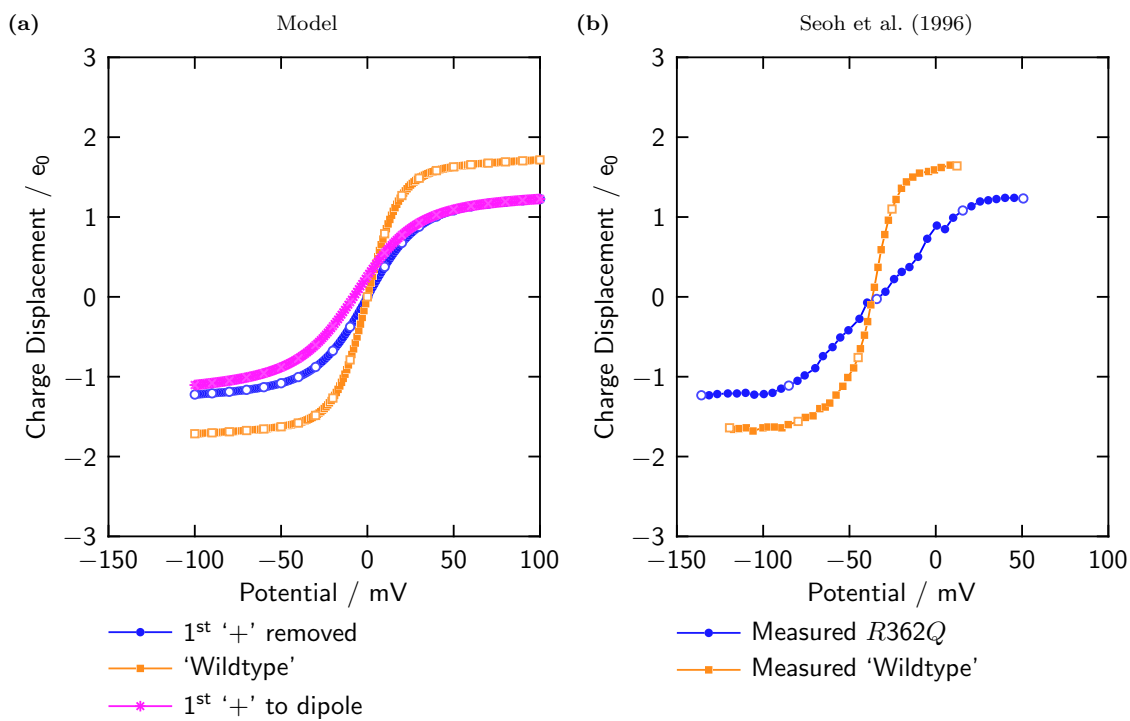

  \figtype{s4mp_QR}
  \setfiggroup{3d}{over}
  \linebox{
    \subfloat[\hspace{3em}Model]{
      \label{fig:m0:t}
      \label{fig:m:the}
      \getfiggraph{1_2_3_14_0_1}{displacementV}
    }
    \hspace{\stretch{1}}
    \subfloat[\hspace{3em}\citet{seoh:1996}]{
      \label{fig:m0:m}
      \label{fig:m:exp}
      \getfigseoh{m0}{displacementV}
    }
  }
  \caption[\emph{Shaker}: $R362Q$]{\textbf{Effective gating charge is
      reduced by mutating the outermost arginine.} The slope of the QV
    relation, the effective gating charge, is reduced relative to the
    wildtype for both the computed and experimental results. Note that
    \citet[Fig.~2~E]{seoh:1996} report for this case normalized charge
    displacement, so total gating charge is not comparable with
    wildtype results. Dipole and deletion representations are
    similar.}
  \label{fig:m0}
  \label{fig:m-first}
  \label{fig:m:s4-first}
\end{figure*}

\begin{figure*}
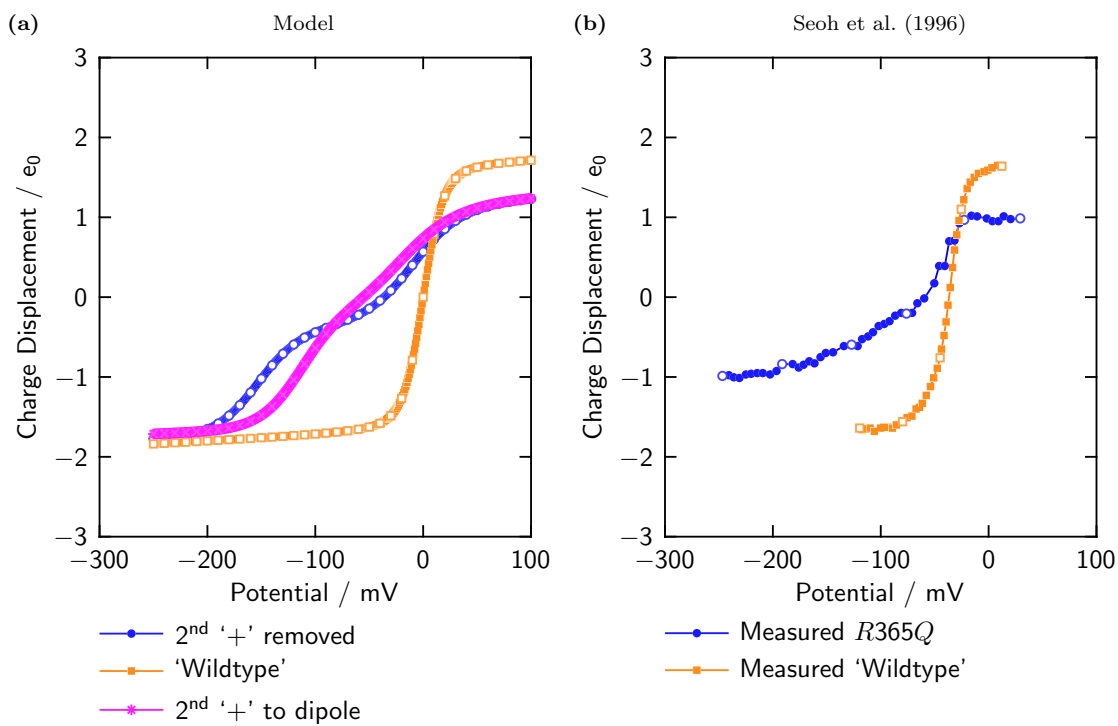

  \centering
  \figtype{s4mp_QR}
  \setfiggroup{3d}{long}
  \linebox{
    \subfloat[\hspace{3em}Model]{
      \label{fig:m1:t}
      \getfiggraph{1_2_3_13_0_1}{displacementV}
    }
    \hspace{\stretch{1}}
    \subfloat[\hspace{3em}\citet{seoh:1996}]{
      \label{fig:m1:m}
      \getfigseoh{m1}{displacementV}
    }
  }
  \caption[\emph{Shaker}: $R365Q$]{\textbf{Mutation of the
      2\sscript{nd} arginine significantly reduces effective charge.}
    Both computed and experimental results show a left
    shift. Note the inflection point for the model mutations
    below -100~mV and the sensitivity to charge representation at
    extreme negative voltages.}
  \label{fig:m1}
\end{figure*}

\begin{figure*}
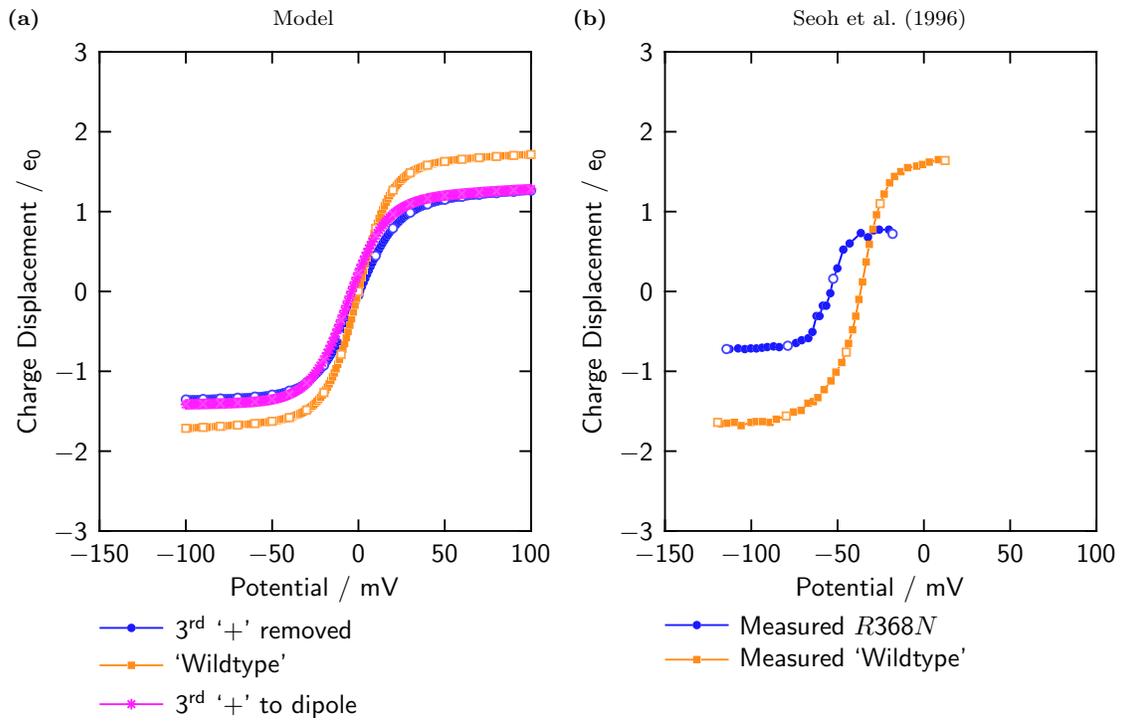

  \centering
  \figtype{s4mp_QR}
  \setfiggroup{3d}{over}
  \linebox{
    \subfloat[\hspace{3em}Model]{
      \label{fig:m2:t}
      \getfiggraph{1_2_3_12_0_1}{displacementV}
    }
    \hspace{\stretch{1}}
    \subfloat[\hspace{3em}\citet{seoh:1996}]{
      \label{fig:m2:m}
      \getfigseoh{m2}{displacementV}
    }
  }
  \caption[\emph{Shaker}: $R368N$]{\textbf{Mutation of the
      3\sscript{rd} arginine reduces total gating charge.} In this
    case, experimental results show a left shift that is not apparent
    in the computational results, combined with an even more
    significant total reduction in charge. $R368N$ also shows a left
    shift in open probability (data not shown). Dipole and
    deletion representations are similar.}
  \label{fig:m2}
\end{figure*}

\begin{figure*}
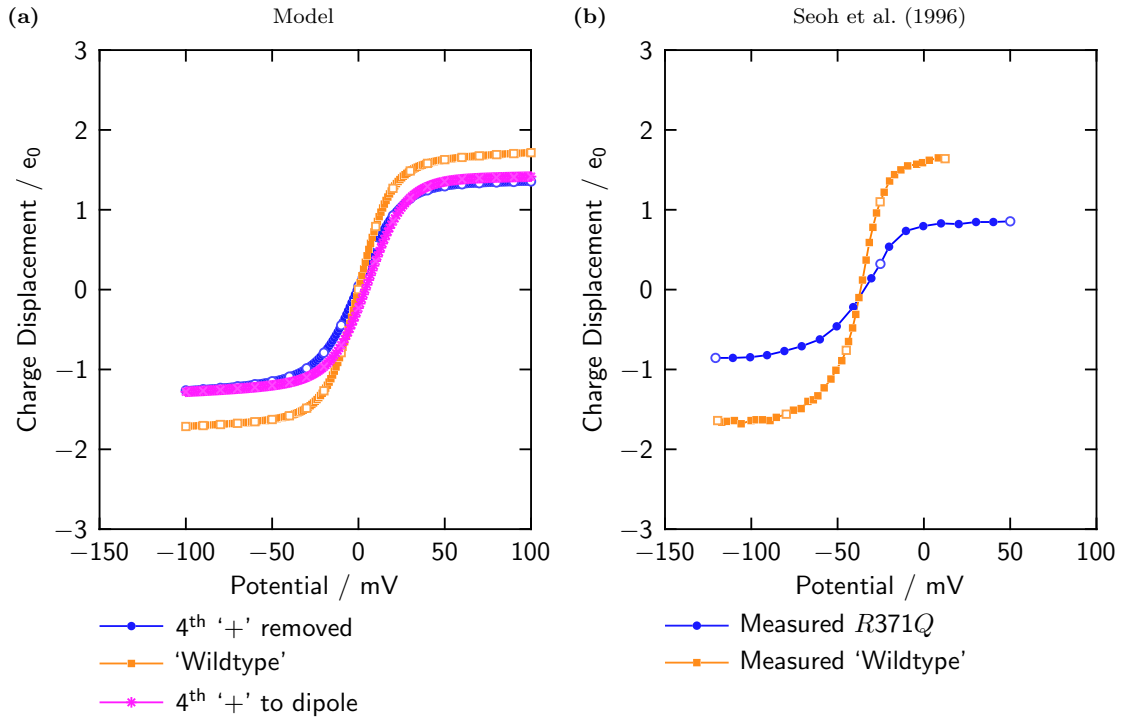

  \centering
  \figtype{s4mp_QR}
  \setfiggroup{3d}{over}
  \linebox{
    \subfloat[\hspace{3em}Model]{
      \label{fig:m3:t}
      \getfiggraph{1_2_3_11_0_1}{displacementV}
    }
    \hspace{\stretch{1}}
    \subfloat[\hspace{3em}\citet{seoh:1996}]{
      \label{fig:m3:m}
      \getfigseoh{m3}{displacementV}
    }
  }
  \caption[\emph{Shaker}: $R371Q$]{\textbf{Mutation of the
      4\sscript{th} arginine reduces total gating current.} Neither
    results shows a shift in the charge/voltage
    relationship. Dipole and deletion representations are
    similar.}
  \label{fig:m3}
  \label{fig:m:s4-last}
\end{figure*}
}

\subsection{Mutation of a negative charge}

The results of neutralizing one of a group of putative
counter-charges, two glutamate residues of \SII and an aspartate
residue of \SIII, are shown in Figs.~\ref{fig:c0} through
\ref{fig:c2}. In the model, these mutations are mimicked by deleting
the outermost ($E283$), central ($D316$), and innermost ($E293$)
counter-charge. Since these assignments are tentative, testing the
model against experiments also tests for the adequacy of the guessed
counter-charge to residue assignments.

The comparisons of predicted and experimental mutant effects are,
again, surprisingly good. The varying directions and degrees of shift
along the voltage axis are predicted quite well. Because the model
assigns the first and third in the group of counter-charges to the
residues $E283$ \& $E293$ and because of the symmetry in the model's
charge configuration, the simulations predict that the Q/V curves for
$E283Q$ and $E293Q$ should reflect such symmetry, but in fact the
experimental curves are not symmetric. In particular $E293Q$ has a
substantially smaller total gating charge than $E283Q$. Thus the \VS
of the channel has an asymmetry in its structure or operates under
asymmetrical constraints that are not included in these \VS
models. $E283Q$ and $D316N$ reveal activation curves (not shown) of
the ionic current that are strongly shifted to positive voltages. The
charge/voltage curves reported by \citet{seoh:1996} do not extend over
this voltage range, so they can not reveal the full gating charge.

\addtotoks\figures{
\begin{figure*}
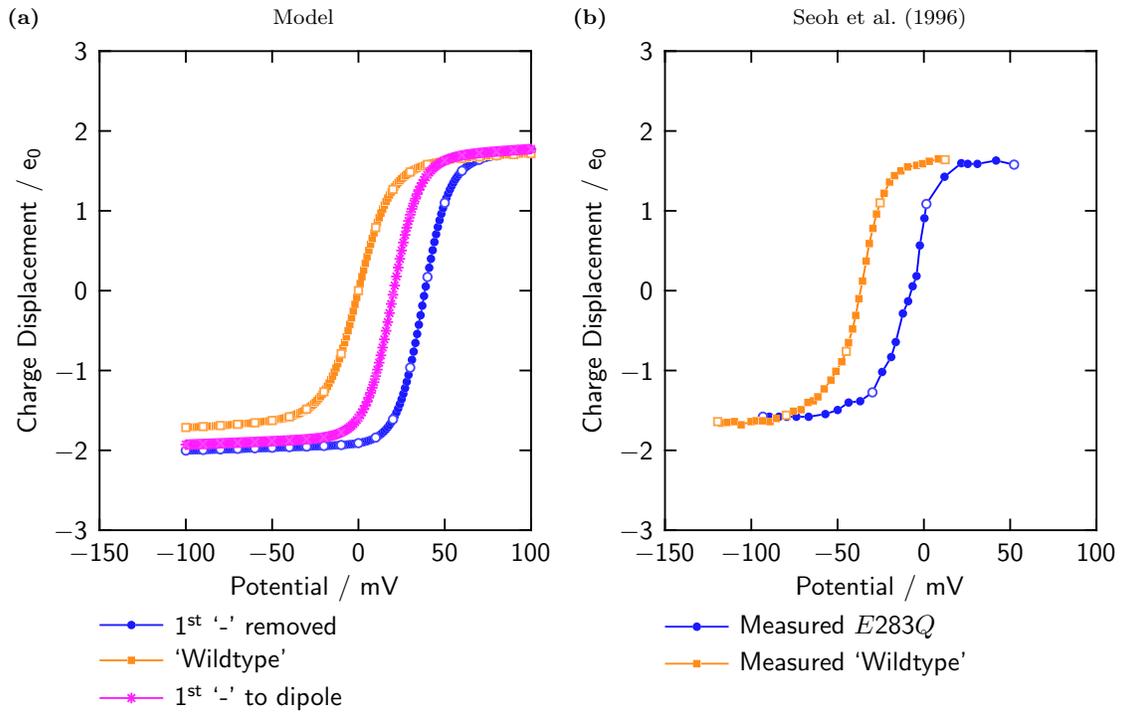

  \centering
  \figtype{s4mp_QR}
  \setfiggroup{3d}{over}
  \linebox{
    \subfloat[\hspace{3em}Model]{
      \label{fig:c0:t}
      \getfiggraph{1_2_5_8_0_1}{displacementV}
    }
    \hspace{\stretch{1}}
    \subfloat[\hspace{3em}\citet{seoh:1996}]{
      \label{fig:c0:m}
      \getfigseoh{c0}{displacementV}
    }
  }
  \caption[\emph{Shaker}: $E283Q$]{\textbf{Mutation of the outermost
      counter-charge produces a right shift with no gating charge
      reduction.} Experimental results also show a right shift for the
    open probability (data not shown). Both representations of
    the mutation are right-shifted and have similar slopes, but the
    magnitude of the right-shift is distinct.}
  \label{fig:c0}
\end{figure*}

\begin{figure*}
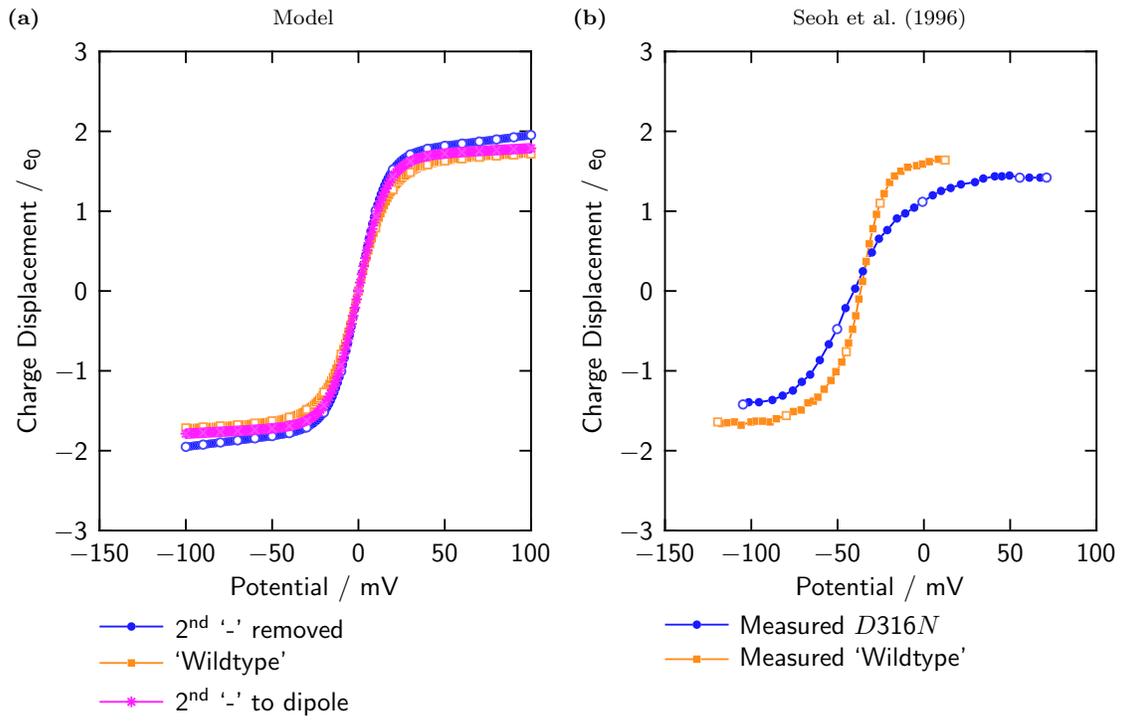

  \centering
  \figtype{s4mp_QR}
  \setfiggroup{3d}{over}
  \linebox{
    \subfloat[\hspace{3em}Model]{
      \label{fig:c1:t}
      \getfiggraph{1_2_6_8_0_1}{displacementV}
    }
    \hspace{\stretch{1}}
    \subfloat[\hspace{3em}\citet{seoh:1996}]{
      \label{fig:c1:m}
      \getfigseoh{c1}{displacementV}
    }
  }
  \caption[\emph{Shaker}: $D316N$]{\textbf{Mutation of the middle
      counter-charge has a mild effect.} Experimental results show a
    small reduction in gating charge \& slope, while computational
    results predict a slight increase for both due to the exclusion of
    \SIV from central positions. Dipole and deletion
    representations are similar.}
  \label{fig:c1}
\end{figure*}

\begin{figure*}
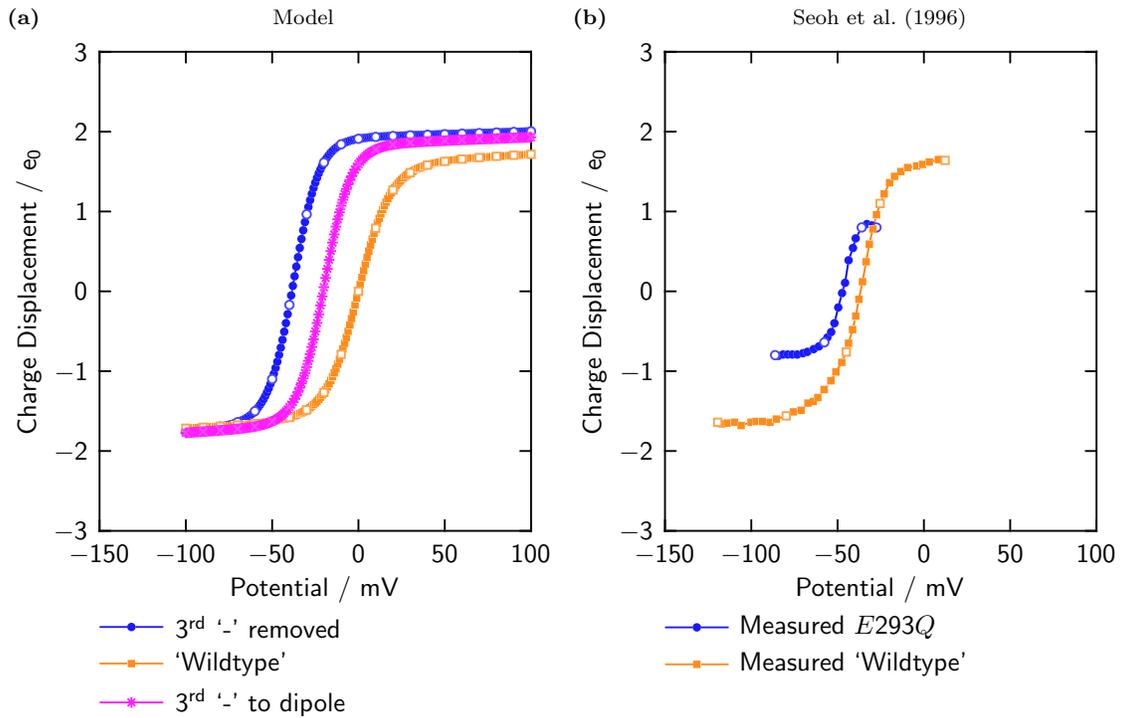

  \vskip-\baselineskip
  \centering
  \figtype{s4mp_QR}
  \setfiggroup{3d}{over}
  \linebox{
    \subfloat[\hspace{3em}Model]{
      \label{fig:c2:t}
      \getfiggraph{1_2_8_8_0_1}{displacementV}
    }
    \hspace{\stretch{1}}
    \subfloat[\hspace{3em}\citet{seoh:1996}]{
      \label{fig:c2:m}
      \getfigseoh{c2}{displacementV}
    }
  }
  \caption[\emph{Shaker}: $E293Q$]{\textbf{Mutation of the innermost
      counter charge produces a left shift.} Computational results are
    symmetrical with \ref{fig:c0}~\subref{fig:c0:t}, while
    experimental results show a large reduction in gating charge which
    is not apparent in $E283Q$. Experimental results also show a left
    shift for open probability, as opposed to to the right shift for
    $E283Q$ (data not shown). Both representations of the
    mutation are left-shifted and have similar slopes, but the
    magnitude of the left-shift is distinct.}
  \label{fig:c2}
\end{figure*}
}

\subsection{Combined mutations of a positive and a negative charge}

The two double mutants investigated by \citet{seoh:1996} are
represented in Figs.~\ref{fig:m4-c1} and \ref{fig:m4-c2}. The
charge/voltage curves of the mutants are shifted to negative voltages;
the activation curves are also strongly altered (with a left shift in
the case of $K374Q+D316N$). The \VS of these mutant channels appears
to have difficulty in moving into fully-closed positions. The
predicted curves reveal limitations of the model. A double mutant
lacking two \VS charges is expected to be particularly sensitive to
the geometry assigned to charges due to the positions and orientations
of potential gaps between charges and counter-charges, apart from the
already mentioned distinction regarding gating load.

\addtotoks\figures{
\begin{figure*}
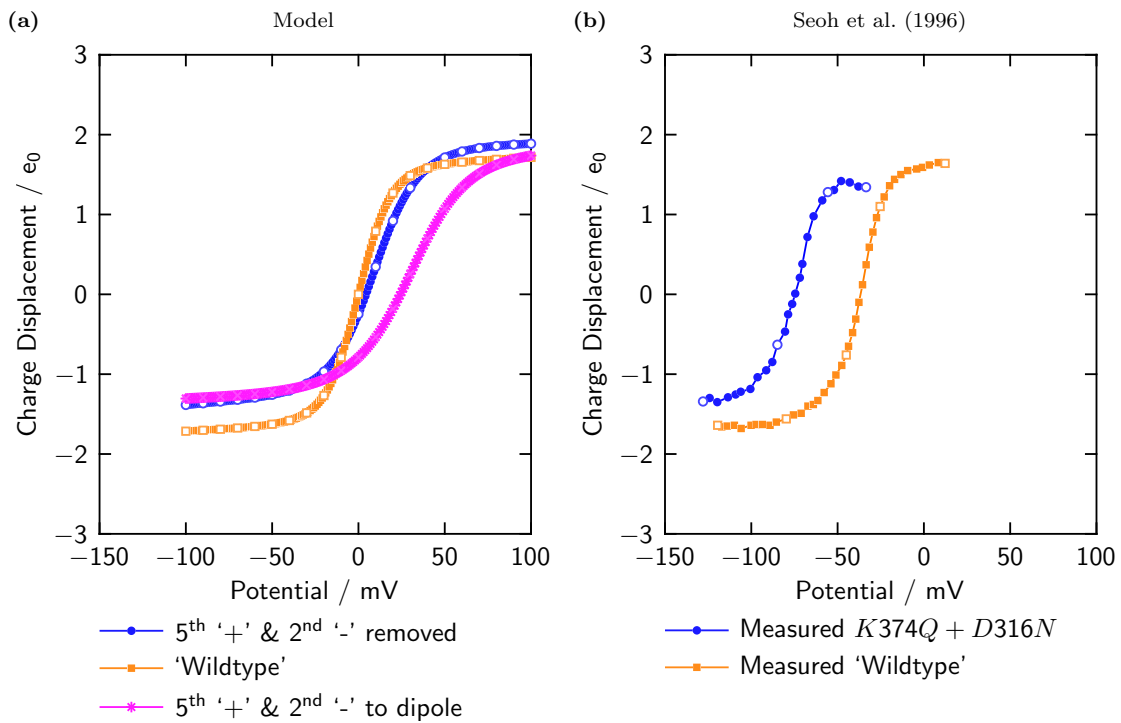

  \centering
  \figtype{s4mp_QR}
  \setfiggroup{3d}{over}
  \linebox{
    \subfloat[\hspace{3em}Model]{
      \label{fig:m4-c1:t}
      \getfiggraph{1_2_6_10_0_1}{displacementV}
    }
    \hspace{\stretch{1}}
    \subfloat[\hspace{3em}\citet{seoh:1996}]{
      \label{fig:m4-c1:m}
      \getfigseoh{m4-c1}{displacementV}
    }
  }
  \caption[\emph{Shaker}: $K374Q+D316N$]{\textbf{Mutating the
      outermost lysine \& central counter-charge produces a small
      reduction in total charge displacement.} Experimental results
    show a left shift that computational results do not
    reproduce. $K374Q+D316N$ also shows a left shift in open
    probability (data not shown). The dipole representation displays a
    much larger right-shift and a slightly larger reduction in total
    charge displacement relative to the deletion representation.}
  \label{fig:m4-c1}
  \vskip-\baselineskip
\end{figure*}

\begin{figure*}
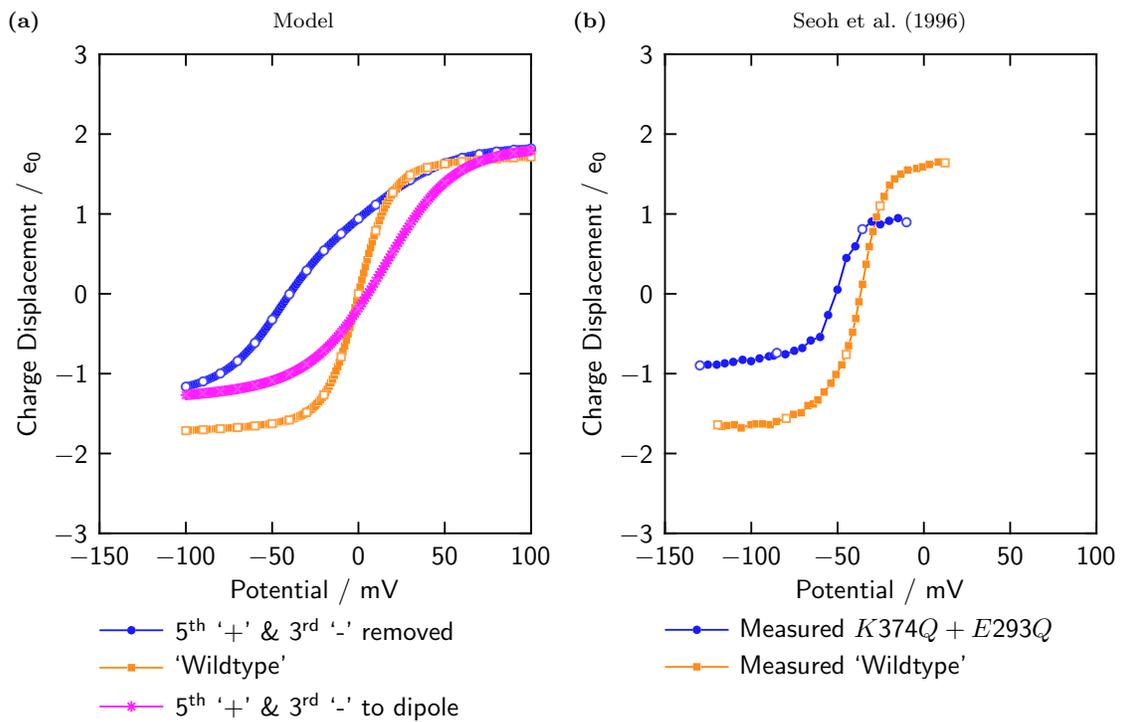

  \centering
  \figtype{s4mp_QR}
  \setfiggroup{3d}{over}
  \linebox{
    \subfloat[\hspace{3em}Model]{
      \label{fig:m4-c2:t}
      \getfiggraph{1_2_8_10_0_1}{displacementV}
    }
    \hspace{\stretch{1}}
    \subfloat[\hspace{3em}\citet{seoh:1996}]{
      \label{fig:m4-c2:m}
      \getfigseoh{m4-c2}{displacementV}
    }
  }
  \caption[\emph{Shaker}: $K374Q+E293Q$]{\textbf{Mutating the
      outermost lysine \& the innermost counter-charge produces a
      reduction in total gating current.} The computational results
    for deletion, however, predict a reduction in slope that is
    replaced by a larger reduction in total gating charge in the
    experimental results. $K374Q+E293Q$ also shows a left shift in
    open probability (data not shown). The dipole representation fails
    to predict the left shift of the experimental results reproduced
    by the deletion representation; however, the dipole representation
    reproduces the higher slope of the experimental mutant.}
  \label{fig:m4-c2}
  \label{fig:m-last}
\end{figure*}
}

\subsection{Investigation of a mutant lacking functional expression}

The double mutations of $K374Q+E293Q$ and $K374Q+D316N$ by
\citet{seoh:1996} were partially motivated by the lack of functional
expression by $K374Q$ mutants. Since ionic conductance was blocked at
all potentials, charge displacement per channel could not be
estimated. One could speculate as to the cause of this --- whether the
\VS proper was non-functioning or whether some folding pathology
blocked proper expression. Experimentally, there is limited
accessibility for non-expressive behavior; however, computational
exploration is still possible. In Fig.~\ref{fig:m4}, a computational
analog of $K374Q$ is tested: \acsfont{(1)} charge displacement is
greatly reduced in panel~\subref{fig:m4:disp}, with some increase
extending towards large positive potentials; \acsfont{(2)} \SIV
position is constrained to the intracellular, closed positions in
panel~\subref{fig:m4:zv}; and \acsfont{(3)} no path for fully moving
to the open position is apparent from the potential energy landscapes
in panels~\subref{fig:m4:0} \& \subref{fig:m4:100}. The failure of
$K374Q$ to function biologically is consistent with the predicted
electrostatic limitations of $K374Q$ as a \VS.

\addtotoks\figures{
\begin{figure*}
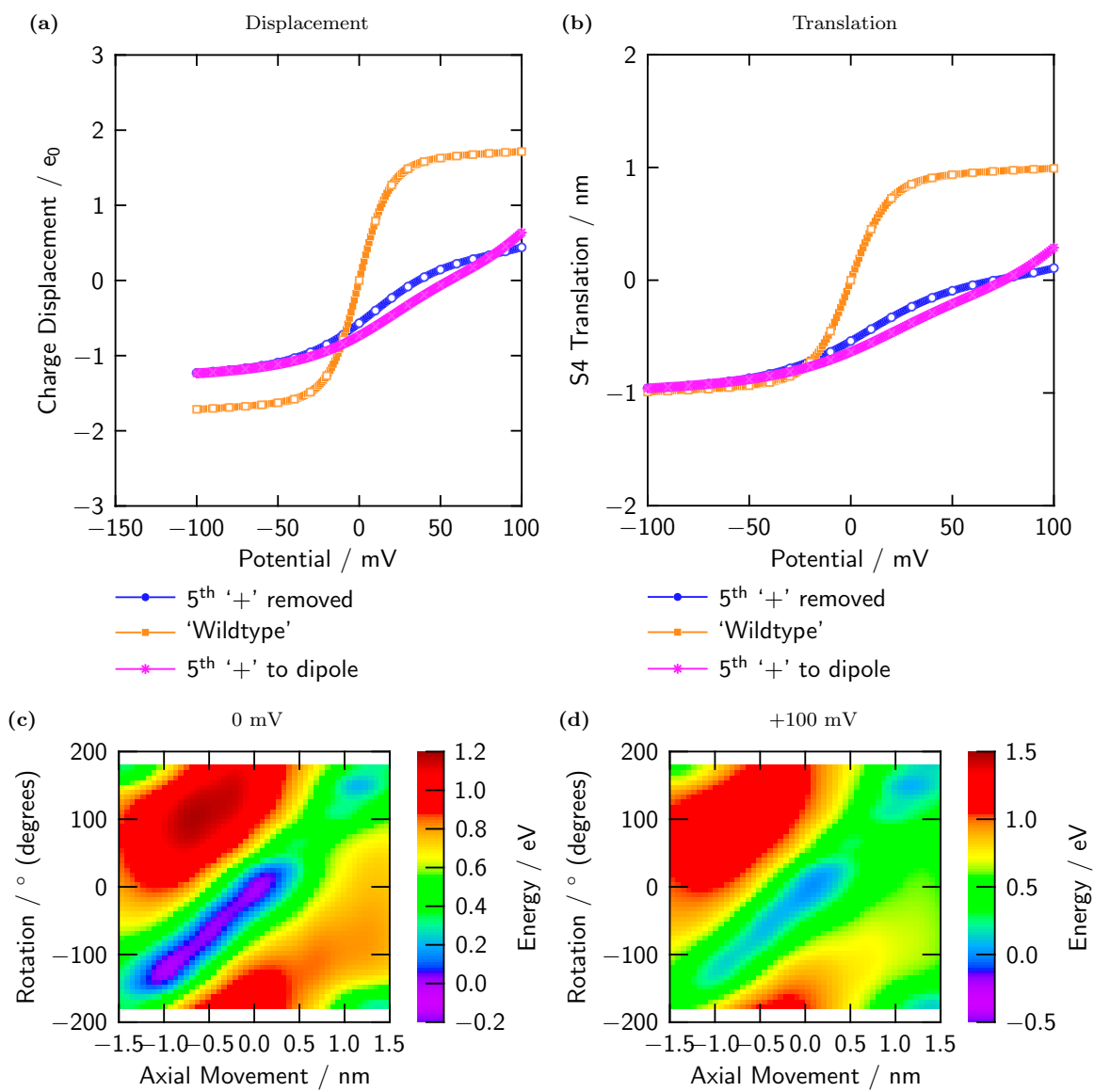

  \centering
  \figtype{s4mp_QR}
  \setfiggroup{3d}{over}
  \linebox{
    \subfloat[\hspace{2.5em}Displacement]{
      \label{fig:m4:disp}
      \getfiggraph{1_2_3_10_0_1}{displacementV}
    }
    \hspace{\stretch{1}}
    \subfloat[\hspace{3em}Translation]{
      \label{fig:m4:zv}
      \getfiggraph{1_2_3_10_0_1}{ZV}
    }
  }
  \setfiggroup{2d}{potential}
  \linebox{
    \subfloat[0~mV]{
      \label{fig:m4:0}
      \getfiggraph{1_2_3_10_0_1}{0/e}
    }
    \hspace{\stretch{1}}
    \subfloat[+100~mV]{
      \label{fig:m4:100}
      \getfiggraph{1_2_3_10_0_1}{100/e}
    }
  }
  \caption[\emph{Shaker}: $K374Q$ (Imaginary)]{\textbf{The
      non-functional $K374Q$ mutant is electrostatically incompetent.}
    Charge displacement in panel~\subref{fig:m4:disp} is marginal,
    only reaching 0~$e_0$ at highly positive potentials. The
    associated translation in panel~\subref{fig:m4:zv} tracks the
    gating charge, never moving past the central position until the
    internal relative electrode potentials approaches +100~mV. In
    panels~(\subref*{fig:m4:0} \& \subref*{fig:m4:100}), a large
    energy barrier at 1~nm is apparent even at
    +100~mV. Panels~\subref{fig:m4:0} and \subref{fig:m4:100}
    were produced with the deletion representation. Both mutant
    representations show similar results.}
  \label{fig:m4}
\end{figure*}
}

\subsection{Summary of mutation simulations}\enlargethispage{\baselineskip}

These model studies of charge mutants show that the quality of the
model predictions varies in a pattern. \SIV charge mutants are
described well, \SII and \SIII (counter-charge) mutants less well, and
double mutants least well. \SIV charges likely form a regular array of
charges because of the \SIV's helical structure. Thus the model
assumption of uniform spacing of the \SIV charges is probably
sound. The arrangement of putative counter-charges provided by the
\SII and \SIII segments has a much larger range of uncertainty, of
which I have explored only a small subrange --- more exploration is
needed.

\newgeometry{vmargin=0.5in,footskip=\baselineskip}
\usetoks\figures
\restoregeometry
\chapter{Perspectives}

\noindent I have developed a computational approach for studying the
electrostatics of the voltage sensor (\VS) controlling conduction in
voltage-dependent ion channels. The \VS is described in a microscopic
physical model that is reduced in detail to those features whose
relevance for \VS function is to be investigated. The microscopic
model is complemented by a simulation system that establishes
voltage-clamp conditions and records gating charge movements in a
manner analogous to and comparable with a macroscopic experimental
setup. Using efficient computational methods that allow a
statistical-mechanical analysis of \VS behavior, characteristics that
are experimentally accessible (such as the charge/voltage relation)
are computed for different models of the \VS. My approach thus
substantially extends the computational means for studying
structure-function relationships of the \VS system while making
comparisons with experimental results. The presented results provide
the following perspectives on the \VS.

\paragraph{How do charged residues `contribute' to gating charge? }
Published experiments have sought answers to this question by
neutralizing a formally charged residue of the \VS and measuring the
slope of voltage dependence of activation \citep{stuehmer:1989} or
recording ensemble gating charge while counting channels with an
independent method such as noise analysis of ionic current
\citep{aggarwal:1996,seoh:1996,baker:1998,ledwell:1999}. The results
of the latter are expressed as the change (typically reduction) in
gating charge per channel. My computational studies show how charge
neutralization can modify gating charge in a manner amenable to
several modes of analysis: \acsfont{(1)} the \SIV charge motion
carries less charge as a direct consequence of removing one of its
charges --- the simplest mode; \acsfont{(2)} the range of \SIV charge
motion becomes electrostatically restricted, leading \emph{all} \SIV
charges to travel a shorter distance, as they are part of a solid
body; and \acsfont{(3)} the electrostatic potential energy landscape
for \SIV travel is altered in such a manner that the probability
distribution of positions is altered, with consequences for the shape
of Q/V relationships. Modes \acsfont{(2)} and \acsfont{(3)} may apply
regardless of whether the neutralized residue is mobile or fixed in
position. Under the latter conditions, the total gating charge can be
reduced by an amount greater than the neutralized charge since the
motion of all charges is modified by altering one charge.

\paragraph{Counter-charges to the positively charged \SIV residues are
  essential.} Much discussion of how the \SIV helix with its array of
positively charged residues might be stabilized in the membrane has
focused on the polarizability of the \SIV matrix and its
environment. The primary requirements for allowing the \SIV charges to
act as a \VS are the proper number and positions of
counter-charges. The importance of negatively charged residues for \VS
function is well established; the respective residues of the \SII and
\SIII segments are highly conserved \citep{islas:1999,tao:2010}, and
neutralization mutants show strong alterations of function
\citep{seoh:1996}.

Computationally, I find that a viable \VS model includes
counter-charges. Removing counter-charges from the model creates a
landscape of electrostatic potential energy that tends to exclude the
\SIV charged region from the membrane region --- a large barrier
develops from charge induced on the bath interface by buried,
unbalanced \SIV charges. This happens in the presence of deep gating
pores that reduce the number of buried \SIV charges to no more than
three. My computations show that much of the observed consequences of
neutralization mutants on \SII, \SIII, and \SIV can be understood on
electrostatic grounds as resulting from the charge/counter-charge
interactions among the buried residues and the charges that they
induce on dielectric boundaries. The computational results for a
reduced electrostatic model suggest that the charge/counter-charge
interactions in the \VS warrant detailed investigation.

\balance
\paragraph{Electrostatics dominate.} The fact that a model solely including
electrostatic interactions reproduces many experimental phenomena
reveals the primary importance of electrostatics in the \VS
system. Reduced models focused on electrostatic interactions can
thus provide essential insights into experimental behavior.

My simulations assemble a handful of positive and negative charges,
enclose them in a reflective dielectric boundary, reshuffle their
configurations with an external field and provide an experimental
window for viewing. The result is a kaleidoscope. A ``simple''
electrostatic system produces complex phenomena which appear to be
more easily understood by constructing the mechanism of the
kaleidoscope than by inferring the mechanism via induction from the
phenomena viewed through it.

\newpage\nobalance
\appendix

\bibliography{all}

\end{document}